\documentclass[acmtog,nonacm=true]{acmart}

\usepackage{booktabs} %
\usepackage{xcolor}
\usepackage{comment}
\usepackage{graphicx} %
\usepackage{caption}
\usepackage{subcaption}
\usepackage{tikz}
\usetikzlibrary{arrows.meta}
\usepackage{adjustbox}
\usepackage{overpic}
\usepackage{multirow}
\usepackage[normalem]{ulem}
\usepackage{array}

\usepackage[ruled]{algorithm2e} %

\SetAlFnt{\small}
\SetAlCapFnt{\small}
\SetAlCapNameFnt{\small}
\SetAlCapHSkip{0pt}

\newcommand{\bias}{w}
\newcommand{\prompt}[1]{\textit{``#1''}}

\newcommand{\matQ}{\mathbf{Q}}
\newcommand{\matK}{\mathbf{K}}
\newcommand{\matV}{\mathbf{V}}
\newcommand{\matB}{\mathbf{B}}

\acmJournal{TOG}

\begin{document}
\title{Generative Detail Enhancement for Physically Based Materials}

\author{Saeed Hadadan}
\affiliation{%
  \institution{University of Maryland, College Park,}
  \institution{NVIDIA}  
  \state{MD}
  \postcode{20740}
  \country{USA}}
\email{saeedhd@umd.edu}

\author{Benedikt Bitterli}
\affiliation{%
 \institution{NVIDIA}
 \state{WA}
 \country{USA}}
\email{bbitterli@nvidia.com}

\author{Tizian Zeltner}
\affiliation{%
 \institution{NVIDIA}
 \country{Switzerland}}
\email{tzeltner@nvidia.com}

\author{Jan Novák}
\affiliation{%
 \institution{NVIDIA}
 \country{Czech Republic}}
\email{jnovak@nvidia.com}

\author{Fabrice Rousselle}
\affiliation{%
 \institution{NVIDIA}
 \country{Switzerland}}
\email{frousselle@nvidia.com}

\author{Jacob Munkberg}
\affiliation{%
 \institution{NVIDIA}
 \country{Sweden}}
\email{frousselle@nvidia.com}

\author{Jon Hasselgren}
\affiliation{%
 \institution{NVIDIA}
 \country{Sweden}}
\email{frousselle@nvidia.com}

\author{Bartlomiej Wronski}
\affiliation{%
 \institution{NVIDIA}
 \country{USA}}
\email{bwronski@nvidia.com}

\author{Matthias Zwicker}
\affiliation{%
 \institution{University of Maryland, College Park}
 \state{MD}
 \country{USA}}
\email{zwicker@cs.umd.edu}

\begin{abstract}
    We present a tool for enhancing the detail of physically based materials using an off-the-shelf diffusion model and inverse rendering.
    Our goal is to increase the visual fidelity of existing materials by adding, for instance, signs of wear, aging, and weathering that are tedious to author.
    To obtain realistic appearance with minimal user effort, we leverage a generative image model trained on a large dataset of natural images.
    Given the geometry, UV mapping, and basic appearance of an object, we proceed as follows:
    We render multiple views of the object and use them, together with an appearance-defining text prompt, to condition a diffusion model. 
    The generated details are then backpropagated from the enhanced images to the material parameters via inverse rendering.
    For inverse rendering to be successful, the generated appearance has to be consistent across all the images. 
    We propose two priors to address the multi-view consistency of the diffusion model. First, we ensure that the noise that seeds the diffusion process is itself consistent across views by integrating it from a view-independent UV space.
    Second, we enforce spatial consistency by biasing the attention mechanism via a projective constraint so that pixels attend strongly to their corresponding pixel locations in other views. 
    Our approach does not require any training or finetuning of the diffusion model, is agnostic to the used material model, and the enhanced material properties, i.e., 2D PBR textures, can be further edited by artists. We demonstrate prompt-based material edits exhibiting high levels of realism and detail.
\end{abstract}

\begin{teaserfigure}
  \centering
  \setlength{\tabcolsep}{0.0016\textwidth}%
  \renewcommand{\arraystretch}{0.5}%
  \footnotesize
  \begin{tabular}{cccccc}
    \multicolumn{2}{c}{(\textbf{a}) Initial 3D asset}&
    \multicolumn{2}{c}{2D diffusion w/ multi-view visual prompting, ``... ancient Greek vase ...''}&
    \multicolumn{2}{c}{(\textbf{d}) Enhanced 3D asset}\\[-1.4pt]
    \cmidrule(lr){1-2}
    \cmidrule(lr){3-4}
    \cmidrule(lr){5-6}\\[-5.5pt]
    Textures & Renderings & (\textbf{b}) Baseline & (\textbf{c}) Ours & Renderings & Textures\\
    \begin{tabular}{c}
      \includegraphics[width=0.092\textwidth]{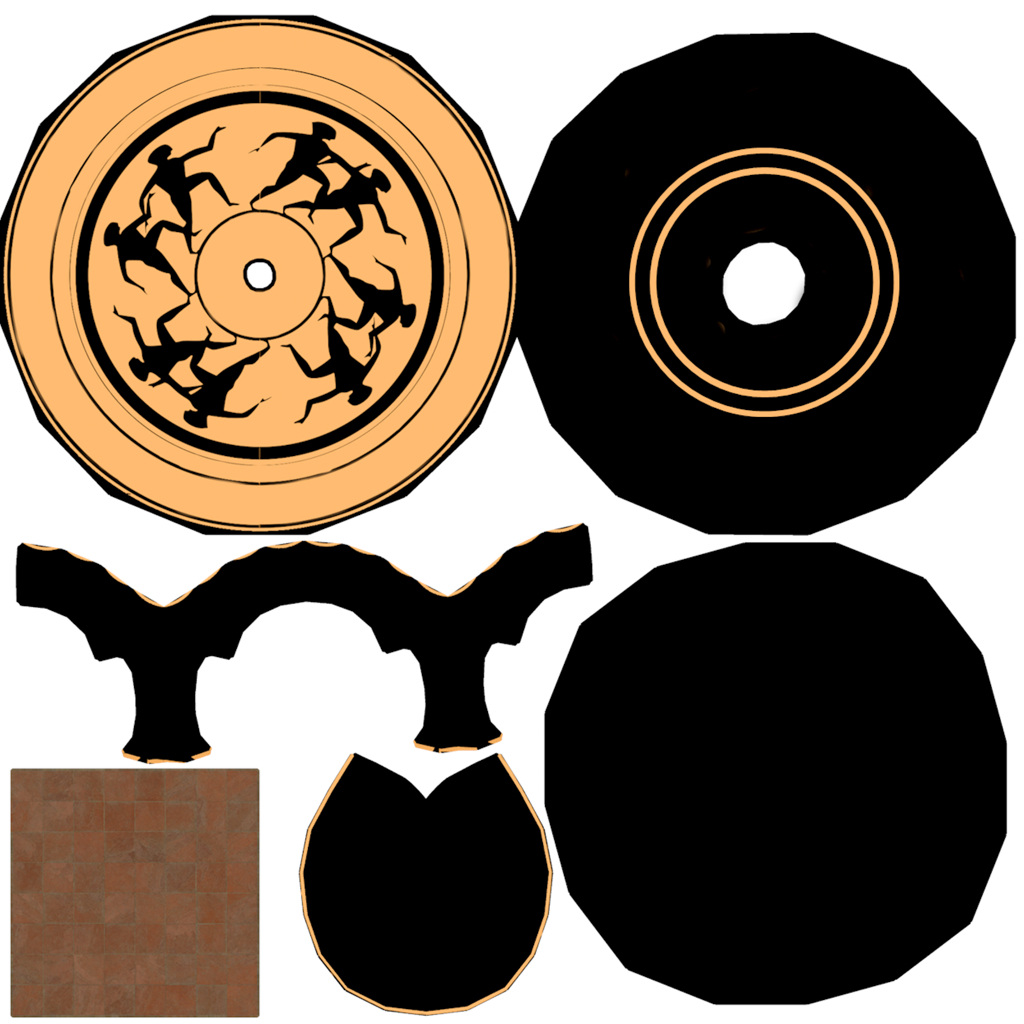}\\
      \includegraphics[width=0.092\textwidth]{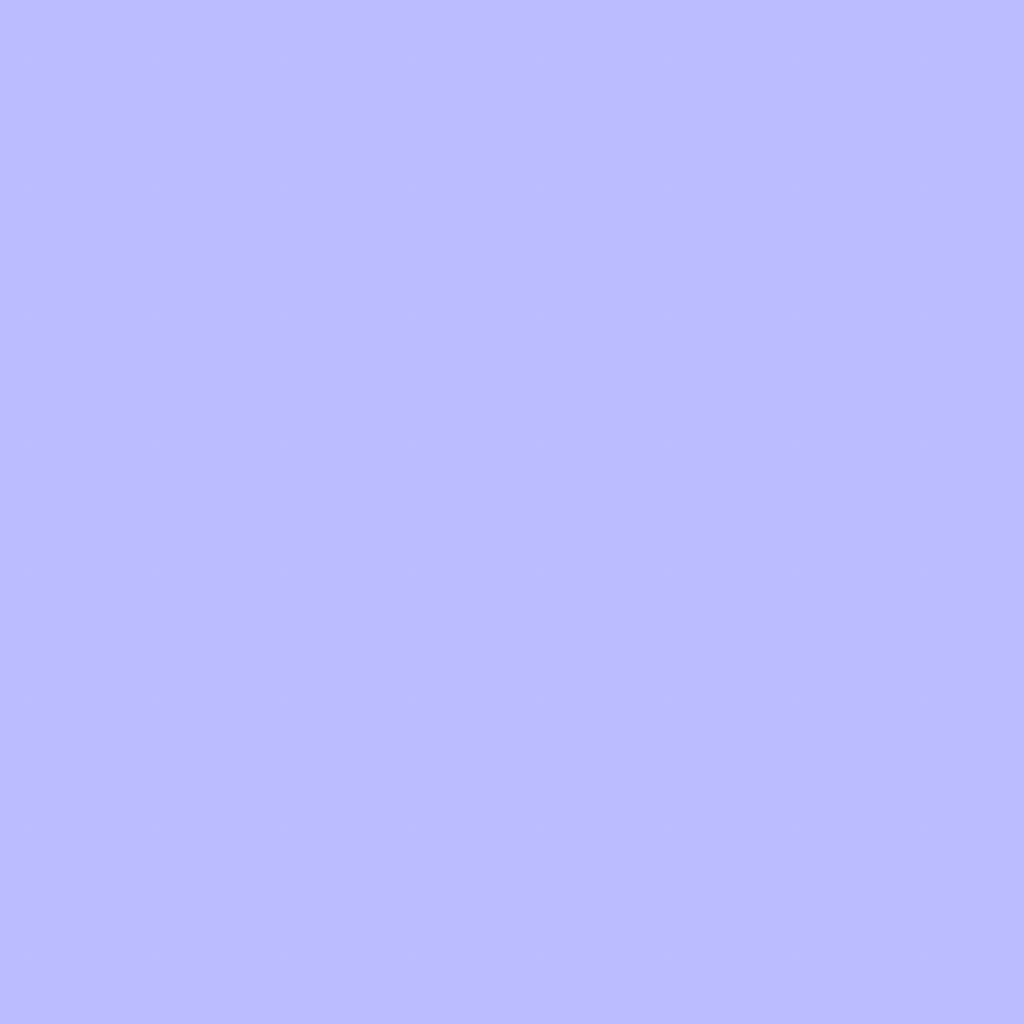}\\
      \includegraphics[width=0.092\textwidth]{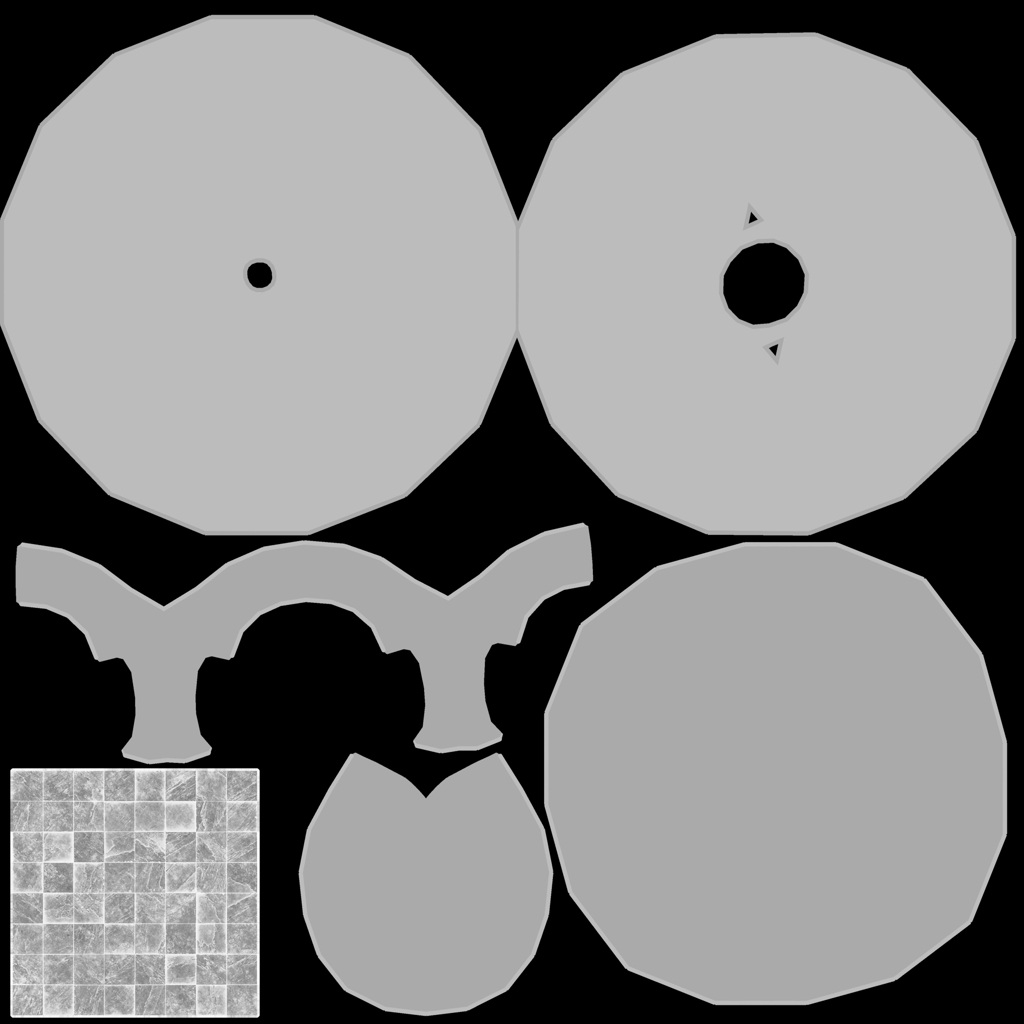}
    \end{tabular}
    &
    \begin{tabular}{cc}
      \begin{overpic}[width=0.092\textwidth, trim=210 0 190 0, clip]{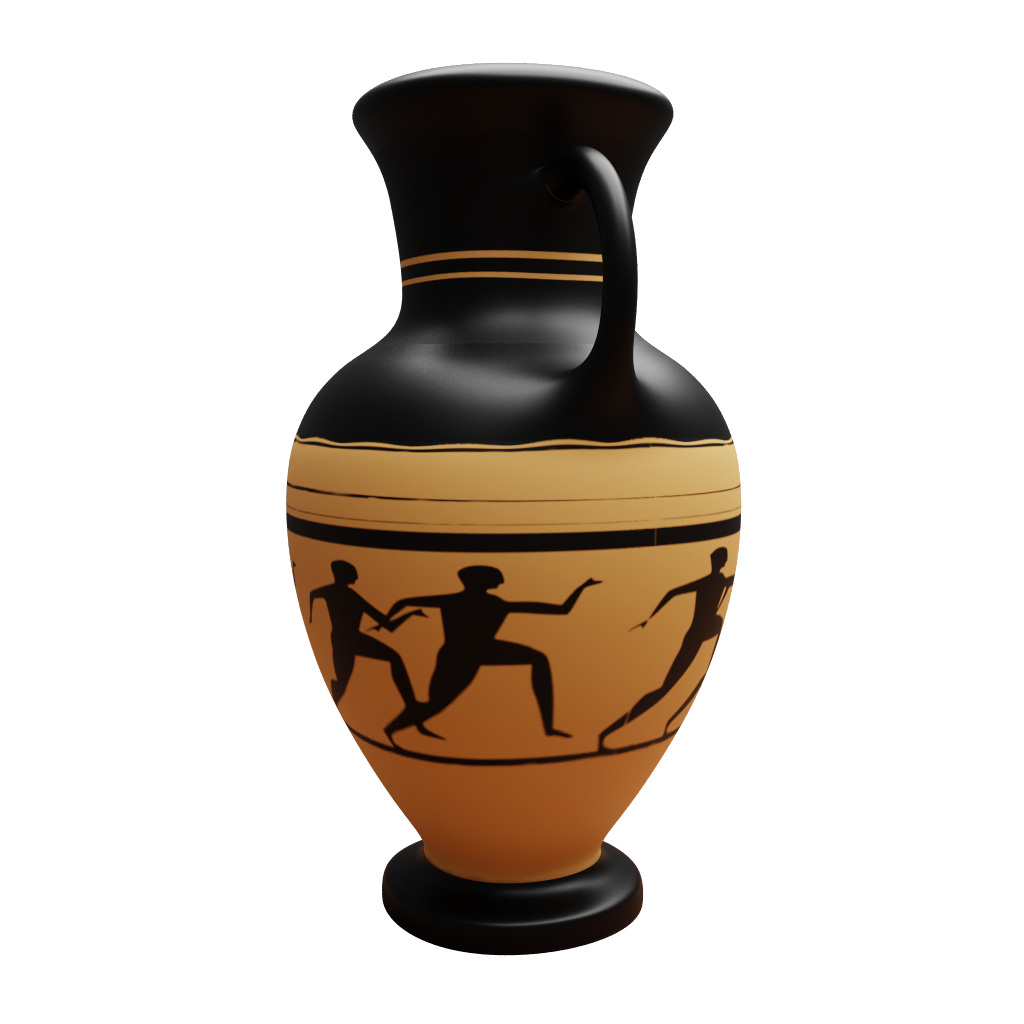}
      \end{overpic}
      &
      \begin{overpic}[width=0.092\textwidth, trim=210 0 190 0, clip]{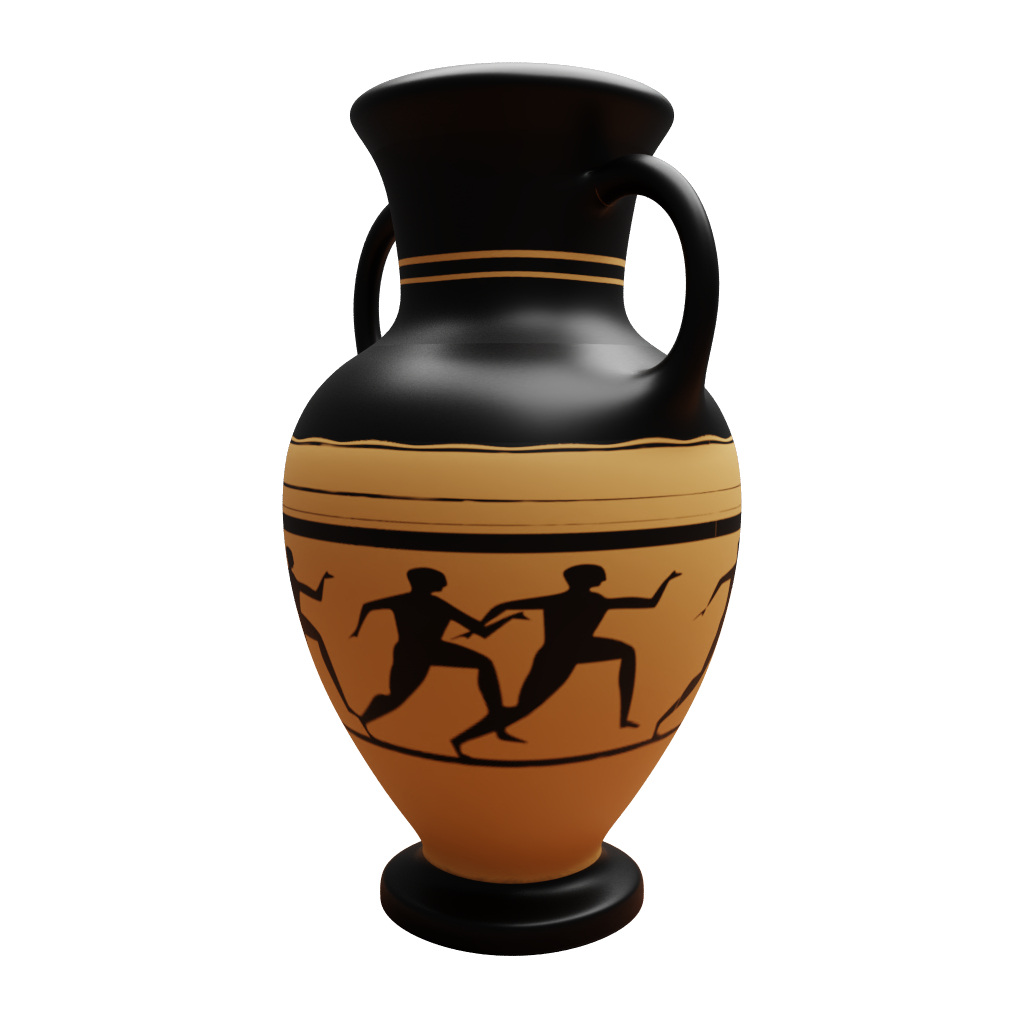}
      \end{overpic}
      \\
      \begin{overpic}[width=0.092\textwidth, trim=410 370 390 340, clip]{figures/results/greekvase/masked/initial_rendering_004.jpg}
      \end{overpic}
      &
      \begin{overpic}[width=0.092\textwidth, trim=500 370 300 340, clip]{figures/results/greekvase/masked/initial_rendering_005.jpg}
      \end{overpic}
    \end{tabular}
    &
    \begin{tabular}{cc}
      \begin{overpic}[width=0.092\textwidth, trim=210 0 190 0, clip]{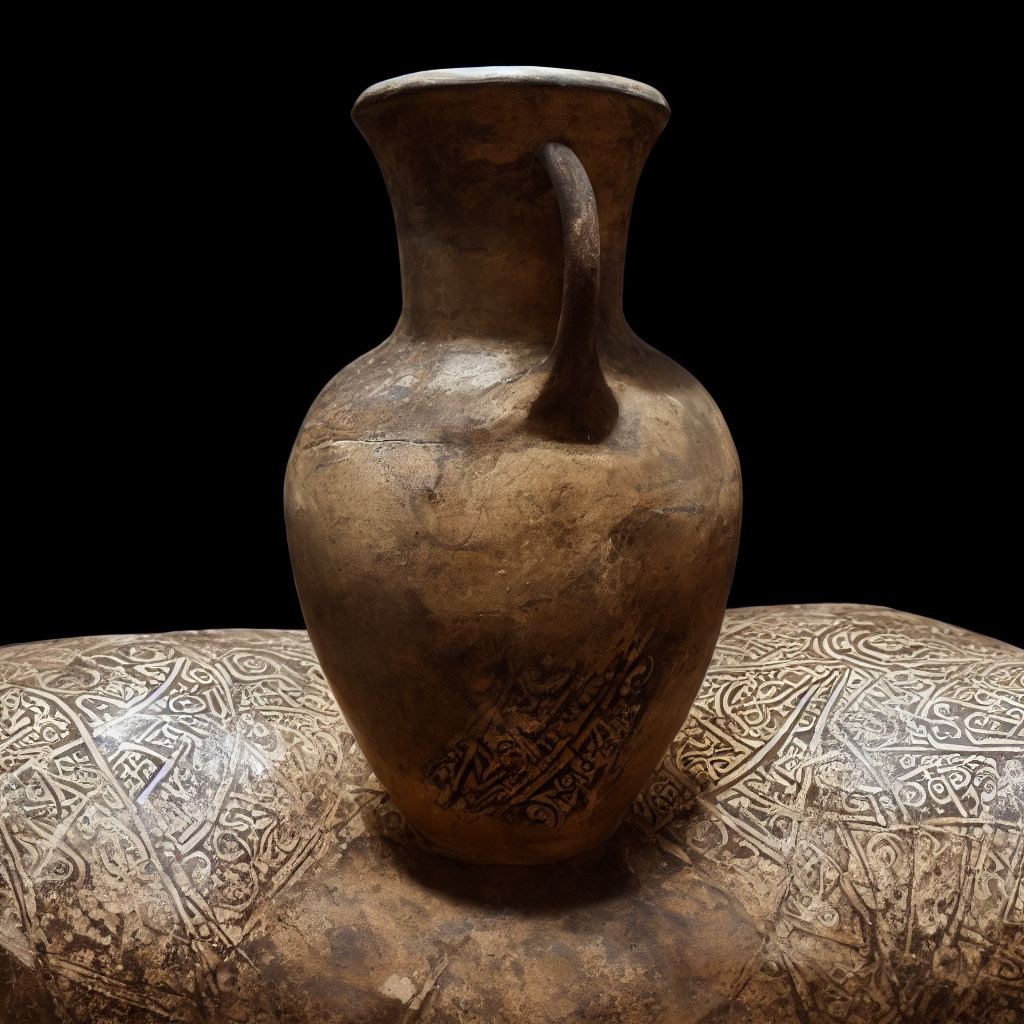}
      \end{overpic}
      &
      \begin{overpic}[width=0.092\textwidth, trim=210 0 190 0, clip]{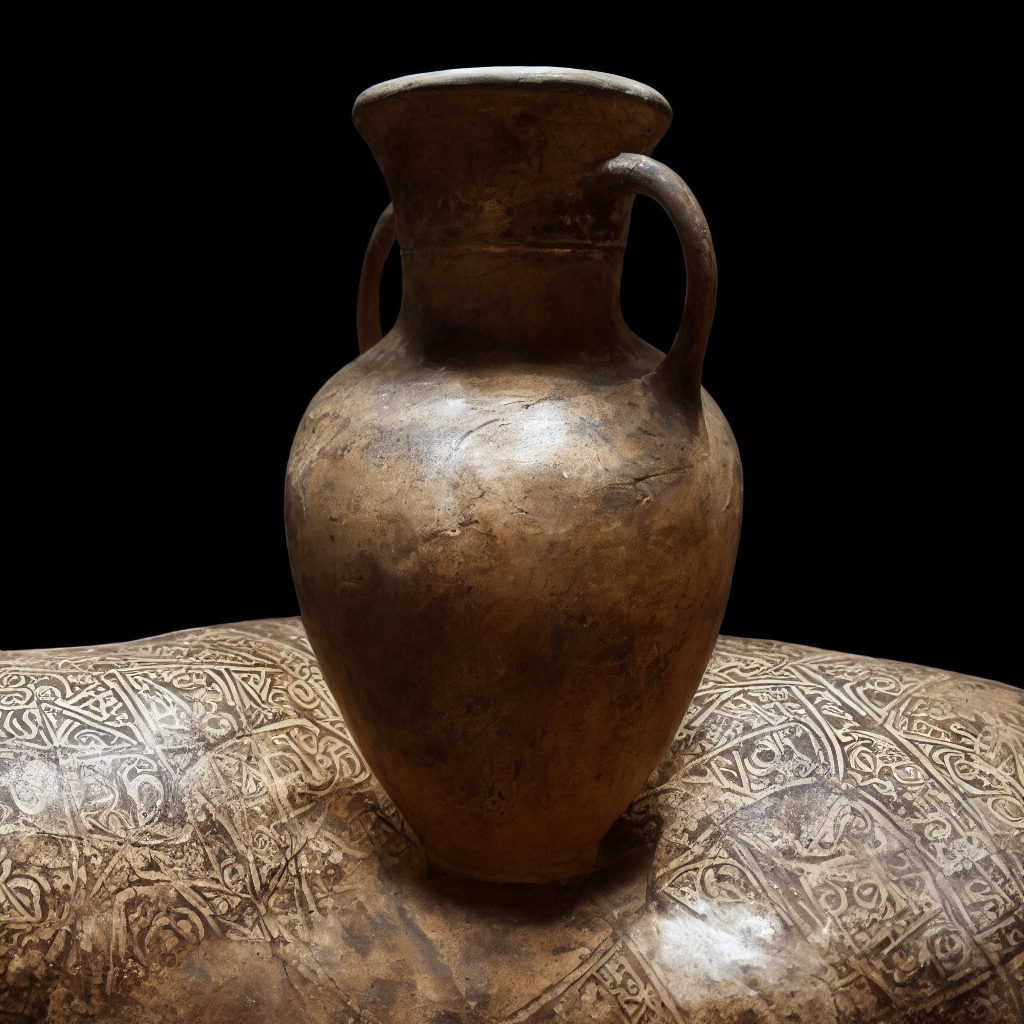}
      \end{overpic}
      \\
      \begin{overpic}[width=0.092\textwidth, trim=410 370 390 340, clip]{figures/results/greekbase/0000.png}
      \end{overpic}
      &
      \begin{overpic}[width=0.092\textwidth, trim=500 370 300 340, clip]{figures/results/greekbase/0001.png}
      \end{overpic}
    \end{tabular}
    &
    \begin{tabular}{cc}
      \begin{overpic}[width=0.092\textwidth, trim=210 0 190 0, clip]{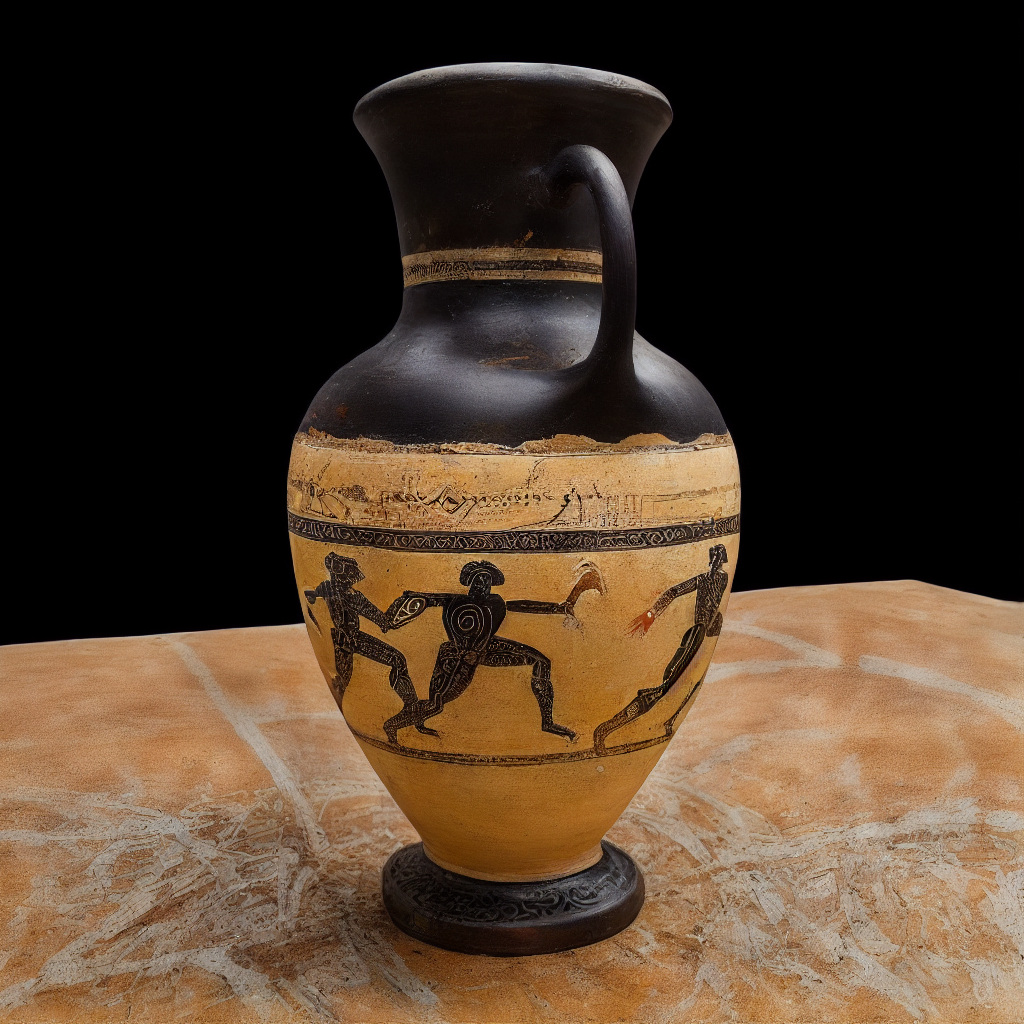}
      \end{overpic}
      &
      \begin{overpic}[width=0.092\textwidth, trim=210 0 190 0, clip]{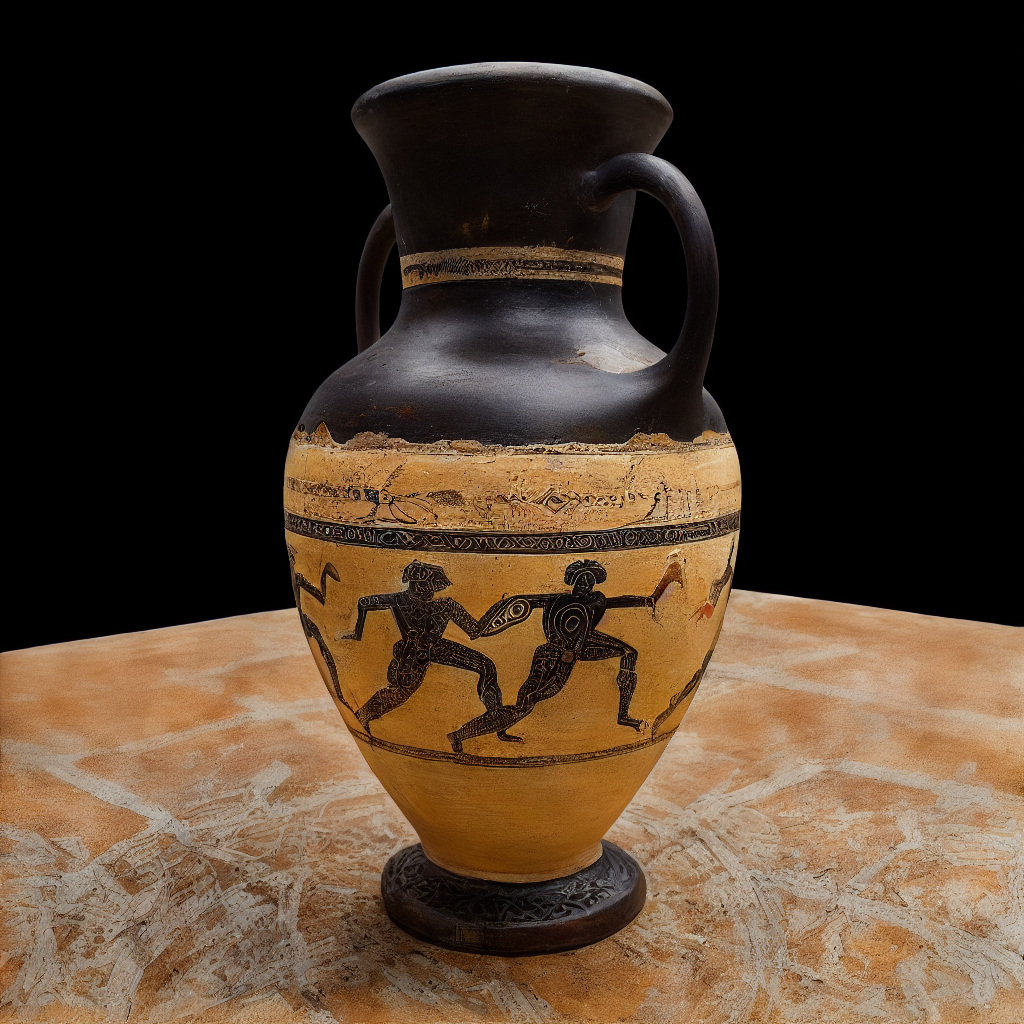}
      \end{overpic}
      \\
      \begin{overpic}[width=0.092\textwidth, trim=410 370 390 340, clip]{figures/results/greekvase/diffused_004.jpg}
      \end{overpic}
      &
      \begin{overpic}[width=0.092\textwidth, trim=500 370 300 340, clip]{figures/results/greekvase/diffused_005.jpg}
      \end{overpic}
    \end{tabular}
    &
    \begin{tabular}{cc}
      \begin{overpic}[width=0.092\textwidth, trim=210 0 190 0, clip]{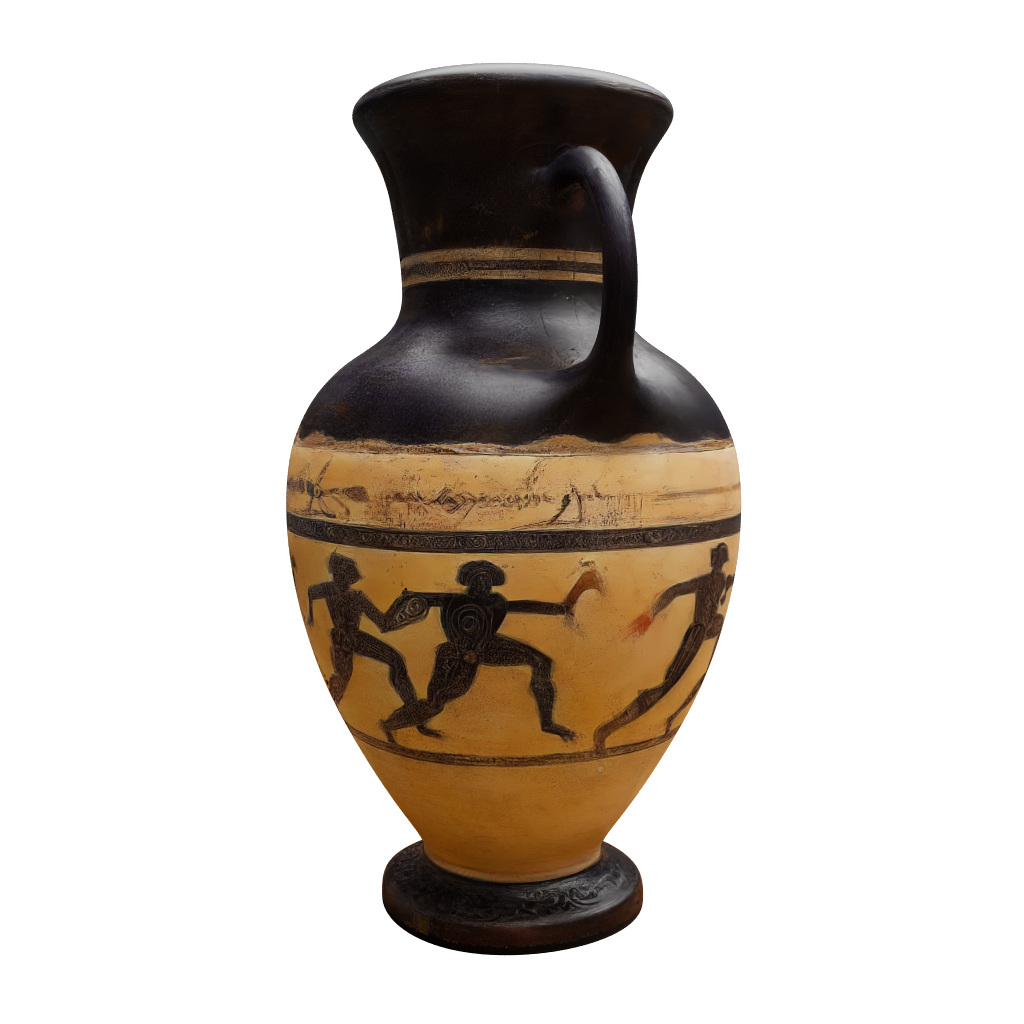}
      \end{overpic}
      &
      \begin{overpic}[width=0.092\textwidth, trim=210 0 190 0, clip]{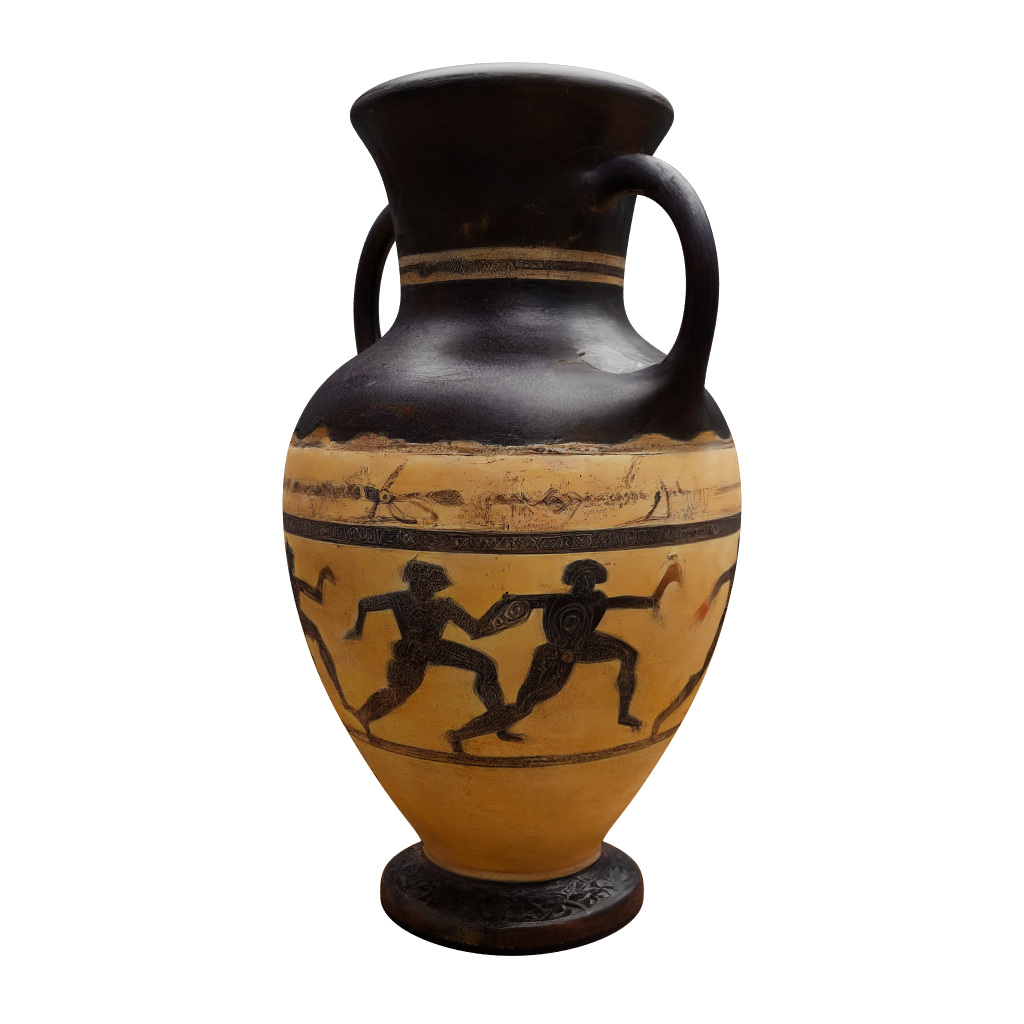}
      \end{overpic}
      \\
      \begin{overpic}[width=0.092\textwidth, trim=410 370 390 340, clip]{figures/results/greekvase/masked/recovered_default_004.jpg}
      \end{overpic}
      &
      \begin{overpic}[width=0.092\textwidth, trim=500 370 300 340, clip]{figures/results/greekvase/masked/recovered_default_005.jpg}
      \end{overpic}
    \end{tabular}
    &
    \begin{tabular}{c}
      \includegraphics[width=0.092\textwidth]{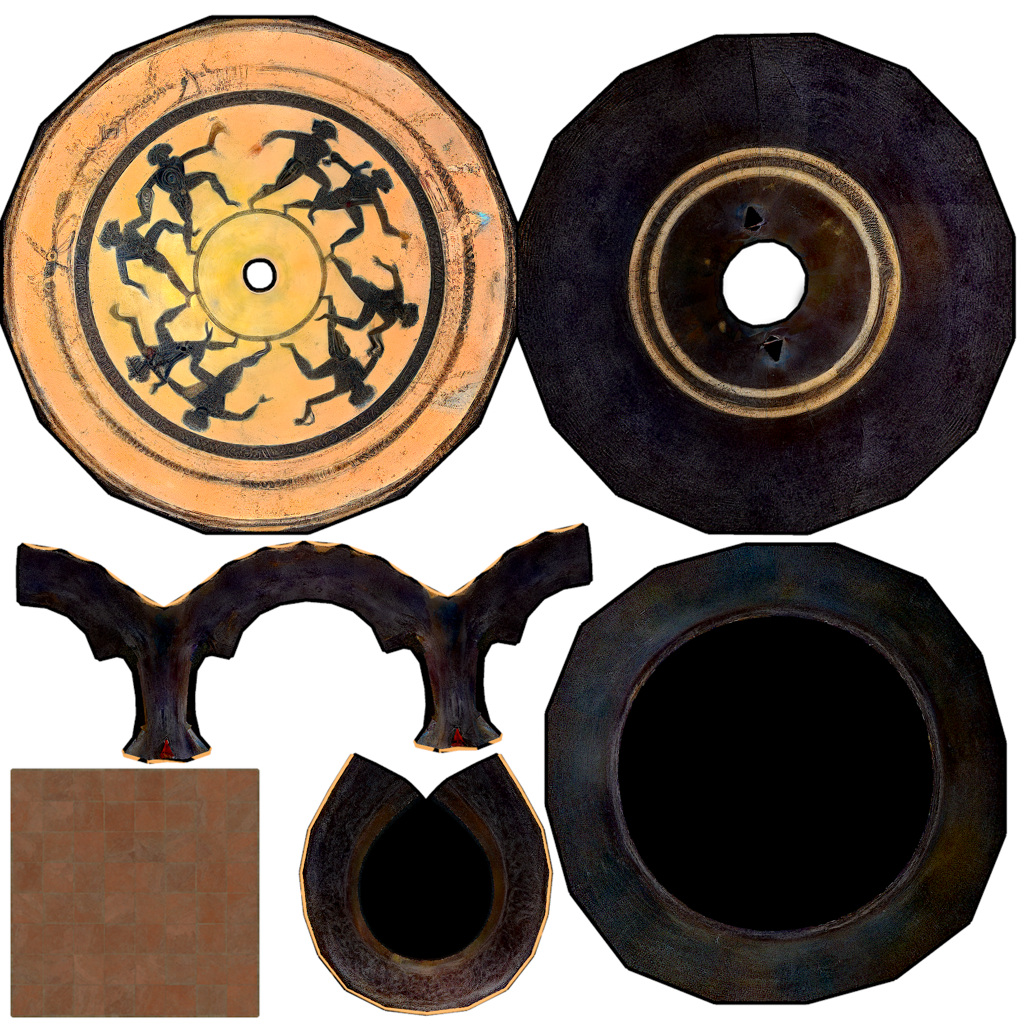}\\
      \includegraphics[width=0.092\textwidth]{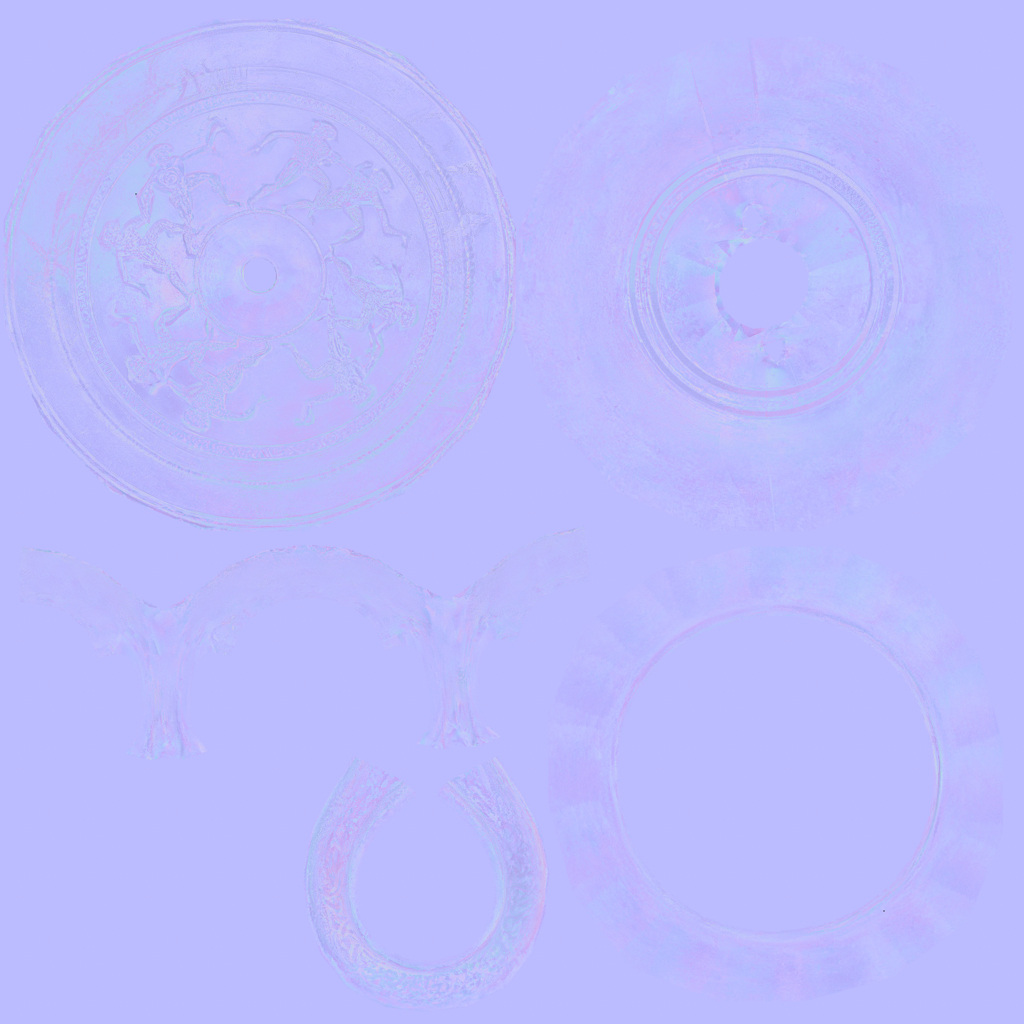}\\
      \includegraphics[width=0.092\textwidth]{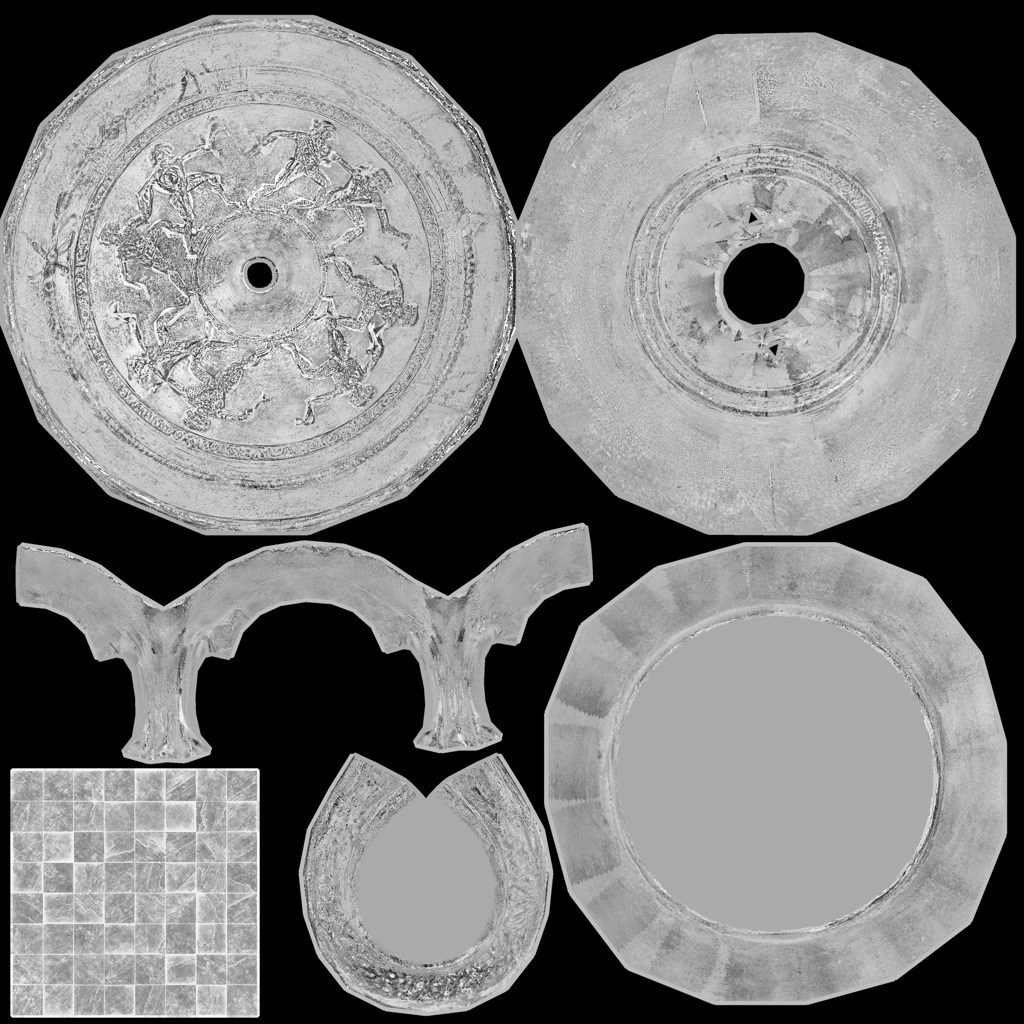}
    \end{tabular}    
  \end{tabular}
  \caption{
    We enhance material definitions of existing 3D assets (\textbf{a}) by applying effects specified by text prompts such as aging, weathering, etc. 
    Conditioned on a set of renderings, we synthesize the corresponding visuals in 2D using a diffusion model building on multi-view visual prompting~\cite{flashtex} (\textbf{b}). We improve the multi-view consistency of the generator using two key additions---view-correlated noise and attention biasing (\textbf{c})---that enable succesful inverse-rendering of the visual enhancements back to the original material textures (\textbf{d}).
  }
  \vspace*{5mm}
  \label{fig:teaser}
\end{teaserfigure}


\maketitle

\section{Introduction}
\label{sec:intro}

Depicting rich 3D worlds is a driving goal of computer graphics.
While achieving this goal is possible today for experienced artists with expert tools, the pareto principle applies:
The creative step of authoring the overall look of an asset takes little time in comparison to the disproportionate effort of infusing details and imperfections of the real world.
Our goal is therefore to create a tool that enhances 3D objects with appearance details requiring comparatively minimal effort from the artist.

For this, we turn to diffusion models~\cite{ho2020diffusion} that are capable of producing realistic visuals and can be conditioned using text prompts and guiding images.
A key consideration, however, is the amount of training data available.
While datasets containing 3D objects and materials exist~\cite{objaverse,vecchio2024matsynth}, they cannot compete with natural image datasets in size and diversity, which directly impacts the model capabilities.
We therefore build our tool using an off-the-shelf diffusion model that was trained on an internet-scale image set.

We combine the diffusion model with a physically based renderer to enable two key editing features: 1) specifying the initial look of the object, and 2) outputing a material representation that is complient with traditional authoring workflows.
Our algorithm works as follows.
We start by rendering the original 3D asset from multiple views.
Then we condition the diffusion model on a concatenation of these views, and a text prompt describing the desired detail enhancements.
Since our goal is to merely enhance the appearance, we propose a specific way of using two 
publicly available ControlNets~\cite{controlnet} to condition the model on the asset geometry and initial appearance.
Finally, the differences between the original renderings and the diffusion-generated views are back-propagated to the material parameters via inverse rendering.

The main challenge of multi-view generation with diffusion models is the consistency of individual details in all relevant views.
We address this challenge with two contributions. First, we seed the diffusion model with noise that is itself consistent across the views. 
We adopt the idea of integral noise~\cite{chang2024how} and project a common UV-space noise pattern into each view in a variance-preserving manner.
Second, we bias the attention maps in the diffusion model to encourage pixels to attend to their corresponding locations in different views.
We compute the correspondences by reprojecting points between the views using ray tracing operations.

Our approach has several benefits.
It avoids the impractical creation of a new dataset and/or retraining a large diffusion model; we rely on an off-the-shelf model that can be easily replaced by another one.
We preserve the original geometry and artistic intent, while modifying the material texture maps according to the user's target prompts. 
Because we only enhance the input material, the inverse rendering is more likely to succeed than if starting the optimization from scratch.
Lastly, the input and output of our model are in the form of a classical 3D representation (e.g, triangles, textures) and thus perfectly multi-view consistent. This also allows users to further edit the appearance, integrate it into larger scenes, and render with common renderers.

\section{Prior work}
\label{sec:prior}

\begin{figure*}[t]
    \begin{overpic}[width=0.98\textwidth]{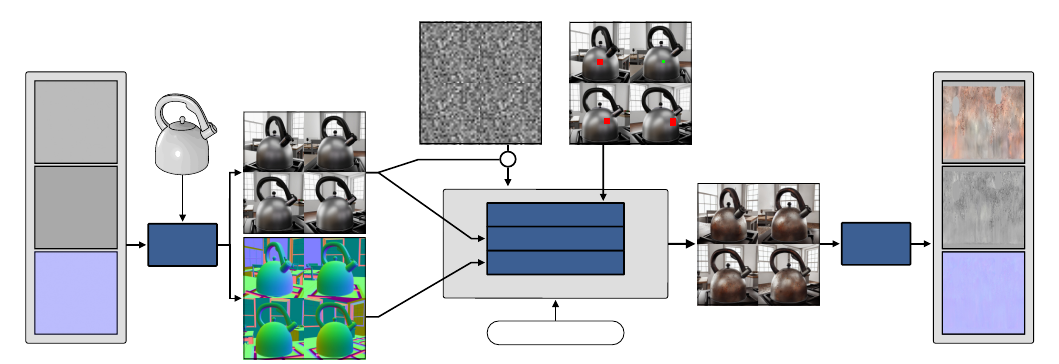}
        \footnotesize
        \put(11.0,27){   \begin{minipage}{2cm}\centering Fixed\\geometry             \end{minipage}}
        \put(1.1,29.5){  \begin{minipage}{2cm}\centering Initial material\\textures  \end{minipage}}
        \put(86.8,29.5){ \begin{minipage}{2cm}\centering Edited material\\textures   \end{minipage}}
        \put(20.0,25.5){ \begin{minipage}{3cm}\centering Rendered color \&\\normal grids\end{minipage}}
        \put(62.7,18){ \begin{minipage}{3cm}\centering Generated images              \end{minipage}}
        \put(43.7,0.4){  \begin{minipage}{3cm}\centering Text prompt                 \end{minipage}}
        \put(43.7,2.4){  \begin{minipage}{3cm}\centering \prompt{a rusty kettle}     \end{minipage}}
        \put(11.3,10.8){   \begin{minipage}{2cm}\centering \textcolor{white}{\textbf{Render}}          \end{minipage}}
        \put(76.7,10.8){  \begin{minipage}{2cm}\centering \textcolor{white}{\textbf{Inverse\\render}} \end{minipage}}
        \put(36.4,33.8){ \begin{minipage}{3cm}\centering View-correlated noise\\(Section \ref{sec:noise_warping}) \end{minipage}}
        \put(50.7,33.8){ \begin{minipage}{3cm}\centering Attention bias\\(Section \ref{sec:attention_bias})  \end{minipage}}
        \put(43.7,13.7){  \begin{minipage}{3cm}\centering \textcolor{white}{\textbf{Stable Diffusion}}\end{minipage}}
        \put(43.7,11.4){  \begin{minipage}{3cm}\centering \textcolor{white}{\textbf{ControlNet tile}}\end{minipage}}
        \put(43.7,9.1){  \begin{minipage}{3cm}\centering \textcolor{white}{\textbf{ControlNet normal}}\end{minipage}}
        \put(43.7,6.8){ \begin{minipage}{3cm}\centering Diffusion model (Section \ref{sec:second-controlnet}) \end{minipage}}
    \end{overpic}
    \vspace{-2mm}
    \caption{Pipeline overview: Given a 3D asset including fixed geometry and initial material textures, we render color and normal images from multiple viewpoints (4 out of 16 views shown above). We then apply enhancements based on text prompts using a multi-view diffusion model designed to produce view-consistent outputs that edit the input images in a controllable manner. We achieve this by leveraging three distinct techniques, including suitable publicly available ControlNets, view-correlated noise, and cross-view attention bias. We finally obtain the edited material textures using inverse rendering. }
    \label{fig:overview}
\end{figure*}
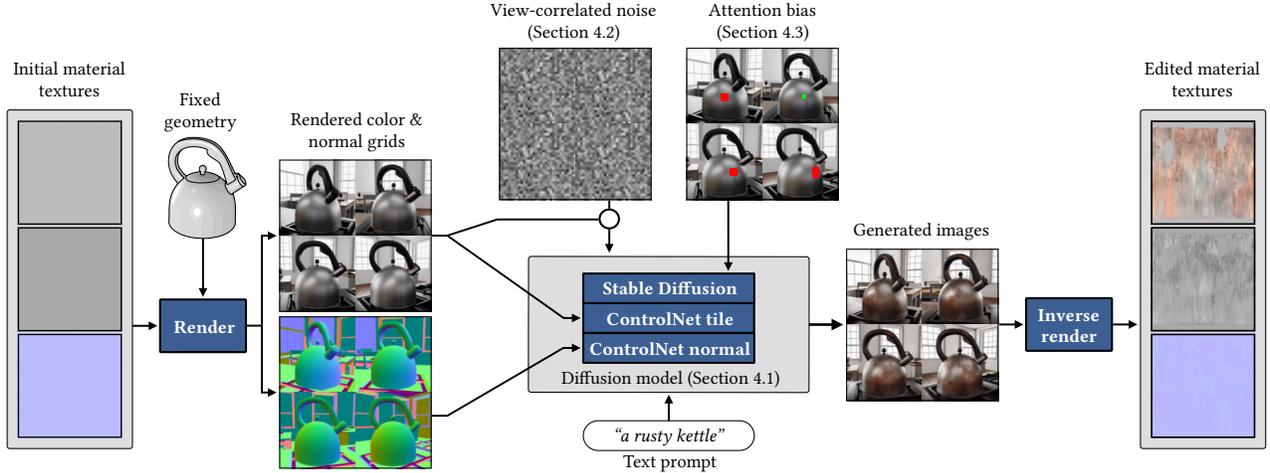

For brevity, we only cover methods directly related to enforcing multi-view consistency in diffusion denoisers and to editing of physically based materials using generative models.

\paragraph{Multi-view consistency via latents sharing}
TexPainter \cite{TexPainter} denoises multi-view latents, and correlates the views in a shared texture space.
Tex4D~\cite{bao2024tex4d} extends the idea to the temporal domain using a video diffusion model.
These methods however do not easily generalize to view dependent PBR materials.
\citet{spiderman} and \citet{diffusion_handles} instead operate on intermediate features of the network to enforce 3D consistent transformations in the output images.

Another approach is to reuse the diffusion model's input noise accross views as proposed by~\cite{chang2024how,daras2024warped}.
Our work extends this idea by anchoring the noise field in UV space for a more robust handling of disocclusions.

\paragraph{Multi-view consistency via view correspondences}
Cross-frame attention modules have been devised for known depth maps~\cite{mvdiffusion}, poses~\cite{cerkezi2023multiview}, or epipolar constraints~\cite{spad}.
These methods however require large-scale training.
Our method exploits known geometry and is training-free.

\paragraph{Text-guided 3D generation}
\citet{poole2023dreamfusion} pioneered generation of 3D models 
using text-to-image diffusion and score distillation sampling (SDS).
The method has been extended to various representations~\cite{lin2023magic3d, taoran2023gaussiandreamer}, 
with improved objective functions~\cite{wang2023prolificdreamer}, 
sampling~\cite{zhu2023hifa}, and material decomposition~\cite{chen2023fantasia3d,youwang2024paintit}.

Early methods~\cite{richardson2023texture,chen2023text2tex,cao2023textfusion} suffer from over-blurring due to the lack of view consistency. 
Follow-up work improved this using spatial attention~\cite{shi2023MVDream}, video-models~\cite{voleti2024sv3d, wu2024cat4d}, and tiled inputs~\cite{flashtex}.
We use the latter idea in our work.

FlashTex~\cite{flashtex}, DreamMat~\cite{dreammat}, and MaPa~\cite{mapa} specialize in material reconstruction for a 
known scene. They leverage known priors by training a controlnet~\cite{controlnet} from geometry buffers 
(e.g. depth and normal), and lighting rendered with known, constant, materials (e.g. fully diffuse and specular). 
This helps greatly with view dependent shading effects, and separating shadows from material albedo. 
\citet{Vecchio2024controlmat} train a generative model to directly synthesize material maps.
All these methods are costly as they require training with specialized object/material datasets, which we avoid.

\paragraph{Image and appearance editing}
Text-guided diffusion models are often applied to image editing while preserving the semantics of the source image.
This is achieved by manipulating the self-attention layers 
\cite{tumanyan2023pnp}, often combined with DDIM inversion~\cite{mokady2023nulltext,parmar2023pix2pixzero}.
$\text{RGB}{\leftrightarrow}\text{X}$~\cite{RGBX} allows editing the content of images by first extracting irradiance and material maps, manually editing them, and then generating the corresponding realistic image.
We consider a dual problem, where the inputs and outputs are PBR maps, and the intermediate step that enhances details involves generation of an image.

\paragraph{Material upscaling}
\citet{gauthier2024matup} consider a subset of detail enhancement, focusing on increasing the resolution of PBR material textures by inverse rendering upscaled images.
However, they operate on flat-geometry renderings and are therefore unable to synthesize detail in the context of the object geometry and the surrounding scene; this is one of our goals.

\paragraph{Video diffusion models}
Video diffusion models~\cite{blattmann2023svd,hong2023cogvideo,yang2024cogvideox} generate (view-consistent) video from text and image conditioning. Notably, SV3D~\cite{voleti2024sv3d} adapts image-to-video diffusion model for novel multi-view synthesis and 3D generation. However, these models come at significant computational cost, and the generated frames are typically not as detailed as text-to-image models.

\section{Background}
\label{sec:background}

\paragraph{Diffusion models.} Denoising diffusion leverages a bidirectional process
where the forward pass gradually corrupts training data by iteratively adding Gaussian noise until the data becomes pure noise.
The reverse process then learns to denoise the corrupted data through a neural network, which predicts and removes noise step-by-step.
We use a latent space diffusion model~\cite{rombach2022stablediffusion}, which extracts the latent space using a variational autoencoder and performs the denoising steps with a U-Net architecture operating at different scales (see Figure 3 in~\citet{rombach2022stablediffusion}).

\paragraph{ControlNet.} In order to guide the denoising process, the ControlNet model~\cite{controlnet} implements a dual-network architecture.
The first network---a pretrained diffusion model---is locked down to perform the usual denoising task and cloned.
The clone network is connected to the locked network using zero-initialized convolution layers,
and enables precise spatial conditioning of the pretrained denoiser using images.

\paragraph{Attention.}
The attention operation~\cite{Vaswani2017attention} has been incorporated to diffusion models to capture relationships between different activations in the denoising process.
The operation takes its inputs and embeds them in three learned linear spaces, referring to them as key, query, and value.
Denoting the matrices of these embeddings $\matQ$, $\matK$, and $\matV$, respectively
the attention formula
\[
\text{Attention}(\matQ, \matK, \matV) = \text{softmax}\left(\frac{\matQ \matK^\top}{\sqrt{d_k}}\right)\matV,
\]
provides a mechanism for capturing the similarity between queries and keys (where $d_k$ is the dimensionality of the key embedding).

A typical U-Net diffusion model features two types of attention:
\emph{cross-attention} layers that guide the denoising of image regions using a given text-prompt,
and \emph{self-attention} layers that allow regions within the image to influence each other.
In the self-attention module, the $\matQ \matK^\top$ product forms a large attention score matrix,
where entry $[i,j]$ describes how strongly region $i$ attends to region $j$.

\section{Method}
\label{sec:method}

Our method (see \autoref{fig:overview}) comprises three stages: forward rendering, detail generation, and inverse rendering discussed below. We then present our three technical contributions in Sections~\ref{sec:second-controlnet} to~\ref{sec:attention_bias}.

\paragraph{Forward rendering.}
We begin with a user-provided asset comprising of known geometry and the material to be enhanced, as well as a 3D scene providing context for the asset.
We render the scene from a small number of views (between 9 and 16 views in practice) in an orbit around the asset.
The output of the renderer consists of the renderings as well as auxiliary buffers of surface normals, which serve to condition the diffusion model during the next stage.

\paragraph{Detail generation.}

The renderings from the previous stage are passed to an off-the-shelf diffusion model to add detail to the renderings, conditioned on a text prompt and the auxiliary buffers. Although the rendered views could be enhanced individually, we find that mutual view consistency is improved if we use \emph{multi-view visual prompting}~\cite{flashtex}, in which we concatenate all views into a grid (e.g. $3\times 3$ or $4\times 4$) and enhance them simultaneously.

While multi-view prompting improves consistency across the views at a coarse scale, there remains enough variation in fine-scale detail between views to make reconstruction of highly detailed materials challenging.
To remedy this, we propose two modifications to the diffusion model:
using view-correlated input noise that is anchored in 3D space (Section~\ref{sec:noise_warping}) 
and biasing the attention layers with pixel-to-pixel correspondence information (Section~\ref{sec:attention_bias}).

\paragraph{Inverse rendering.}
We finally propagate the detail generated by the diffusion model back to the original material of the 3D asset, leveraging a differentiable renderer~\cite{mitsuba} to minimize the difference between the enhanced views and the rendered material in a stochastic gradient optimization.
In all our results, we optimize the spatially-varying albedo, normal, and roughness textures of a typical PBR material~\cite{Burley2012}, but any differentiable material definition can be used in principle.
We initialize the optimization state with the original textures; this improves the convergence likelihood. %

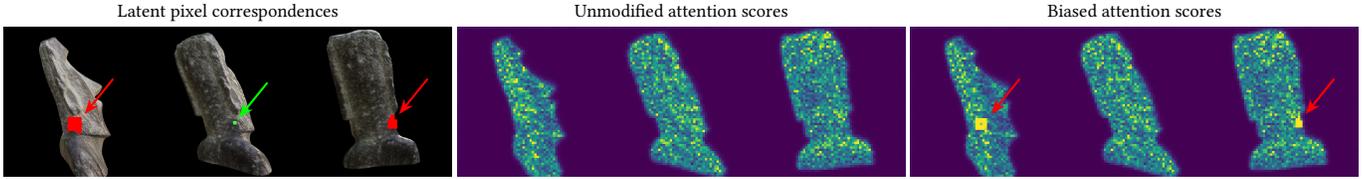
\begin{figure*}[htbp]
    \centering%
    \setlength{\tabcolsep}{0.002\textwidth}%
    \renewcommand{\arraystretch}{1}%
    \footnotesize%
    \begin{tabular}{ccc}
        Latent pixel correspondences
        &
        Unmodified attention scores
        &
        Biased attention scores
        \\[0.8mm]
        \begin{overpic}[width=0.332\textwidth]{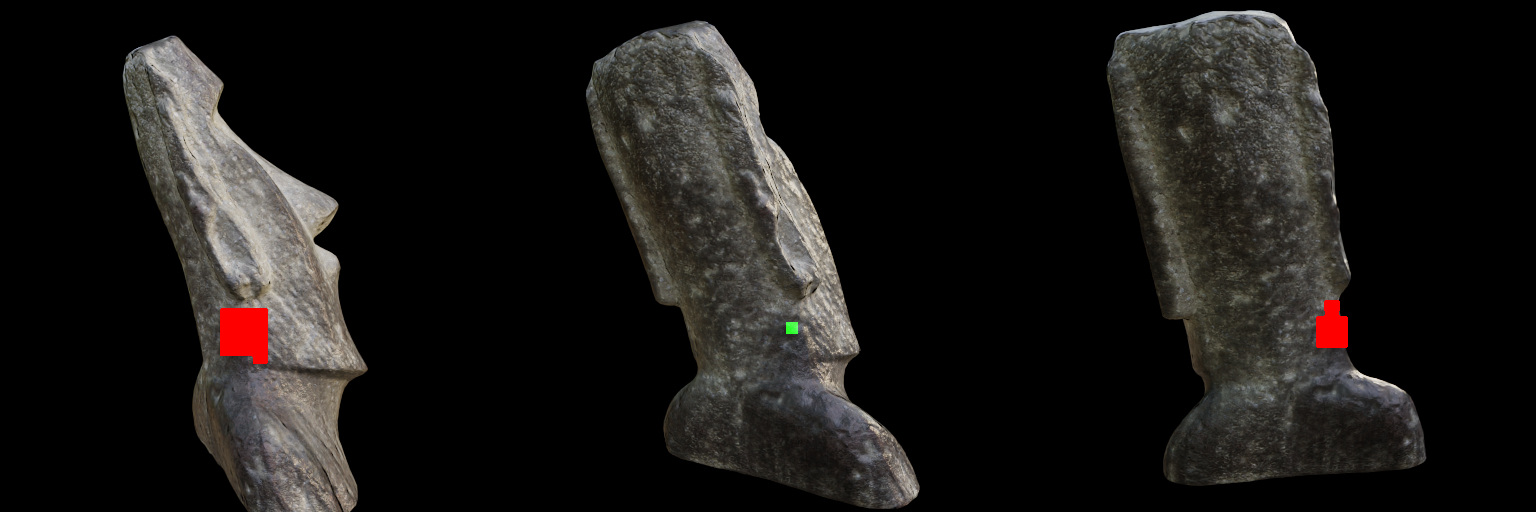}
            \begin{tikzpicture}[overlay, remember picture]
                \draw[-{Stealth[length=2mm]},green,thick,] (3.50,1.25) -- (3.12,0.78);
                \draw[-{Stealth[length=2mm]},red,thick,]   (1.45,1.30) -- (1.07,0.83);
                \draw[-{Stealth[length=2mm]},red,thick,]   (5.63,1.30) -- (5.25,0.83);
            \end{tikzpicture}
        \end{overpic}
        &
        \begin{overpic}[width=0.332\textwidth]{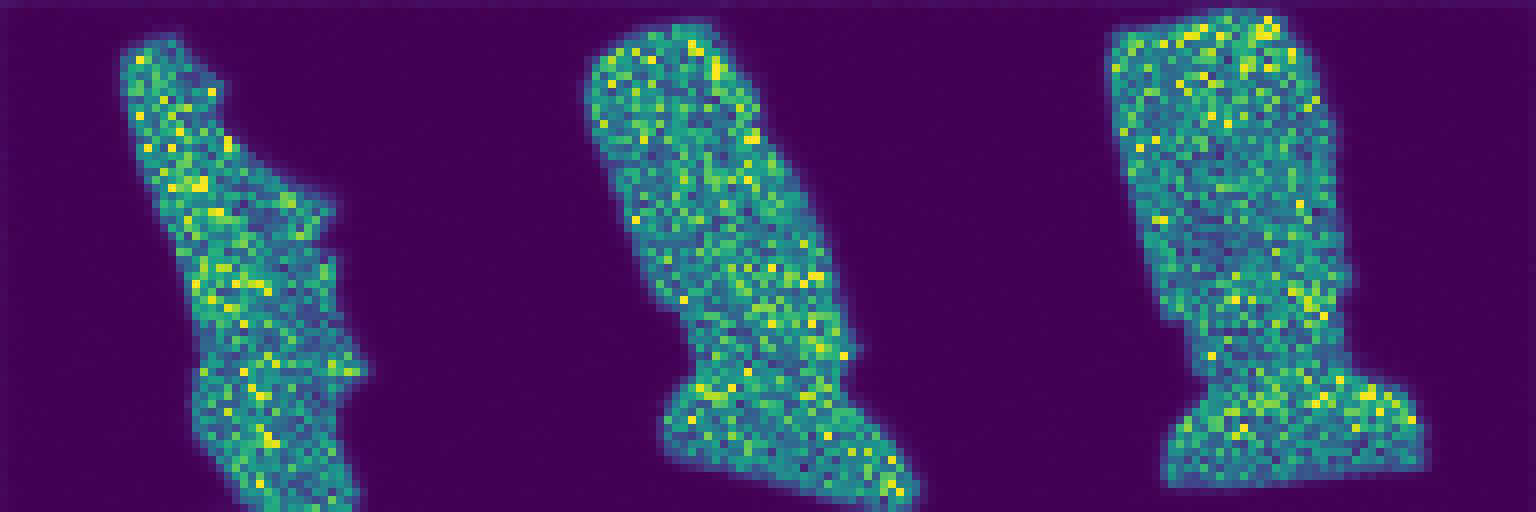}
        \end{overpic}
        &
        \begin{overpic}[width=0.332\textwidth]{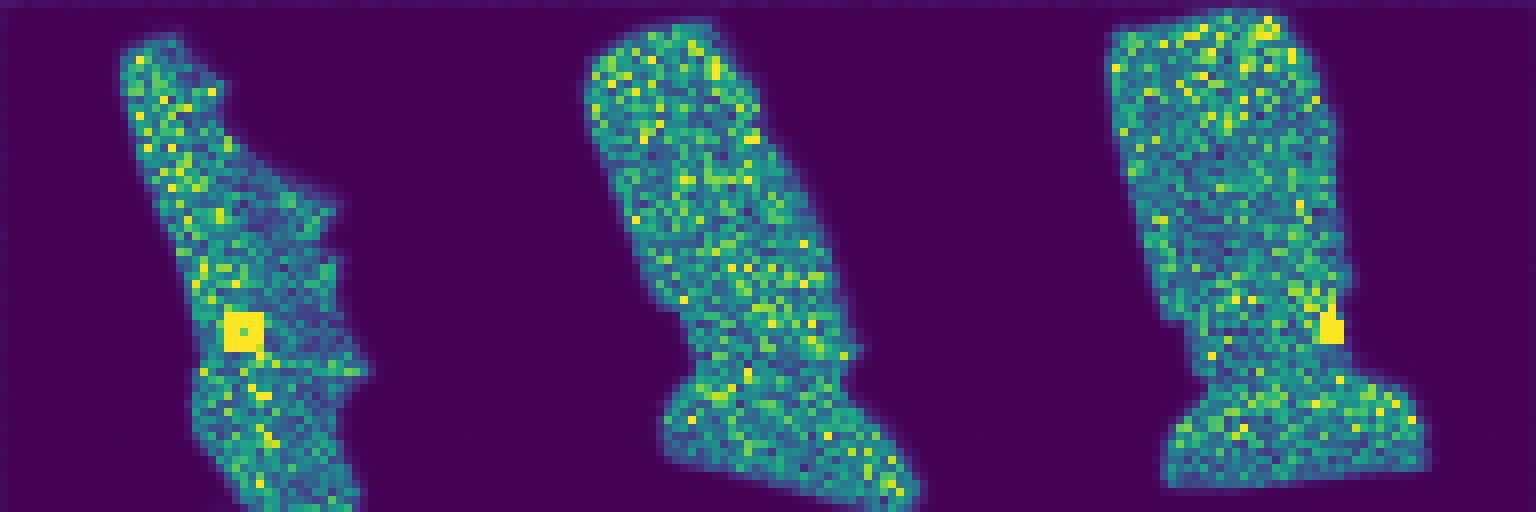}
            \begin{tikzpicture}[overlay, remember picture]
                \draw[-{Stealth[length=2mm]},red,thick,]   (1.45,1.30) -- (1.07,0.83);
                \draw[-{Stealth[length=2mm]},red,thick,]   (5.63,1.30) -- (5.25,0.83);
            \end{tikzpicture}
        \end{overpic}
    \end{tabular}
    \caption{
        Left:
        For one latent pixel (green), we highlight the corresponding neighborhoods in the other conditioning views in red; we use ray tracing to check for occlusions.
        Middle:
        One attention score matrix \emph{row} related to that green latent pixel can be rearranged into an image showing how much it attends to all other latent pixels in one stage of the diffusion model.
        (We crop to three views of the $3 \times 3$ grid. The supplementary document includes an uncropped version.)
        Right:
        We bias the matrix entries in \emph{columns} that correspond to the identified red regions to promote attention---and hence consistency---between these latents.
    }
    \label{fig:attention_viz}
\end{figure*}

\subsection{Structure-preserving detail enhancement}
\label{sec:second-controlnet}

To enhance the rendered views while preserving the character of the input material, we follow the approach of \citet{meng2022sdedit} and add a user-controlled amount of noise to the rendered views to get the initial state for diffusion. This is not enough, however, to preserve the original material and geometry. Hence, we additionally condition the diffusion model with two publicly available ControlNets~\cite{controlnet}: \emph{ControlNet tile}, trained to do super-resolution, which we repurpose to respect the input view while enhancing details, and \emph{ControlNet normal} that helps preserve lighting and curvature details using our auxiliary normal buffer as input.

Together with our other modifications (Sec. \ref{sec:noise_warping} and \ref{sec:attention_bias}) we find this to be effective at achieving consistency with the original material and across views, while avoiding expensive training of task-specific ControlNets as used in prior work~\cite{flashtex}.

\subsection{View-correlated noise prior}
\label{sec:noise_warping}

Although the relationship between the initial noise input and the image output in diffusion is highly non-linear, the two are correlated.
This has previously been exploited for temporal consistency by warping noise by a motion field, and we take a similar approach.
Because diffusion models are highly sensitive to the noise statistics, the noise we produce must be uncorrelated within each view and have uniform variance, or we risk significant artifacts.

Based on \citet{chang2024how}, we propose a simple method to correlate the initial noise of the diffusion model across views while preserving its statistics.
In contrast to their application, we deal with a sparse set of views that do not undergo smooth motion.
It would be challenging to warp an initial noise from a reference view due to the significant amount of disocclusion between views.
Instead, we exploit the known geometry of the asset and anchor a reference noise field in the UV space of the asset.

For each view, we then project the noise from UV space (we use $1024\times 1024$ noise textures) into image space and use this as the initial state for diffusion. Compared to the analytic integration of \citet{chang2024how}, we use a simpler but effective \emph{supersampling} approach.
We subdivide each pixel into a grid of subpixels ($4\times 4$ in our implementation) and project the corners of each subpixel into the UV space of the object:
\begin{center}
    \begin{overpic}[width=0.95\columnwidth]{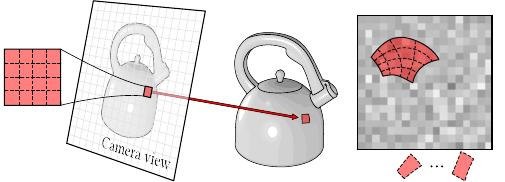}
        \footnotesize
        \put(-6.4,29.1){ \begin{minipage}{2cm}\centering Subdivided\\pixel \end{minipage}}
        \put(71.7,34){   \begin{minipage}{2cm}\centering UV space zoom-in  \end{minipage}}
        \put(73,3){$A_i$\,:}
    \end{overpic}
\end{center}
We then sample the noise field at the center of each projected subpixel, and compute an area-weighted average of the sampled noise values: $\sum_i f_i \cdot A_i$, where $A_i$ is the area of each projected subpixel $i$, and $f_i$ its noise value. Neighboring pixels don't overlap and generally average distinct sets of noise texels, and the resulting noise is independent within each view.

The variance of the projected noise is highly non-uniform, depending on the projected area of each pixel.
To correct this, we could normalize the noise field by an estimate of its variance: $\sqrt{\sum_i \smash{A_i^2}}$.
However, because multiple subpixels may map to the same noise texel, we need to additionally account for the covariance between subpixels. We estimate this with $\mathrm{Cov}_i=\max(A_\mathrm{texel}/A_i-1, 0)$ where $A_\mathrm{texel}=1024^{-2}$ is the area of a noise texel. Intuitively, this counts how many times a distinct noise value is overcounted on average.
The final normalization factor is then $\sqrt{\sum_i A_i^2 (1 + \mathrm{Cov}_i)}$. This matches the variance of projected- and reference noise.

In the case of extreme magnification of the noise texture, individual noise texels may project to multiple pixels and correlate noise within the image.
This is rare and usually caused by missing or degenerate UVs, but it can negatively impact the quality of diffusion.
As a safeguard, we smoothly blend the projected noise value with independent white noise when the pixel area in UV space, $\sum_i A_i$, approaches the area of a noise texel.

\begin{figure*}[htbp]
    \centering%
    \setlength{\tabcolsep}{0.002\textwidth}%
    \renewcommand{\arraystretch}{1}%
    \footnotesize%
    \begin{tabular}{cccccccc}
        &Initial asset& \multicolumn{6}{c}{
            \begin{tikzpicture}
                \draw[<-] (0,0) -- (1,0);
                \draw[-] (2.6,0) -- (6.7,0);
                \draw[-] (8.3,0) -- (12,0);
                \draw[->] (14,0) -- (15,0);
                \node[above] at (1.8,-0.2) {visual fidelity};
                \node[above] at (7.5,-0.2) {$\bias$ parameter};
                \node[above] at (13,-0.2) {view consistency};
            \end{tikzpicture}
        }\\[-4pt]%
        && 0.0 & 0.6 & \textbf{1.2} & \textbf{1.8} & 2.4 & 3.0\\%
        \rotatebox{90}{\hspace*{4em}View 1}&%
        \includegraphics[height=0.14\linewidth, trim=150 0 100 0, clip]{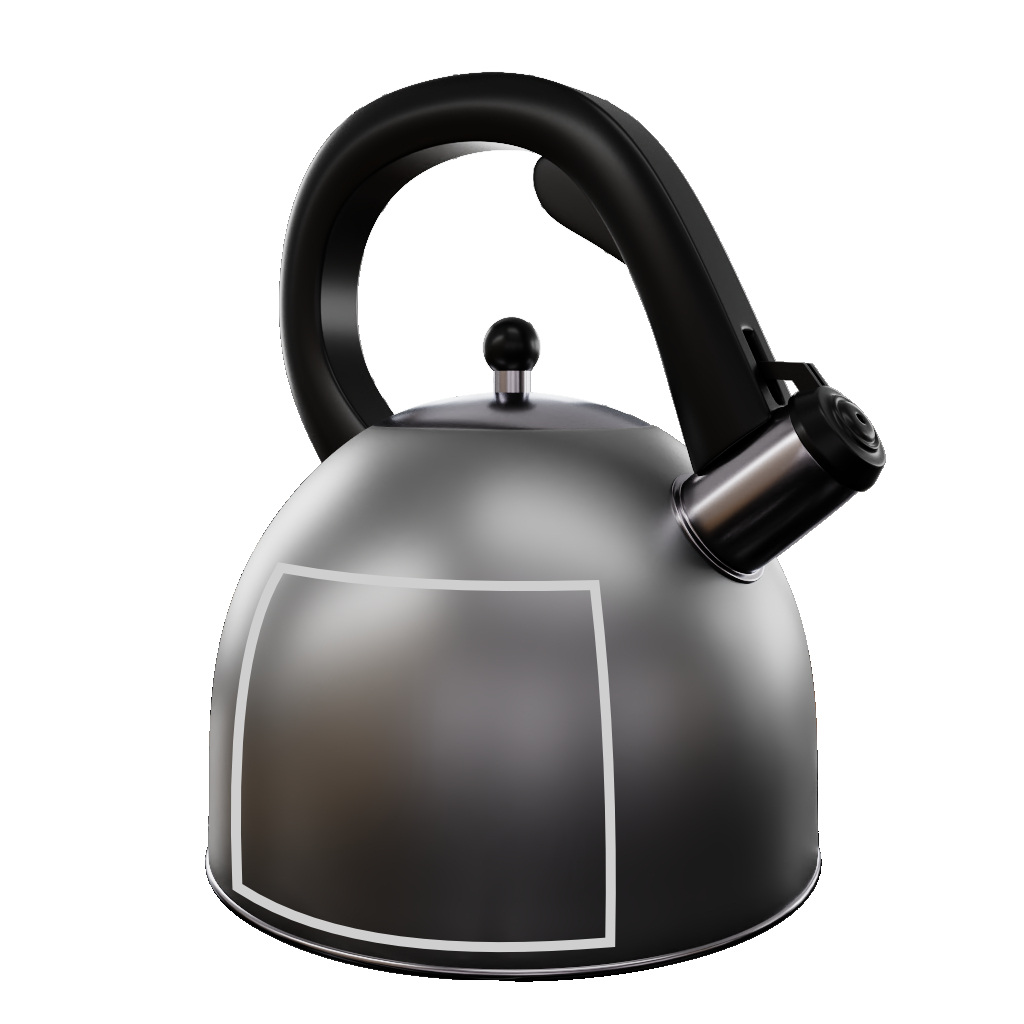}&%
        \includegraphics[height=0.14\linewidth, trim=200 40 300 460, clip]{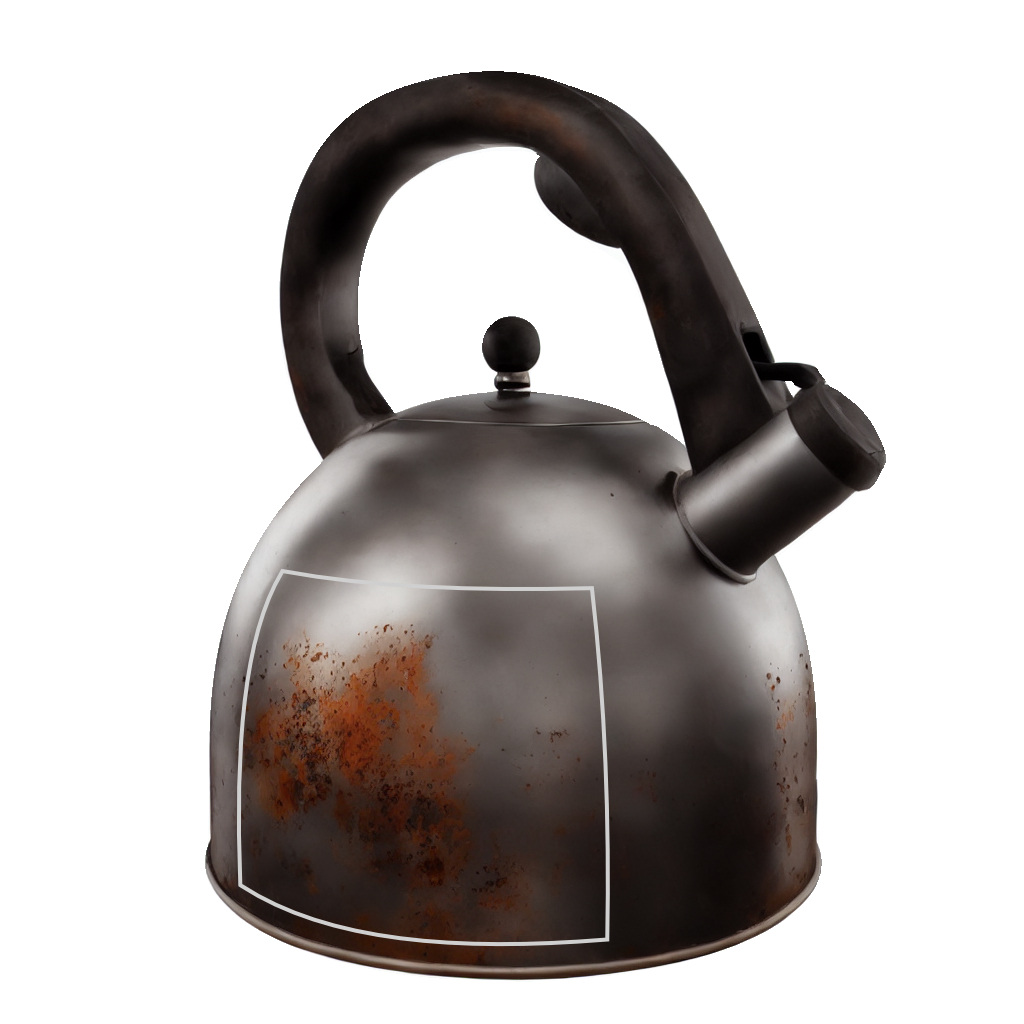}&%
        \includegraphics[height=0.14\linewidth, trim=200 40 300 460, clip]{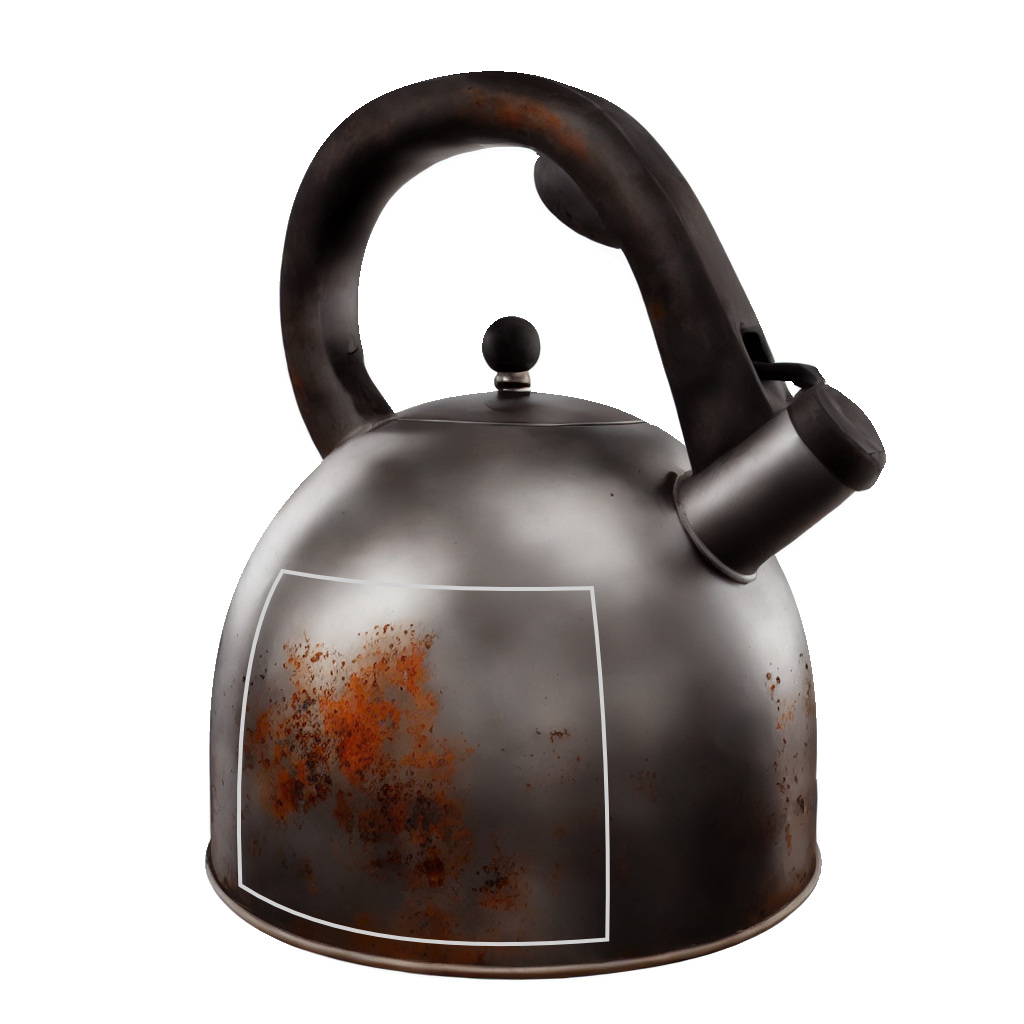}&%
        \includegraphics[height=0.14\linewidth, trim=200 40 300 460, clip]{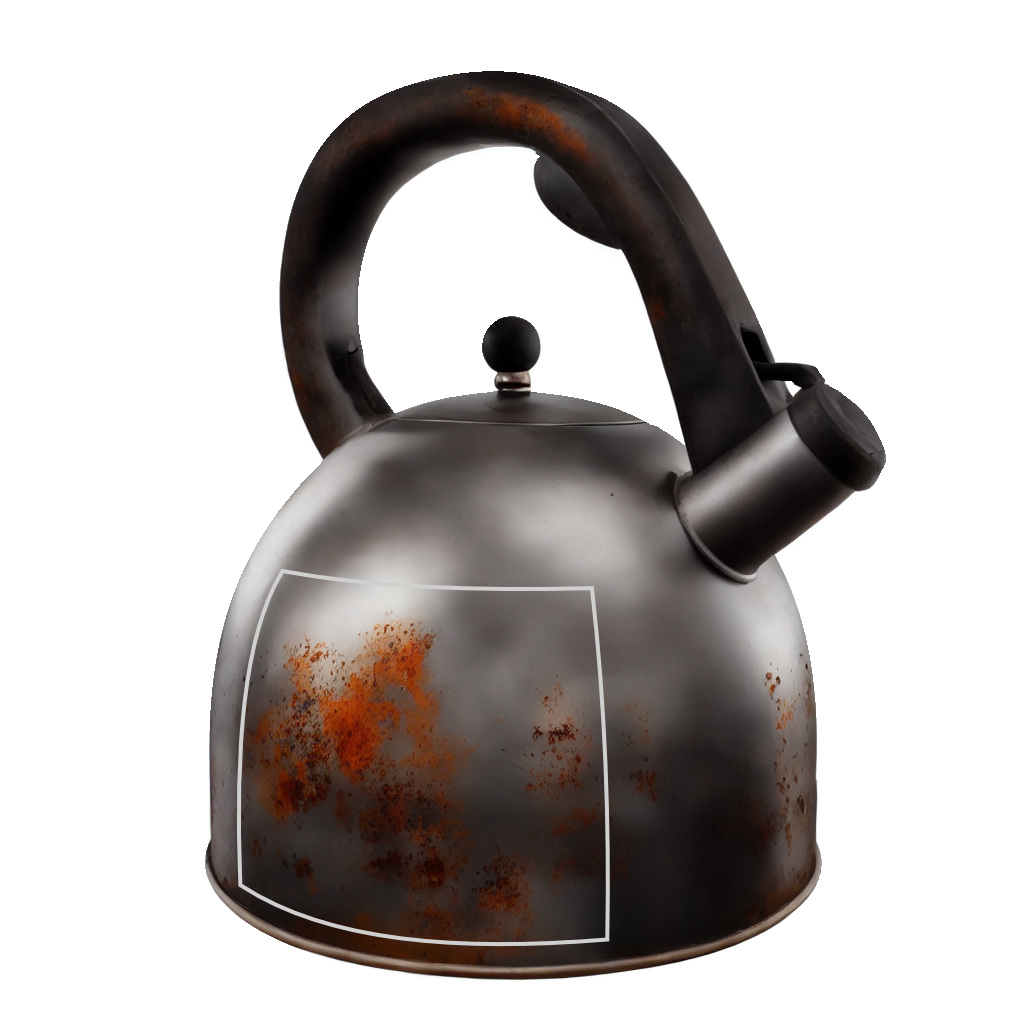}&%
        \includegraphics[height=0.14\linewidth, trim=200 40 300 460, clip]{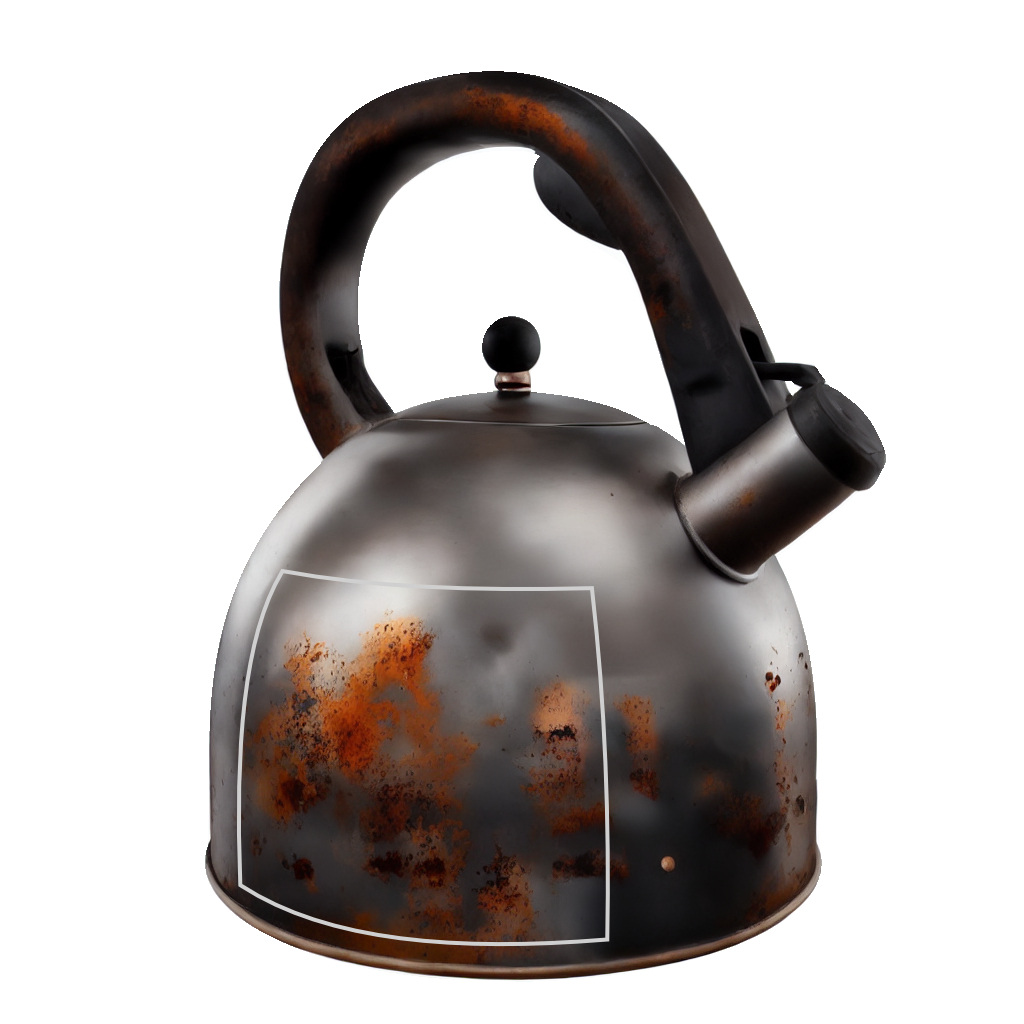}&%
        \includegraphics[height=0.14\linewidth, trim=200 40 300 460, clip]{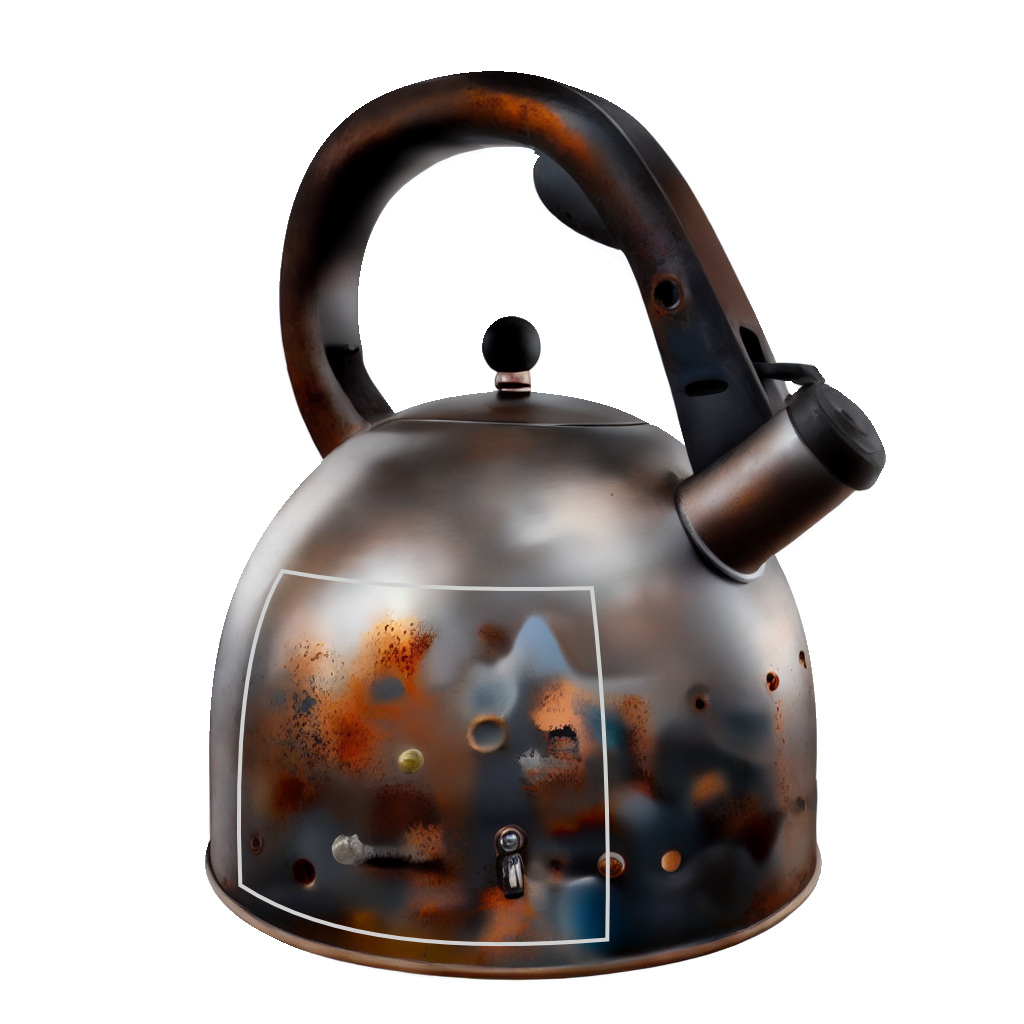}&%
        \includegraphics[height=0.14\linewidth, trim=200 40 300 460, clip]{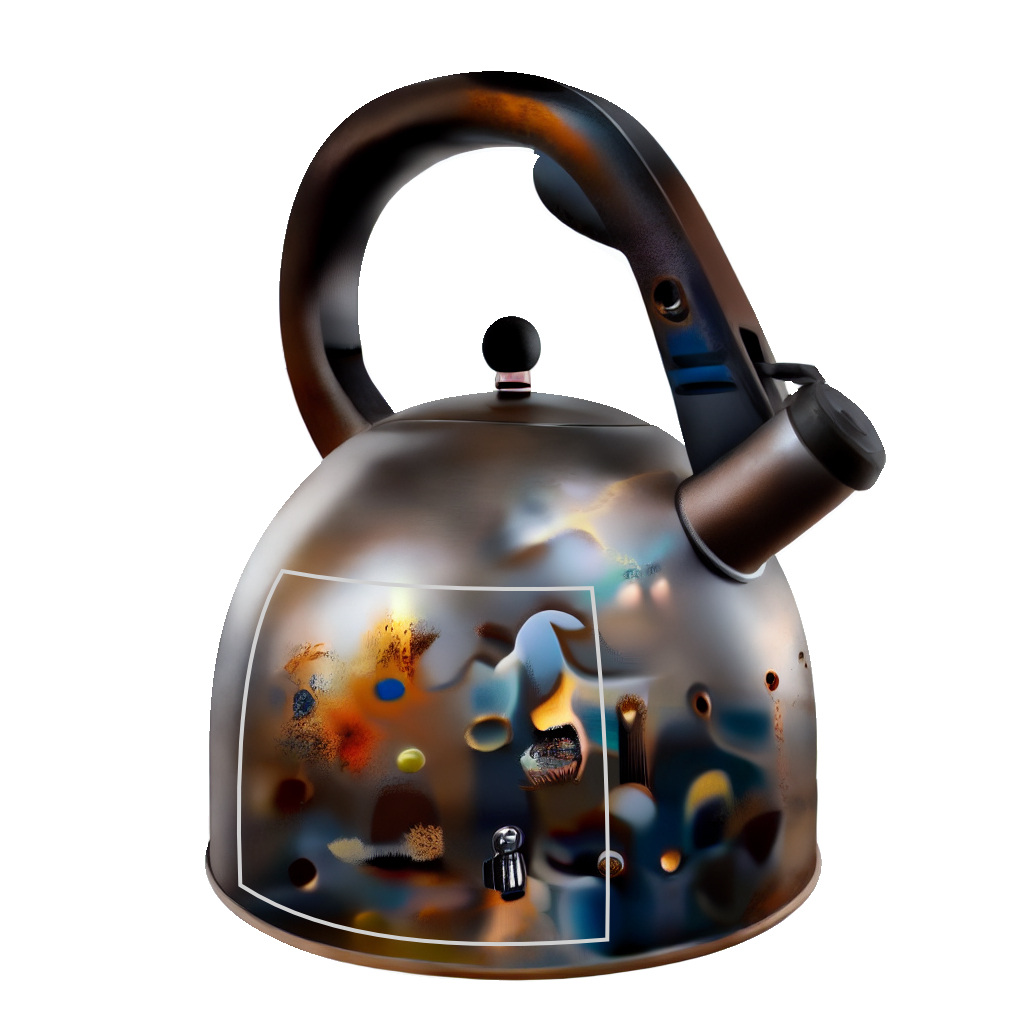}\\%
        \rotatebox{90}{\hspace*{2em}View 2, $+ 40^\circ$}&
        \includegraphics[height=0.14\linewidth, trim=150 0 100 0, clip]{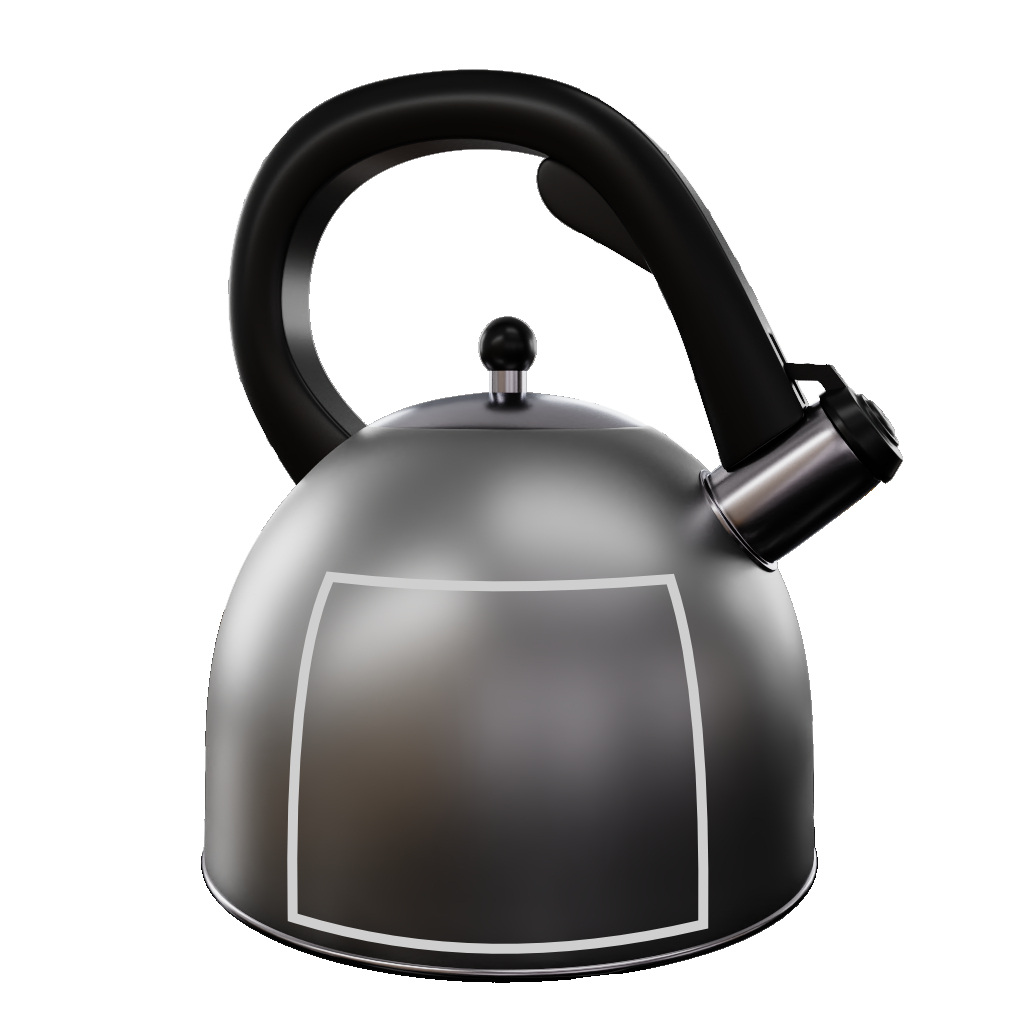}&%
        \includegraphics[height=0.14\linewidth, trim=200 40 300 460, clip]{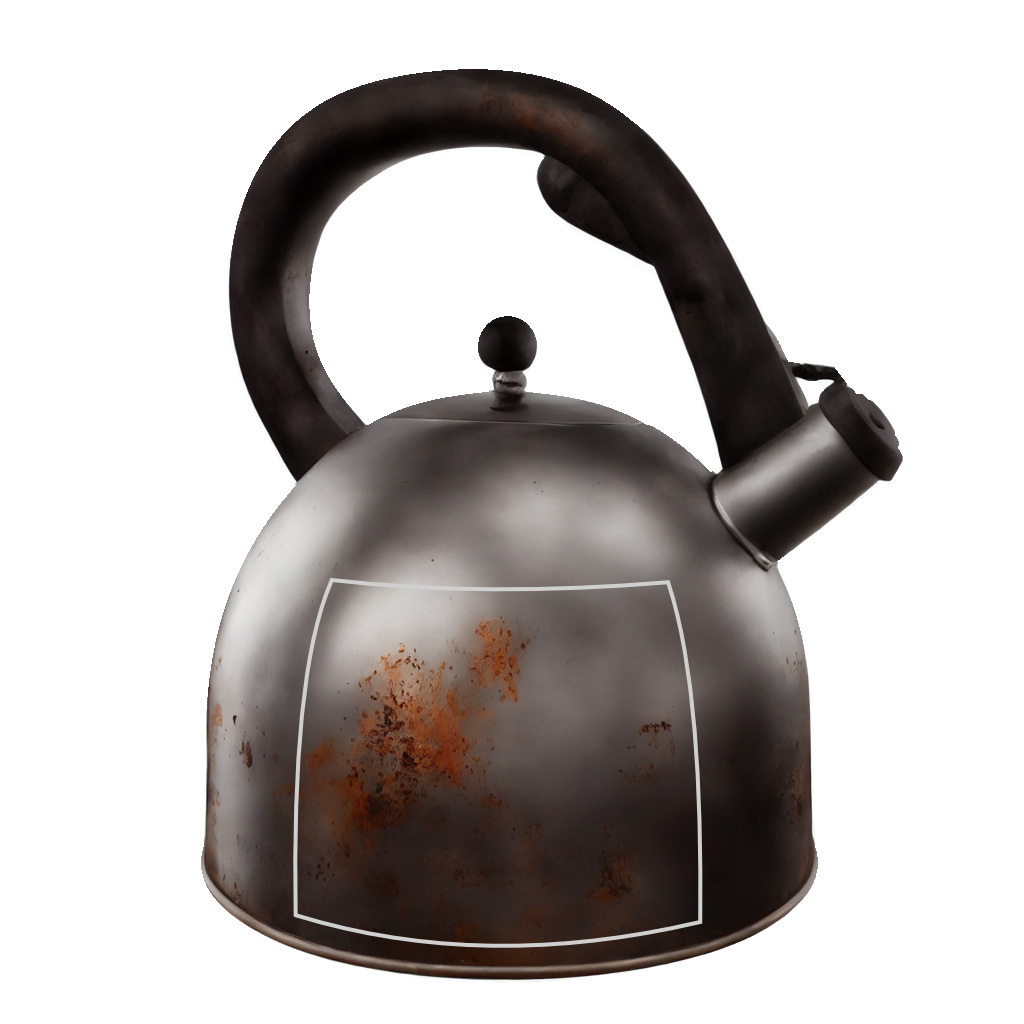}&%
        \includegraphics[height=0.14\linewidth, trim=200 40 300 460, clip]{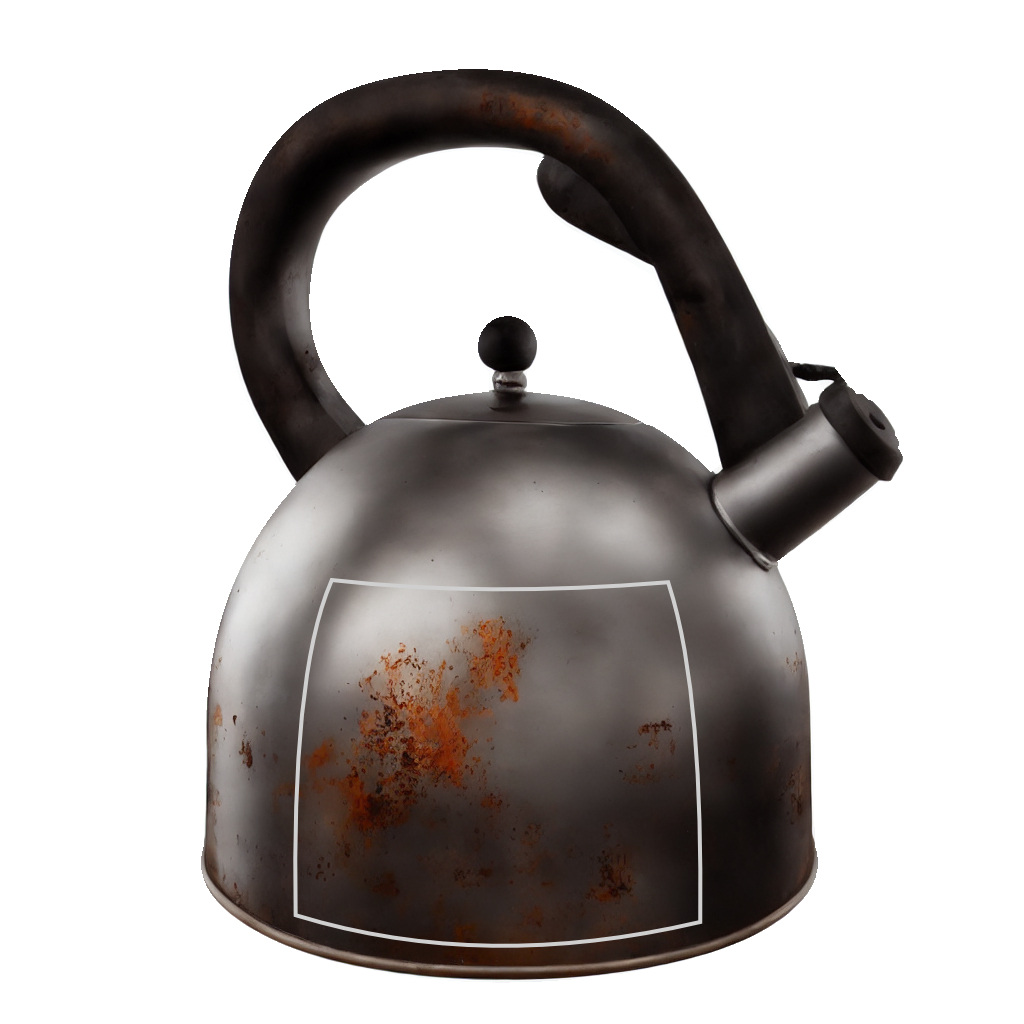}&%
        \includegraphics[height=0.14\linewidth, trim=200 40 300 460, clip]{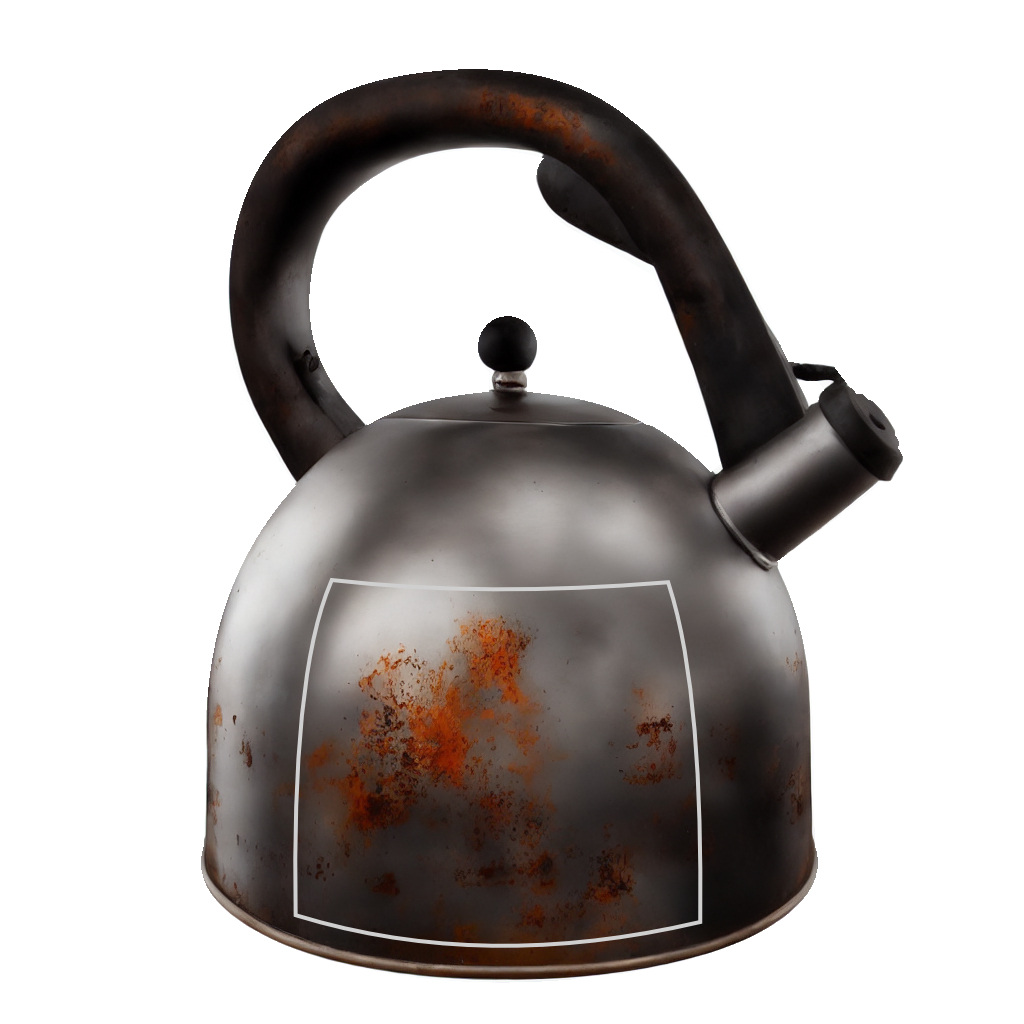}&%
        \includegraphics[height=0.14\linewidth, trim=200 40 300 460, clip]{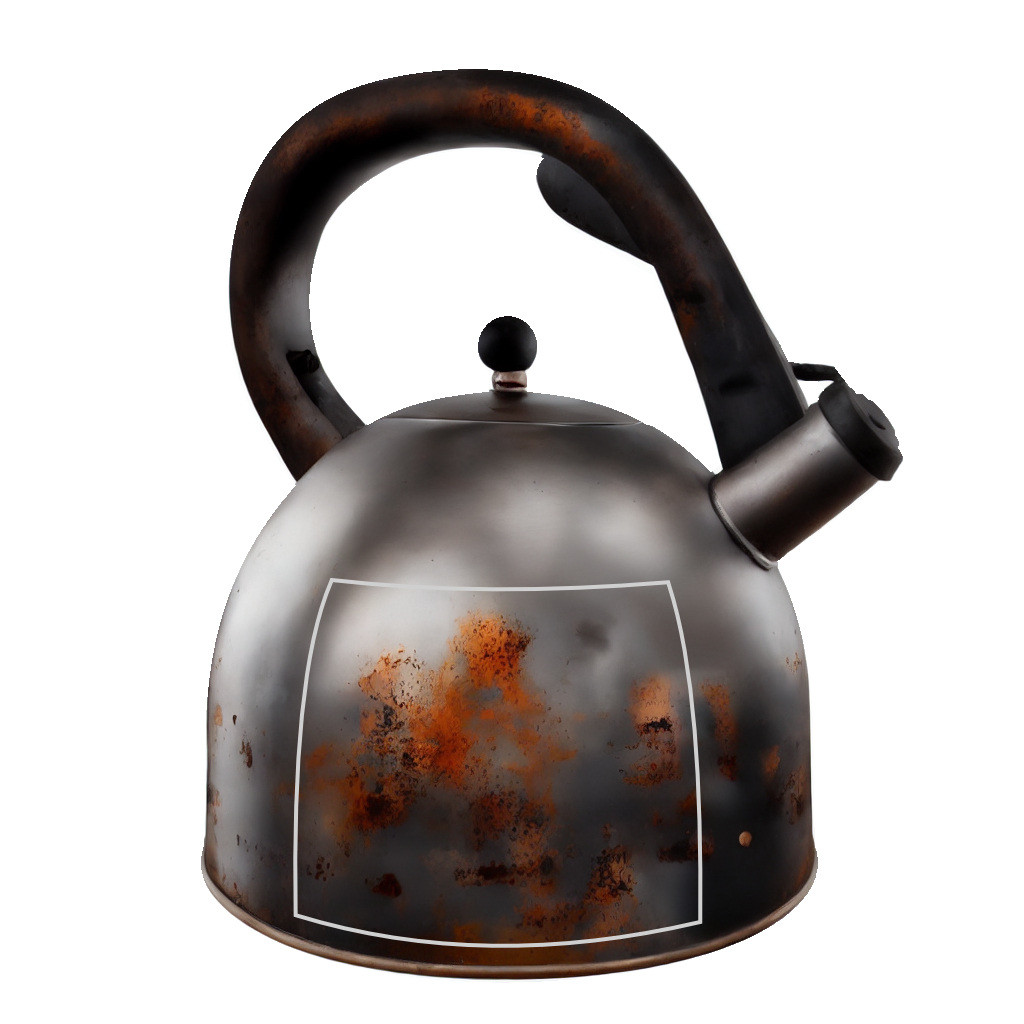}&%
        \includegraphics[height=0.14\linewidth, trim=200 40 300 460, clip]{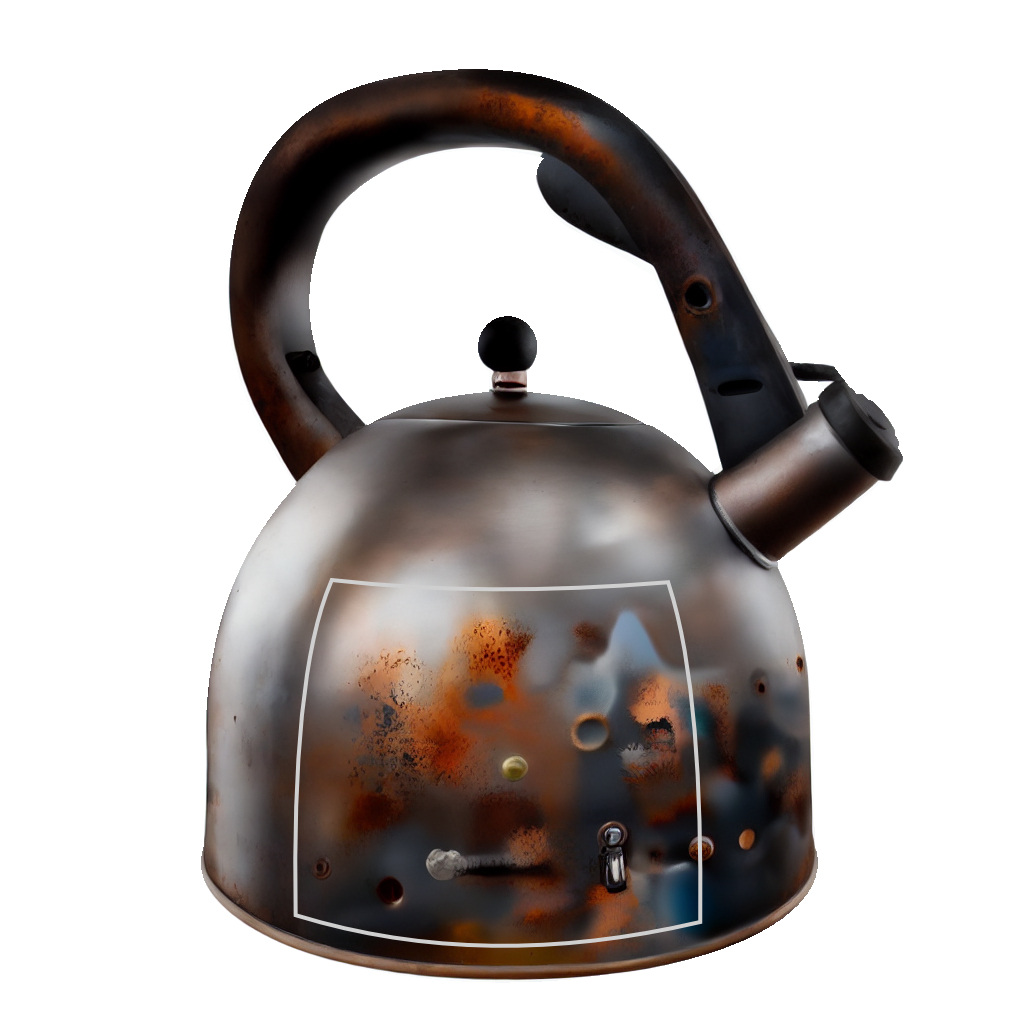}&%
        \includegraphics[height=0.14\linewidth, trim=200 40 300 460, clip]{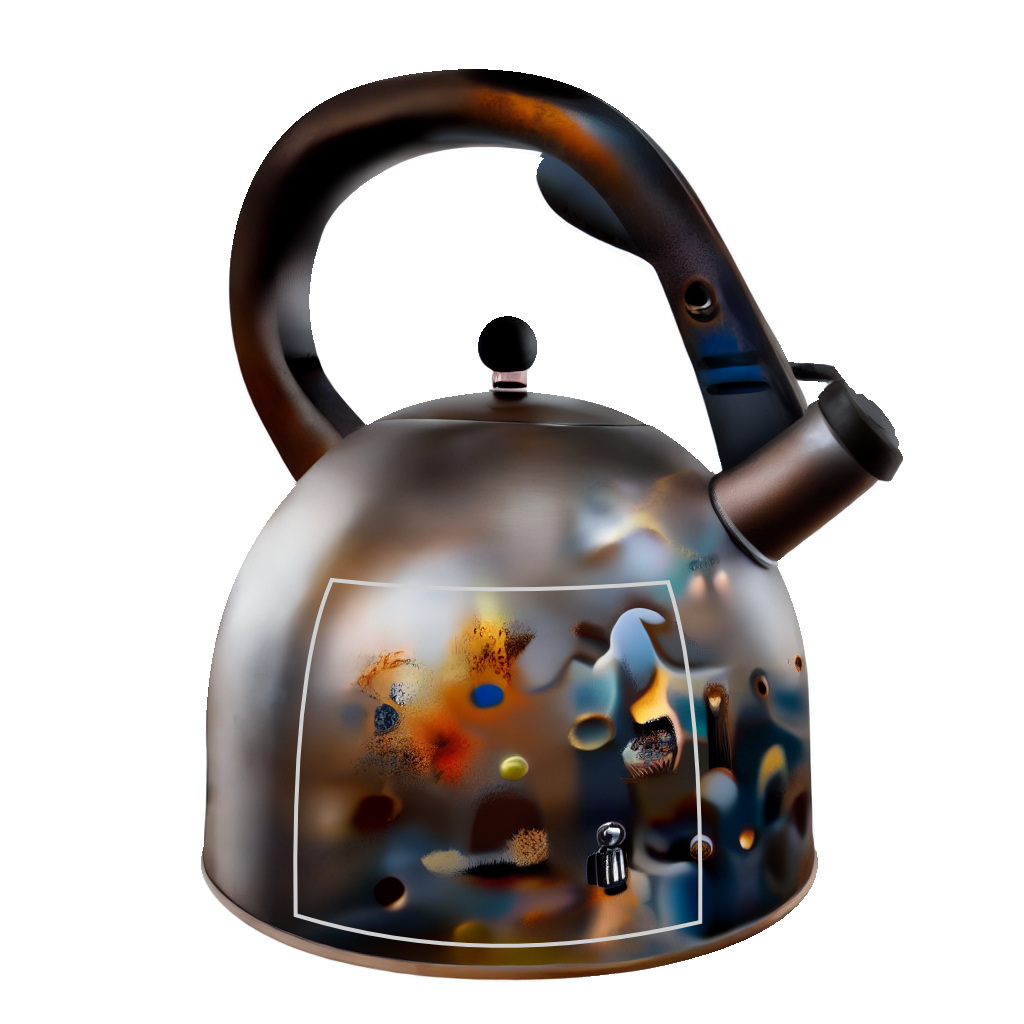}\\%
    \end{tabular}%
    \caption{The bias parameter $\bias$ trades between visual fidelity (left insets) and multi-view consistency (right insets). The first column shows the initial asset in two views that condition the diffusion model. The white shapes outline a UV region, which we analyze in the insets to illustrate the impact of the $\bias$ parameter on generated visuals.
    Values between 1.2 and 1.8 strike a good balance in this particular scene.}
    \label{fig:hparams}
\end{figure*}
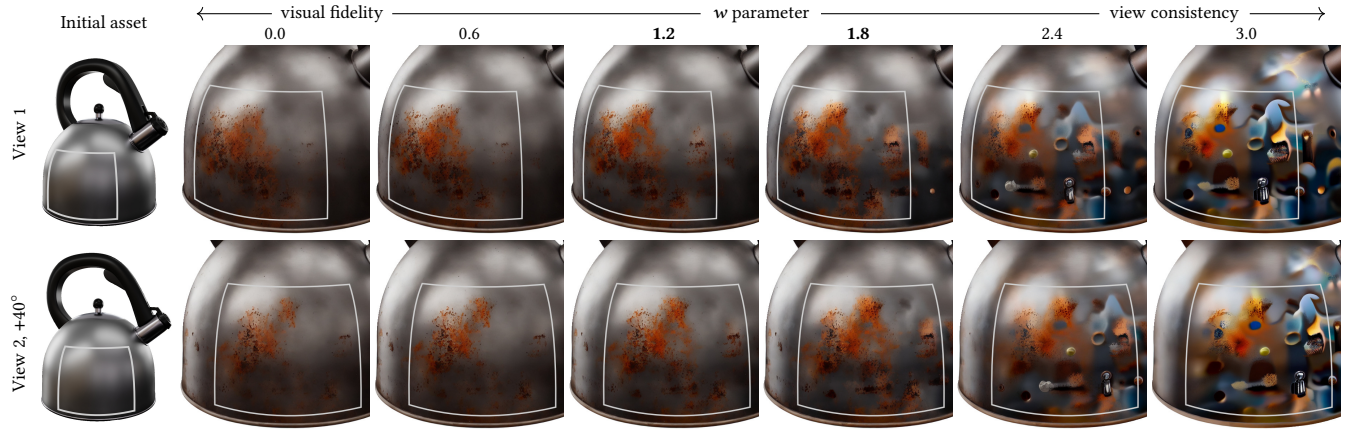

\subsection{Pixel-correspondence attention bias}
\label{sec:attention_bias}

The second technique for improving the multi-view consistency amounts to biasing the self-attention mechanism of the diffusion model according to a reprojection prior.

As described in Section~\ref{sec:background}, the self-attention modules allow image regions to influence each other.
We refer to these regions as \emph{latent pixels} to emphasize that they map to pixels in the input/output image.
The $\matQ \matK^\top$ product in the self-attention module forms a large $N \times N$ score matrix where entry $[i \in Q,j \in K]$ describes how strongly latent pixel $i$ attends to latent pixel $j$; $N$ is the total number of latent pixels.

Our goal is to increase the attention scores between latent pixels in different views that observe the \emph{same surface patch}.
This will increase the chance that these latent pixels will denoise to consistent visuals in the resulting images.
Conceptually, we construct an $N\times N$ \emph{bias matrix} $\matB$,
where any positive value $\matB[i,j]$ will boost the attention of pixel $i$ to pixel $j$.

We determine the values of $\matB$ as follows.
For each pair of pixels $i$ and $j$ (where $i \neq j$) in the $3\times3$ latent image grid,
we cast a ray through the center of latent pixel $j$ into the 3D scene, finding the first hit point~$\mathbf{p}$.
We then project $\mathbf{p}$ onto the image plane containing pixel $i$ and check that $\mathbf{p}$ and the projection $\mathbf{q}$ are mutually visible.
\begin{center}
    \begin{overpic}[width=0.95\columnwidth]{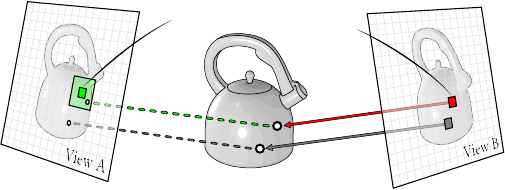}
        \footnotesize
        \put(53.5,15.2){$\mathbf{p}$}
        \put(18.5,14.7){$\mathbf{q}$}
        \put(29.0,34.5){Latent pixel $i$}
        \put(54.0,32.0){Latent pixel $j$}
        \put(60.0,14.9){\rotatebox{7}{$\matB[i,j_0\!] \!=\! \bias$}}
        \put(61.0, 6.3){\rotatebox{7}{$\matB[i,j_1\!] \!=\! 0$}}
        \put(11.0,19.0){$\mathcal{I}$}

    \end{overpic}
\end{center}
If the projection $\mathbf{q}$ is within a neighborhood $\mathcal{I}$ of pixel $i$, we set the value in the bias matrix to a user-defined constant: $\matB[i,j] = \bias$.

The matrix is used to alter the attention operation as:
\[
\text{Attention}(\matQ, \matK, \matV) = \text{softmax}\left(\frac{\matQ \matK^\top + \matB}{\sqrt{d_k}}\right)\matV.
\]
In practice, constructing the full $N \times N$ matrix is often prohibitive, and the bias term has to be evaluated on the fly (see Sec. \ref{sec:implementation}).

The U-Net applies self-attention at multiple scales, and we adjust the size of neighbordhood $\mathcal{I}$ ($9^2, 5^2, 3^2, 1^2$ for layers 1-4, respectively) to map to the same size patch in the original image. Figure \ref{fig:attention_viz} visualizes exemplary attention scores before and after adding the bias for a single specific surface point. The increment $\bias$ is a user-defined constant analysed in Figure~\ref{fig:hparams}. Increasing $\bias$ improves view consistency but eventually generates less compelling appearance as the diffusion starts to lose its global view of the image.

\section{Implementation}
\label{sec:implementation}

We implemented our system in PyTorch~\cite{pytorch} using publicly available models: the \textit{tile}~\cite{controlnet_tile} and \textit{normal}~\cite{controlnet_normal} variants of ControlNet (for color and normal inputs respectively) and the corresponding Hugging Face implementation~\cite{huggingface} of Stable Diffusion 1.5.%

We use the Mitsuba 3 system~\cite{mitsuba} for GPU accelerated (differentiable) rendering.
During inverse rendering, we apply a tonemapping operator~\cite{Reinhard2002} to the high-dynamic range renderings before comparing them with the (low-dynamic range) diffusion-generated target images using a relative L2 loss.
It is important not to clip high values in order to preserve smooth specular highlights and avoid zero-valued gradients during backpropagation.

The ControlNet occasionally fails to preserve the exact silhouette of the rendered 3D geometry causing some background pixels to ``bleed into the object'' during the diffusion. We therefore stop gradient propagation for pixels that are within a fixed distance of the (precomputed) object boundary and downscale the loss at grazing incident angles based on a cosine-factor. Masked points still receive coverage from other views and are not removed from optimization.

\paragraph{Memory considerations and scalability}

Both the main diffusion process and our attention biasing can easily exhaust available GPU memory when operating on large images. 
One memory bottleneck is the variational autoencoder used to convert between images and the spatially-downsampled latent representation. For this conversion we found it necessary to split images back into tiles consisting of the individual views. The latent sampling inside the autoencoder and the main diffusion \& denoising processes can then internally operate on the (re)concatenated grids.

While the pixel-to-pixel correspondences
can easily be precomputed for a given asset (e.g. by storing matching pixel coordinates between each pair of views) we cannot afford to explicitly store the resulting $N\times N$ bias matrix in memory.
This limits our choice of the attention framework. From the ones we tested~\cite{xFormers,flashattention,flashattention2,memeffattention,memeffattention2} only the recent FlexAttention~\cite{flexattention} allowed us to scale to 16 views at resolution $1024^2$ as it can apply attention biases on the fly.
We illustrate how memory consumption and runtime performance scale with varying number of views and resolutions in Table~\ref{tab:runtime}.

\begin{table}[t]
    \centering
    \small
    \caption{Runtime on NVIDIA RTX 5880 and memory usage for different numbers of conditioning views and resolutions.
    Implementing our attention biasing with xFormers (xF) instead of FlexAttention (FA) runs roughly $2\times$ faster, but exceeds available memory with 9 and more views at $1024 \times 1024$.
    }
    \label{tab:runtime}
    \setlength{\tabcolsep}{0.009\textwidth}%
    \begin{tabular}{llrrrrrr}
        \toprule
        && \multicolumn{2}{c}{4 views} & \multicolumn{2}{c}{9 views} & \multicolumn{2}{c}{16 views} \\
        \cmidrule(lr){3-4}
        \cmidrule(lr){5-6}
        \cmidrule(lr){7-8}
        && $512^2$ & $1024^2$ & $512^2$ & $1024^2$ & $512^2$ & $1024^2$ \\
        \midrule
        \multirow{2}{*}{FA} & Runtime (s) & 17 & 187 & 67 & 923 & 187 & 2816 \\
        & Peak memory (GB) & 6.2 & 9.2 & 7.6 & 15.2 & 9.7 & 20.4 \\
        \midrule
        \multirow{2}{*}{xF} & Runtime (s) & 8 & 102 & 34 & - & 100 & - \\
        & Peak memory (GB) & 6.9 & 33.2 & 13.8 & >45 & 33.2 & >45 \\
        \bottomrule
    \end{tabular}
\end{table}

\section{Experiments}
\label{sec:results}

\paragraph{Comparisons}
While our goal of enhancing given 3D assets with existing materials is distinct from recent work leveraging image models for view-consistent editing and generation, it is important to evaluate whether existing works can address our problem. \autoref{fig:baselines} justifies our technique by comparing it to related methods applied in our problem setting. The top half shows image generators that do not rely on the full 3D geometry of a given asset, including SPAD~\cite{spad}, Diffusion Handles~\cite{diffusion_handles}, and $\text{RGB}{\leftrightarrow}\text{X}$~\cite{RGBX}. Because SPAD and Diffusion Handles are not designed to work with the given 3D geometry of an input asset, they struggle to render the asset accurately from multiple viewpoints. On the other hand, $\text{RGB}{\leftrightarrow}\text{X}$ takes accurate scene intrinsics (normals, albedo, roughness, etc.) as input, but it is not equipped to ensure multi-view consistency.

The bottom part of \autoref{fig:baselines} compares our approach to DreamMat~\cite{dreammat} and TexPainter~\cite{TexPainter}, two state-of-the-art material generators for given 3D assets. While they focus on material generation from scratch, our primary goal is to enhance existing materials. Hence our result is more faithful to the initial asset provided as input (shown in \autoref{fig:hparams}), while also providing more realistic fine grained details.

\newcommand{\baselineBlock}[3]{
    \rotatebox{90}{\hspace*{#3}#2}&%
    \includegraphics[width=0.112\textwidth]{figures/baselines/#1_0.jpg}&%
    \includegraphics[width=0.112\textwidth]{figures/baselines/#1_1.jpg}&%
    \includegraphics[width=0.112\textwidth]{figures/baselines/#1_2.jpg}&%
    \includegraphics[width=0.112\textwidth]{figures/baselines/#1_3.jpg}\\%
}

\begin{figure}
  \centering%
  \setlength{\tabcolsep}{0.001\textwidth}%
  \renewcommand{\arraystretch}{0.5}%
  \footnotesize%
  \centering
  \hspace*{-1.2em}
  \begin{tabular}{c@{\hspace{0.25em}}|@{\hspace{0.25em}}c@{\hspace{0.5em}}cccc}
    \multirow{4}{*}{\rotatebox{90}{Image generators\hspace{5em}}}
      &\baselineBlock{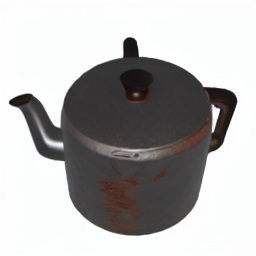}{SPAD}{3em}
      &\baselineBlock{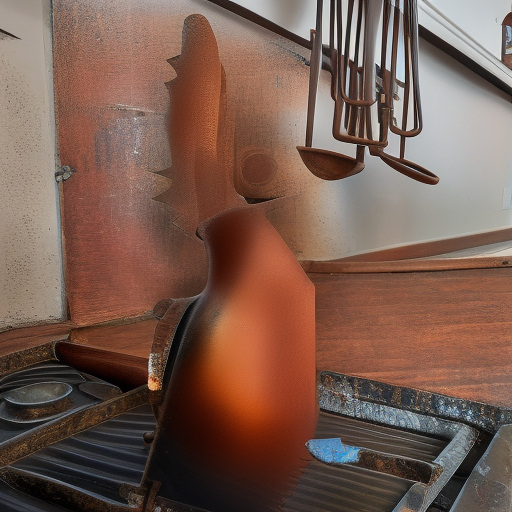}{Diffusion Handles}{0.5em}
      &\baselineBlock{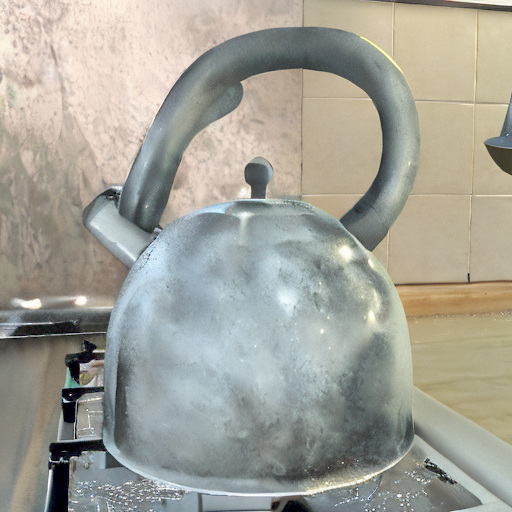}{$\text{RGB}{\leftrightarrow}\text{X}$}{2em}
      &\baselineBlock{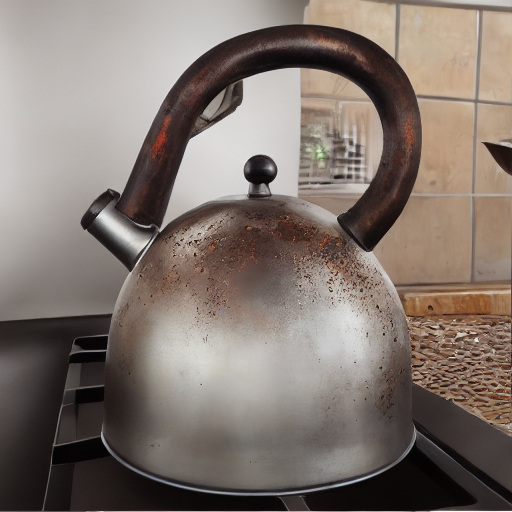}{Ours}{3em}
      \multicolumn{6}{c}{ }\\
      \multirow{3}{*}{\rotatebox{90}{Material generators\hspace{0.5em}}}
      &\baselineBlock{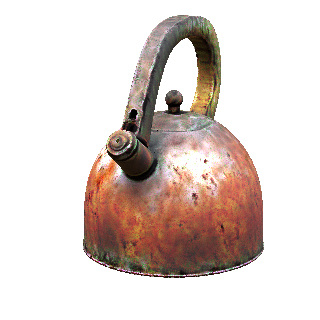}{DreamMat}{2em}
      &\baselineBlock{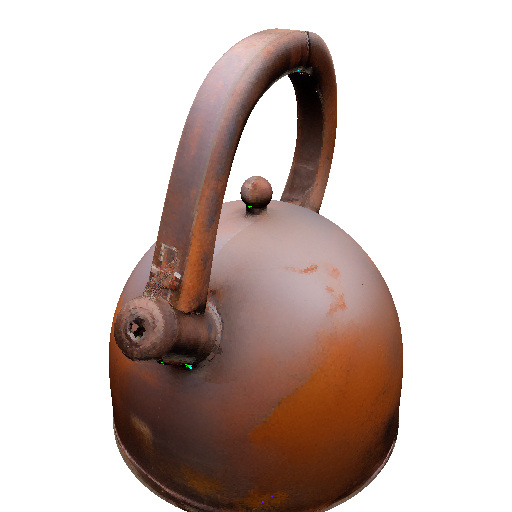}{TexPainter}{1em}
      &\baselineBlock{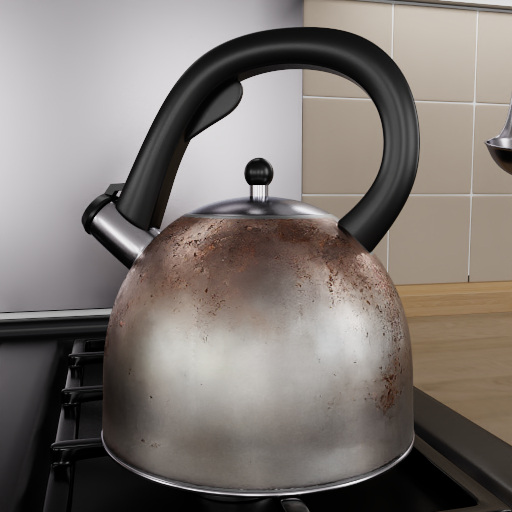}{Ours}{3em}
  \end{tabular}
  \vspace{-2mm}      
  \caption{Comparison to prior work on the prompt \prompt{rusty kettle}. Image generators such as SPAD and Diffusion Handles are not designed to leverage known 3D geometry, which is available in our problem formulation, hence they struggle to accurately render the input from different viewpoints. $\text{RGB}{\leftrightarrow}\text{X}$ takes accurate scene intrinsics as input, but it is not equipped to ensure multi-view consistency (We evaluated the $\text{RGB}{\leftrightarrow}\text{X}$ method in a sequence of $\text{RGB}{\rightarrow}\text{X}{\rightarrow}\text{RGB}$, where inputs are our initial renderings, and outputs are edited renderings). Our problem is more related to material generation techniques such as DreamMat and TexPainter, but we focus on enhancing an existing material.   DreamMat uses a variant of SDS, which tends to output blurrier results.  TexPainter does not allow for view-dependent effects. Both techniques generate their output material from scratch instead of enhancing an input material (shown in \autoref{fig:hparams}).
  }
  \label{fig:baselines}
\end{figure}

\begin{figure*}[htbp]
    \centering%
    \setlength{\tabcolsep}{0.002\textwidth}%
    \renewcommand{\arraystretch}{1}%
    \footnotesize%
    \begin{tabular}{ccccccc}
        Initial asset & \multicolumn{6}{c}{
            \begin{tikzpicture}
                \draw[<-] (0,0) -- (0.45,0);
                \draw[-] (3.2,0) -- (6.0,0);
                \draw[-] (8.8,0) -- (11,0);
                \draw[->] (14.4,0) -- (15,0);
                \node[above] at (1.8,-0.2) {less emphasis on prompt};
                \node[above] at (7.5,-0.2) {Classifer free guidance};
                \node[above] at (13,-0.2) {more emphasis on prompt};
            \end{tikzpicture}
        }\\[-4pt]%
        & 4.0 & \textbf{7.5} & 11.0 & 14.5 & 18.0 & 21.5\\%
        \includegraphics[height=0.14\linewidth]{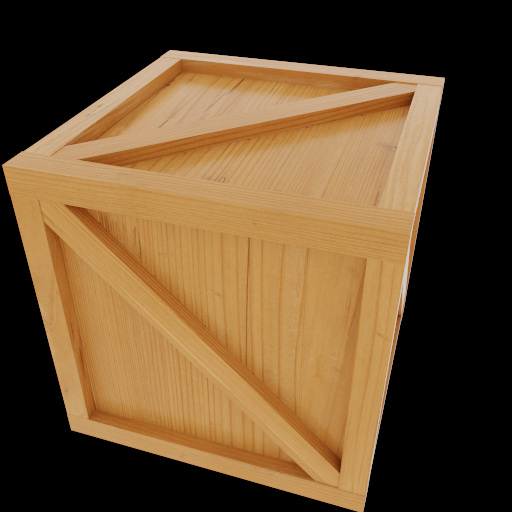}&%
        \includegraphics[height=0.14\linewidth]{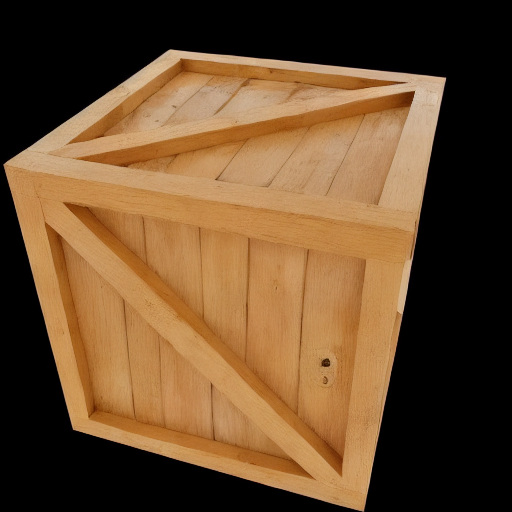}&%
        \includegraphics[height=0.14\linewidth]{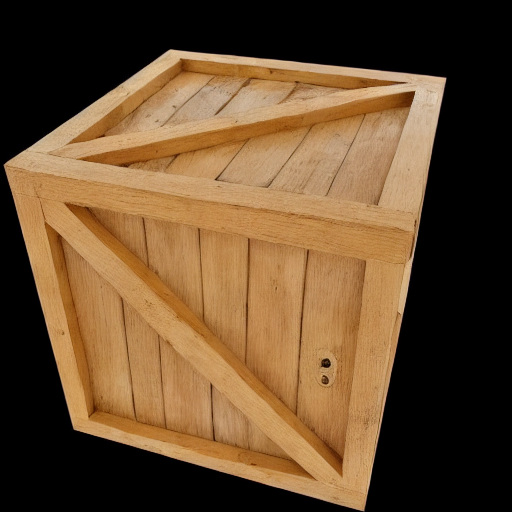}&%
        \includegraphics[height=0.14\linewidth]{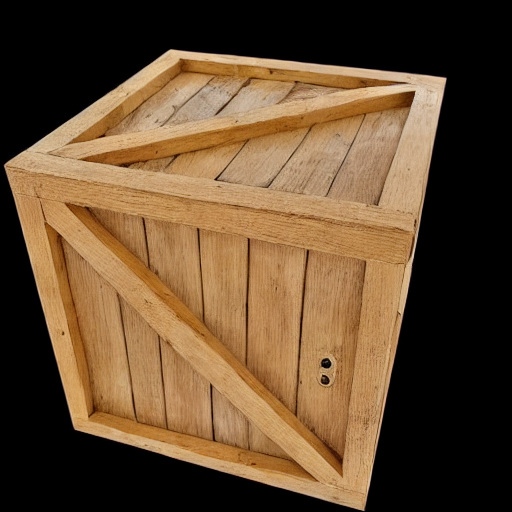}&%
        \includegraphics[height=0.14\linewidth]{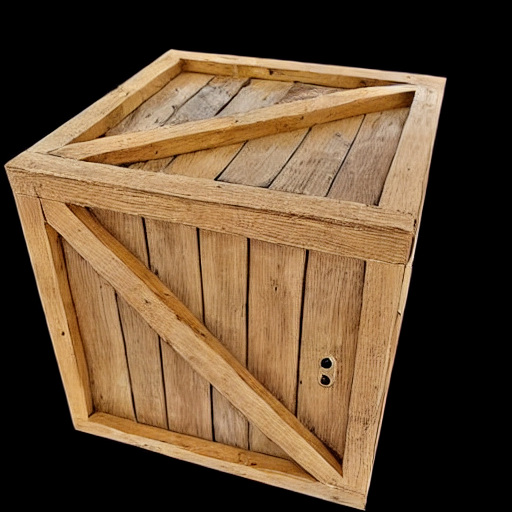}&%
        \includegraphics[height=0.14\linewidth]{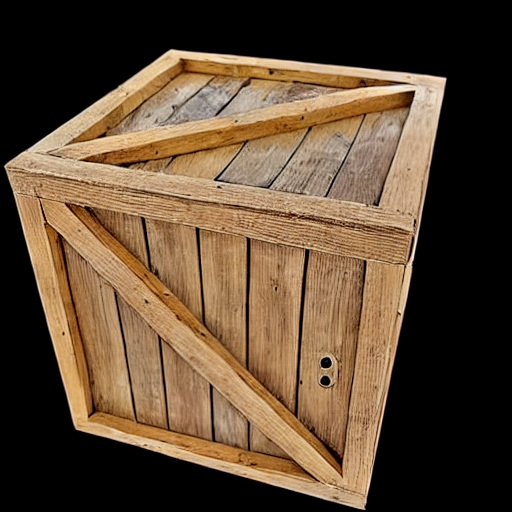}&%
        \includegraphics[height=0.14\linewidth]{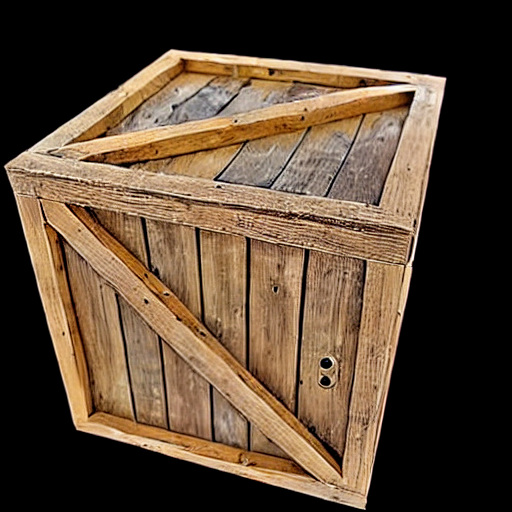}
    \end{tabular}%
    \caption{Controlling the magnitude of detail enhancements using classifier free guidance.
    The scaling factor allows the user to apply more conservative changes and stay closer to the input or obtain more pronounced changes, as desired.}
    \label{fig:hparams_usercontrol}
\end{figure*}
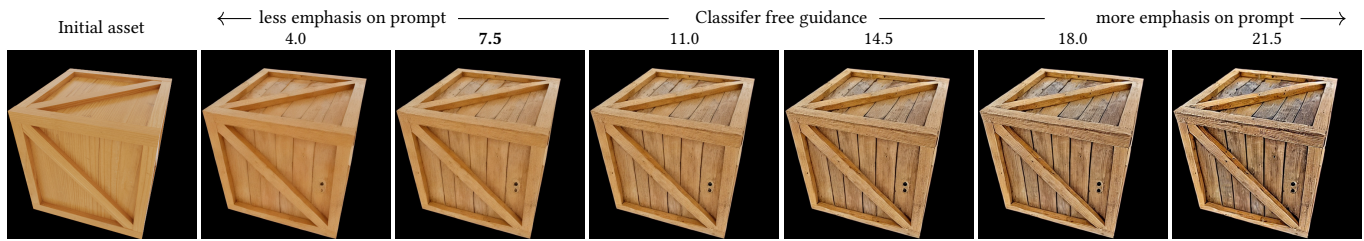

\paragraph{Visual results}
\autoref{fig:results} shows results of our complete pipeline, starting with basic assets, all the way to the recovered material parameters and the corresponding renderings.
The air conditioner result in particular highlights the advantage of using a diffusion model trained on natural images.
The rusting of the blades is distinct from that of the box itself, which aligns with the expectation these components would age differently.
In \autoref{fig:hparams_usercontrol} we show how classifier free guidance~\cite{ho2022cfg} enables the user to control the magnitude of the detail enhancements.
We show further examples in the supplementary document and video.

\paragraph{Ablation}
We validate our contributions and algorithmic choices to achieve view consistency of a pure 2D generative diffusion model by performing an ablation study presented in Figure~\ref{fig:ablation}.
We modify the appearance and detail of two 3D assets: a briefcase and a vase.
Pure ControlNet Tile is able to modify the appearance of the object with respect to the initial material, but fails to achieve view consistency.
Using view-correlated noise (Section.~\ref{sec:noise_warping}) we enhance the detail presence between different camera views, but some of the detail remains misaligned.
Finally, our full model that biases attention (Section.~\ref{sec:attention_bias}) further increases the consistency.

\paragraph{View consistency and inverse rendering}

Intuitively, view consistency between produced outputs is necessary to successfully reconstruct the material maps through inverse rendering.
When different views of the same surface \textit{disagree} and present different details, the inverse rendering process produces blurry results or fails to converge.
We present this effect in Figure~\ref{fig:invrender_inconsistent}.

\section{Limitations and Future work}
\label{sec:limitations}

\begin{figure}[t]
    \centering%
    \setlength{\tabcolsep}{0.002\textwidth}%
    \renewcommand{\arraystretch}{1}%
    \footnotesize%
    \begin{tabular}{cccc}
        \rotatebox{90}{\hspace{1em}Input views}&
        \includegraphics[angle=90,height=1.82cm,trim=100 0 100 50,clip]{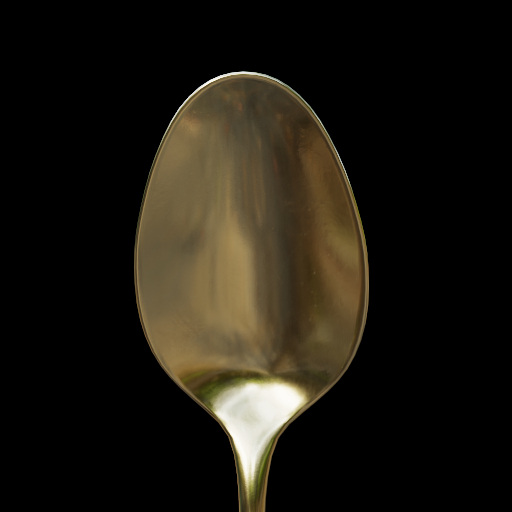}&
        \includegraphics[angle=90,height=1.82cm,trim=100 0 100 50,clip]{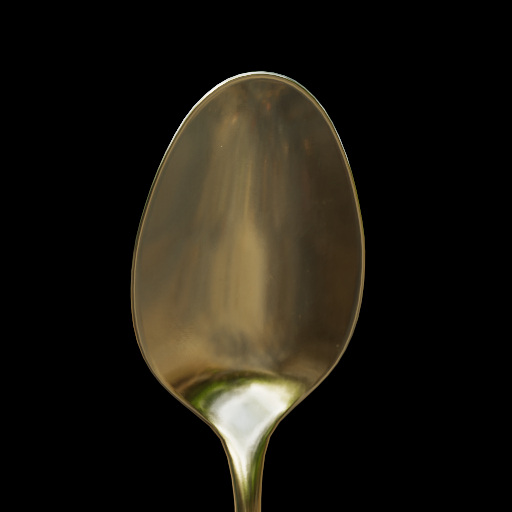}&
        \includegraphics[angle=90,height=1.82cm,trim=100 0 100 50,clip]{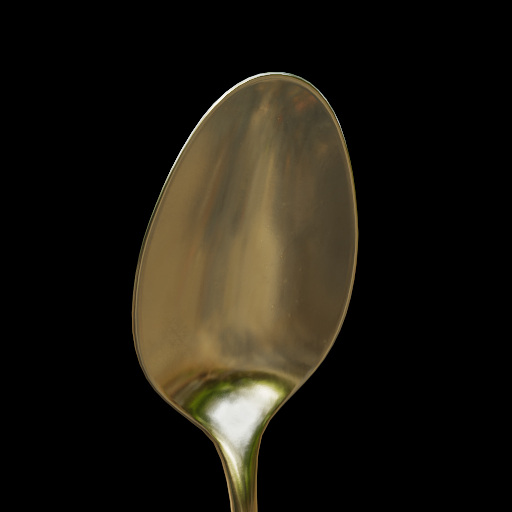}\\
        \rotatebox{90}{\hspace{0.5em}Diffusion result}&
        \includegraphics[angle=90,height=1.82cm,trim=100 0 100 50,clip]{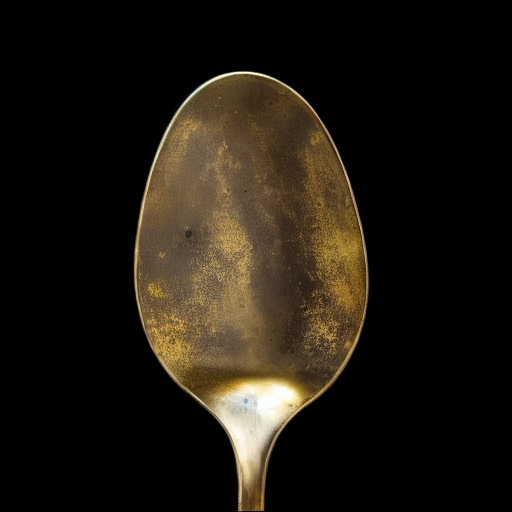}&
        \includegraphics[angle=90,height=1.82cm,trim=100 0 100 50,clip]{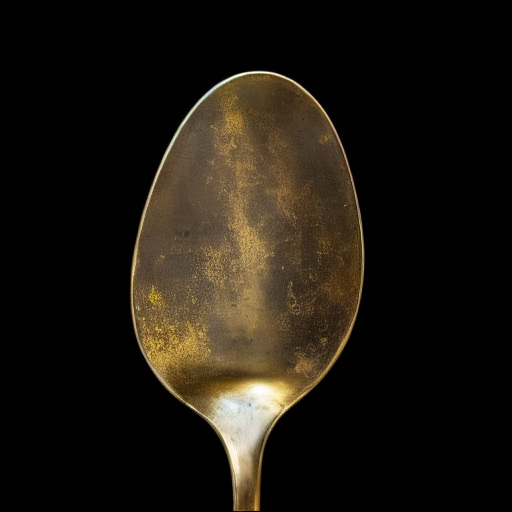}&
        \includegraphics[angle=90,height=1.82cm,trim=100 0 100 50,clip]{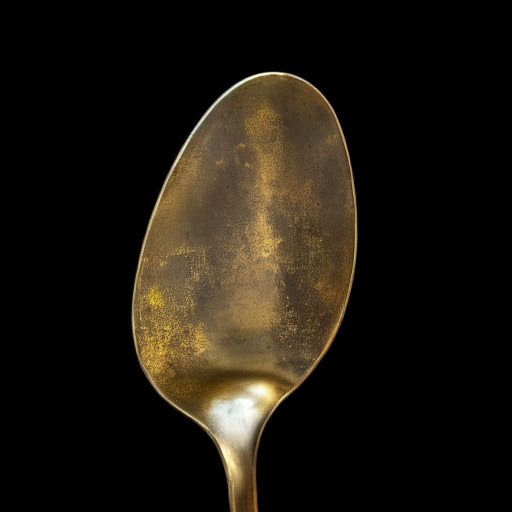}\\
    \end{tabular}
    \vspace{-2mm}
    \caption{
        Environment reflections on the spoon (top) are enhanced with diffuse details that follow the reflections across views (bottom) in a manner that may be inconsistent with user intentions.
    }
    \label{fig:limit_spoon}
\end{figure}

\paragraph*{Additional consistency improvements}
Although our contributions improve multi-view consistency, occasional deviations between views still occur at small scale. 
The inverse rendering typically resolves them by superimposing the conflicting visuals in the material maps.
The final representation is view consistent, by design, but high-frequency view-dependent effects (e.g., mirror reflections) may end up baked in the albedo texture; see \autoref{fig:limit_spoon}.
Since multi-view consistency of diffusion generators is an actively sought property, we believe future work will alleviate this
either in a data-driven fashion, e.g., by employing multi-step optimization and video models, or by imposing additional geometric priors, such as identifying pixel correspondences using manifold walks on specular surfaces~\cite{Jakob2012Manifold}.

\paragraph*{Controlling the diffusion model}
We currently expose only a text prompt as the control over the generated detail. 
Future work could explore more fine-grained user control, such as CLIP priors~\cite{stable_unclip_pipeline, dall_e2} or manipulations using texture exemplars~\cite{texsliders}.
The recently published Stable Diffusion 3.x~\cite{SD3,SD3_5_large} supports more sophisticated text encoders and ControlNets, and swapping them in place of our current diffusion model provides a near-term avenue for better control.
Importantly, artists still retain full editability of the material produced by our tool using traditional workflows.

\paragraph*{Enhancing macro geometry} 
We focused on visual enhancements that can be captured using texture maps
without attempting to refine the input macro geometry. 
While inverse rendering is in principle capable of updating the mesh, 
we leave this for future work.

\paragraph*{Manual hyperparameter tuning}
The attention bias $\bias$ is currently a manually tuned hyperparameter. Although the range of reasonable values is limited ($[0-3.5]$ in our experiments), and $\bias$ can be tuned quickly on low-resolution images, a parameter-free biasing would make the technique more practical.

\section{Conclusion}
\label{sec:conclusion}

We presented a method for enhancing the detail of a classically authored material using a diffusion model.
Our method renders the provided material from multiple views, adds details to the renderings using a diffusion model, and then backpropagates the changes to the material using inverse rendering.
Inverse rendering requires detail to be consistent across views, and we achieve this with two technical contributions: noise correlation by projecting from a reference noise anchored in the UV space, and attention biasing using the known geometry of the object.
This requires no new datasets or expensive retraining and is largely built from off-the-shelf, pre-trained components.

The resulting method serves the important use case of \emph{human-in-the-loop} authoring: Rather than entirely replacing the artist and generating materials from scratch, we allow the artist to maintain creative control using traditional workflows, while reducing the time spent on tedious detailing of assets---analogous to ``auto-complete'' for material detail. Because the input and output of our method are traditional materials, our method can be used at any stage in the authoring process, and the produced enhacements arbitrarily post-processed, blended, and combined. We believe our work builds a solid foundation for future practical tools that will further improve the robustness and controllability of the generative process.

\newcommand{\ablationSceneBlock}[3]{
    \multirow{2}{1em}{\rotatebox{90}{#2\hspace*{-#3}}}
    &
    \includegraphics[width=0.093\textwidth]{figures/ablation/#1/initial_0.jpg}&
    \includegraphics[width=0.093\textwidth]{figures/ablation/#1/initial_1.jpg}&
    \includegraphics[width=0.093\textwidth]{figures/ablation/#1/no_cnet_0.jpg}&
    \includegraphics[width=0.093\textwidth]{figures/ablation/#1/no_cnet_1.jpg}&
    \includegraphics[width=0.093\textwidth]{figures/ablation/#1/white_nop2p_0.jpg}&
    \includegraphics[width=0.093\textwidth]{figures/ablation/#1/white_nop2p_1.jpg}&
    \includegraphics[width=0.093\textwidth]{figures/ablation/#1/warped_nop2p_0.jpg}&
    \includegraphics[width=0.093\textwidth]{figures/ablation/#1/warped_nop2p_1.jpg}&
    \includegraphics[width=0.093\textwidth]{figures/ablation/#1/full_0.jpg}&
    \includegraphics[width=0.093\textwidth]{figures/ablation/#1/full_1.jpg}\\
    &
    \includegraphics[width=0.093\textwidth]{figures/ablation/#1/initial_crop_0.jpg}&
    \includegraphics[width=0.093\textwidth]{figures/ablation/#1/initial_crop_1.jpg}&
    \includegraphics[width=0.093\textwidth]{figures/ablation/#1/no_cnet_crop_0.jpg}&
    \includegraphics[width=0.093\textwidth]{figures/ablation/#1/no_cnet_crop_1.jpg}&
    \includegraphics[width=0.093\textwidth]{figures/ablation/#1/white_nop2p_crop_0.jpg}&
    \includegraphics[width=0.093\textwidth]{figures/ablation/#1/white_nop2p_crop_1.jpg}&
    \includegraphics[width=0.093\textwidth]{figures/ablation/#1/warped_nop2p_crop_0.jpg}&
    \includegraphics[width=0.093\textwidth]{figures/ablation/#1/warped_nop2p_crop_1.jpg}&
    \includegraphics[width=0.093\textwidth]{figures/ablation/#1/full_crop_0.jpg}&
    \includegraphics[width=0.093\textwidth]{figures/ablation/#1/full_crop_1.jpg}\\[0.3em]
}

\begin{figure*}[ht!]
    \centering%
    \setlength{\tabcolsep}{0.001\textwidth}%
    \renewcommand{\arraystretch}{0.5}%
    \footnotesize%
    \centering
    \begin{tabular}{ccc@{\hspace{0.5em}}cc@{\hspace{0.5em}}cc@{\hspace{0.5em}}cc@{\hspace{0.5em}}cc}
        &
        \multicolumn{2}{c}{Conditioning images}&
        \multicolumn{2}{c}{Baseline}&
        \multicolumn{2}{c}{+ ControlNet tile}&
        \multicolumn{2}{c}{+ View-correlated noise}&
        \multicolumn{2}{c}{+ Attention bias (full model)}\\[0.3em]
        \ablationSceneBlock{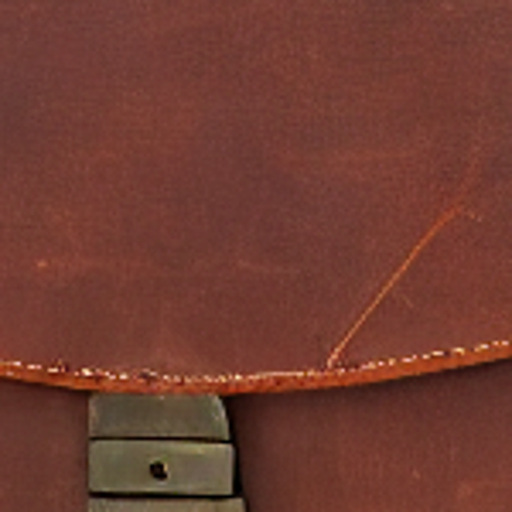}{Briefcase}{1em}
        \ablationSceneBlock{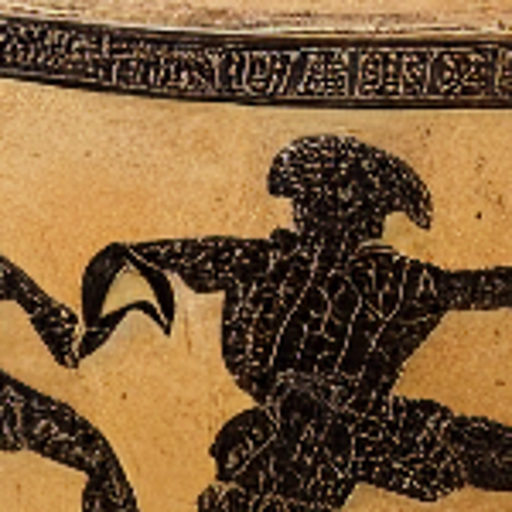}{Greek vase}{2em}
    \end{tabular}
    \vspace{-2mm}
    \caption{We study the impact of individual components of our technique by adding them one-by-one to a multi-view prompting baseline. Using the ControlNet tile significantly improves the overall visual quality. The view-correlated noise and biased attention improve the multi-view consistency.}
    \label{fig:ablation}
\end{figure*}

\begin{figure*}[ht!]
    \centering%
    \setlength{\tabcolsep}{0.001\textwidth}%
    \renewcommand{\arraystretch}{0.5}%
    \footnotesize%
    \centering
    \begin{tabular}{ccc@{\hspace{0.4em}}cc@{\hspace{0.4em}}cc}    
        &
        \multicolumn{2}{c}{(\textbf{a}) ControlNet tile \& normal}&
        \multicolumn{2}{c}{(\textbf{b}) + Multi-view visual prompting}&
        \multicolumn{2}{c}{(\textbf{c}) + View-correlated noise \& attention bias}\\[0.3em]
        \multirow{3}{*}{\rotatebox{90}{Diffusion-generated images\hspace*{-3em}}}&
        \includegraphics[width=0.161\textwidth, trim=0 50 0 0, clip]{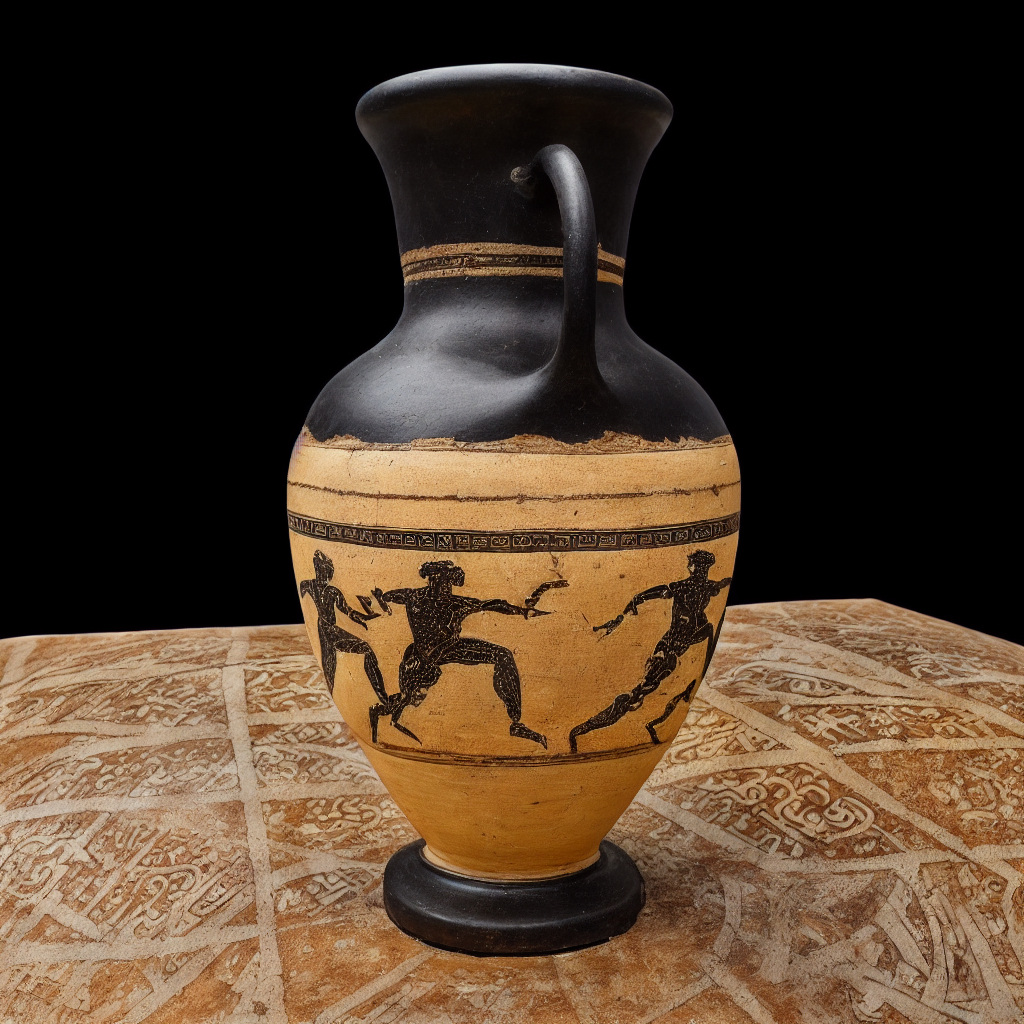}&
        \includegraphics[width=0.161\textwidth, trim=0 50 0 0, clip]{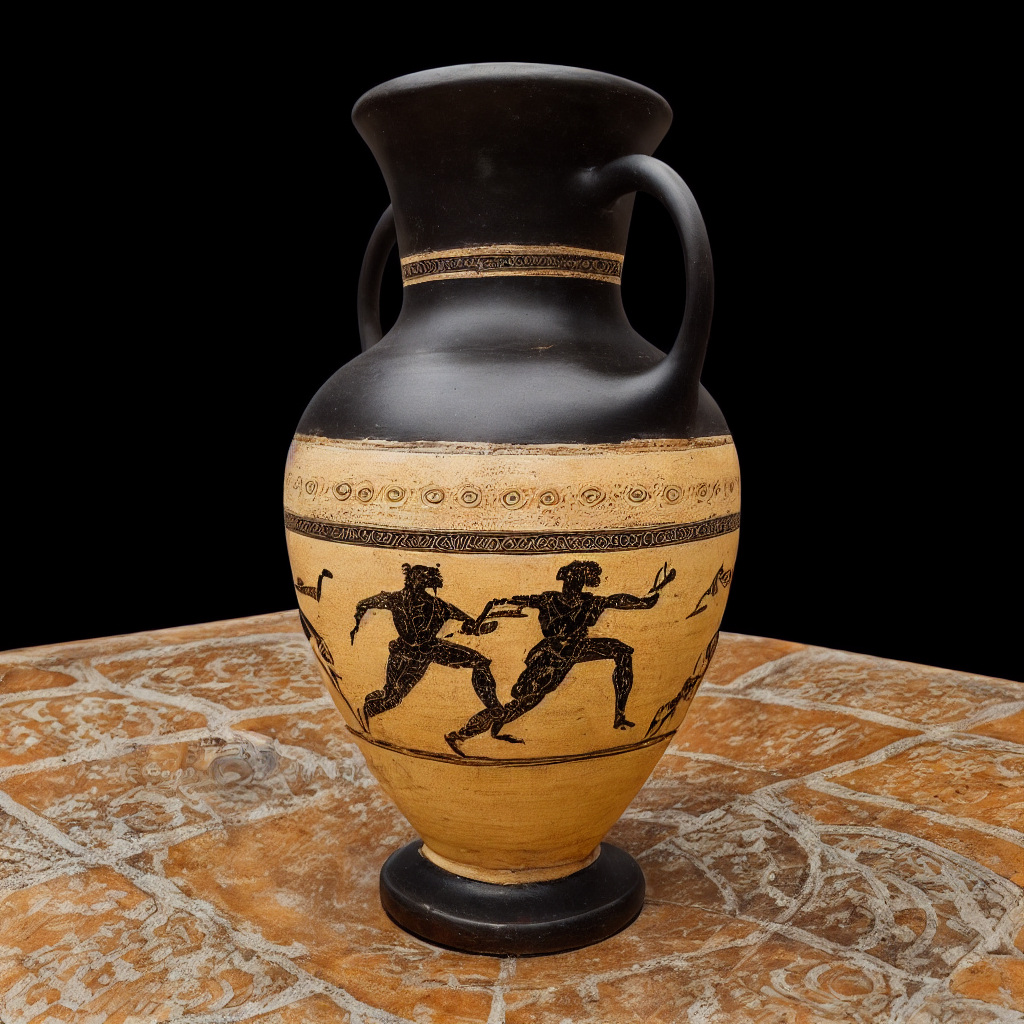}&
        \includegraphics[width=0.161\textwidth, trim=0 50 0 0, clip]{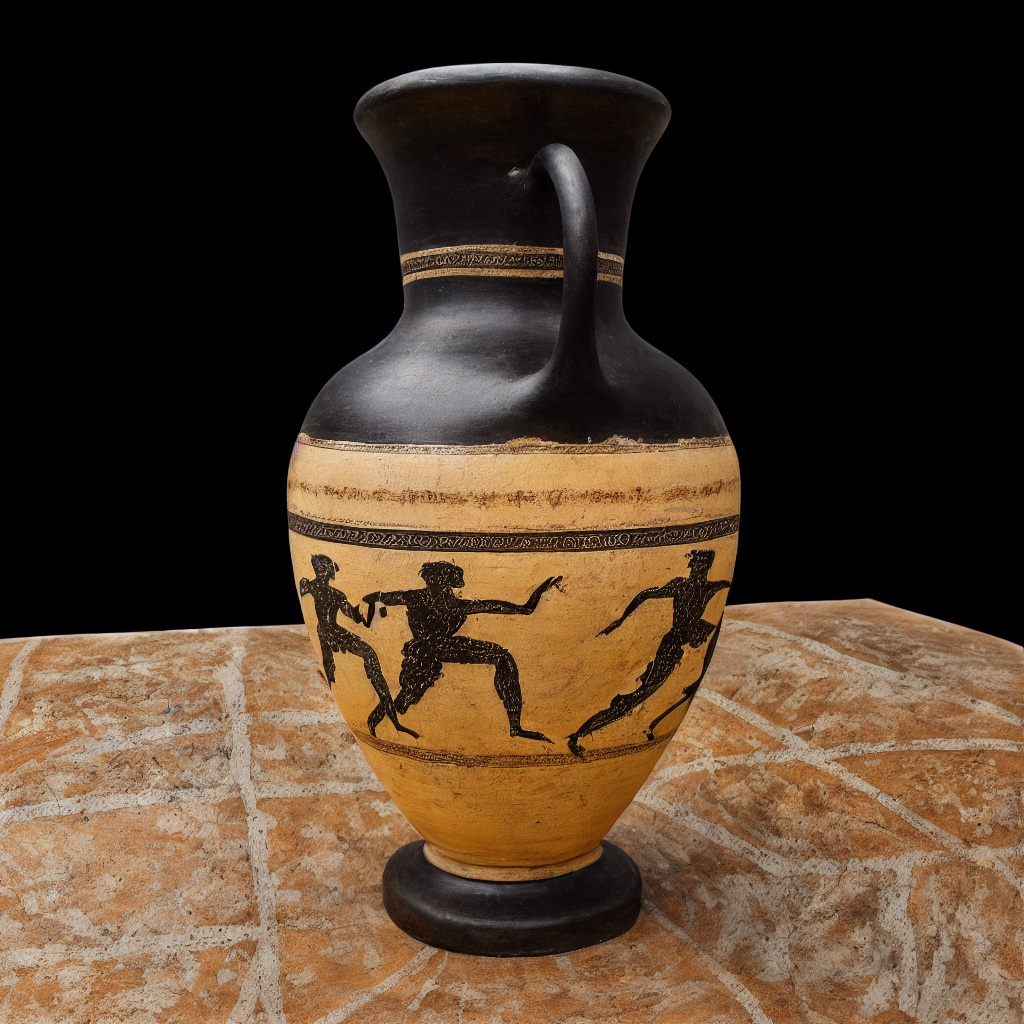}&
        \includegraphics[width=0.161\textwidth, trim=0 50 0 0, clip]{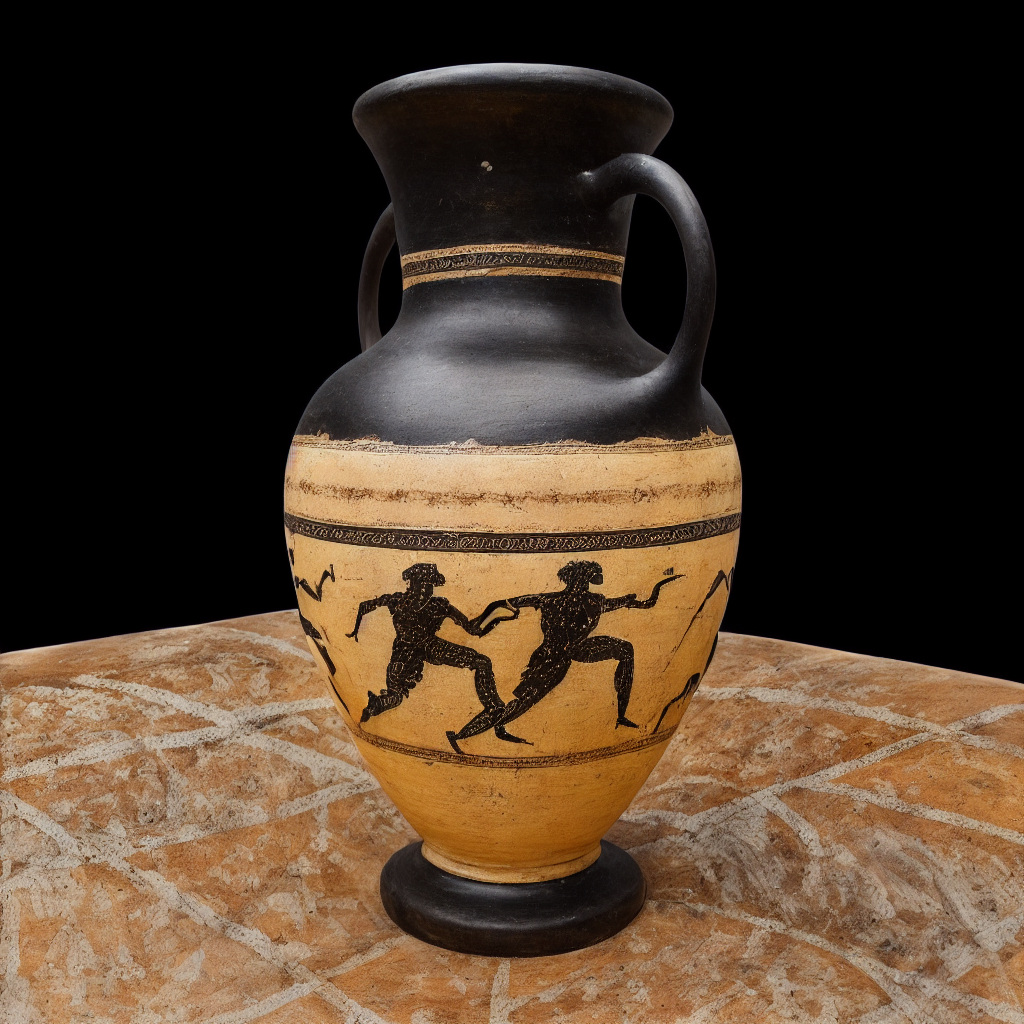}&
        \includegraphics[width=0.161\textwidth, trim=0 50 0 0, clip]{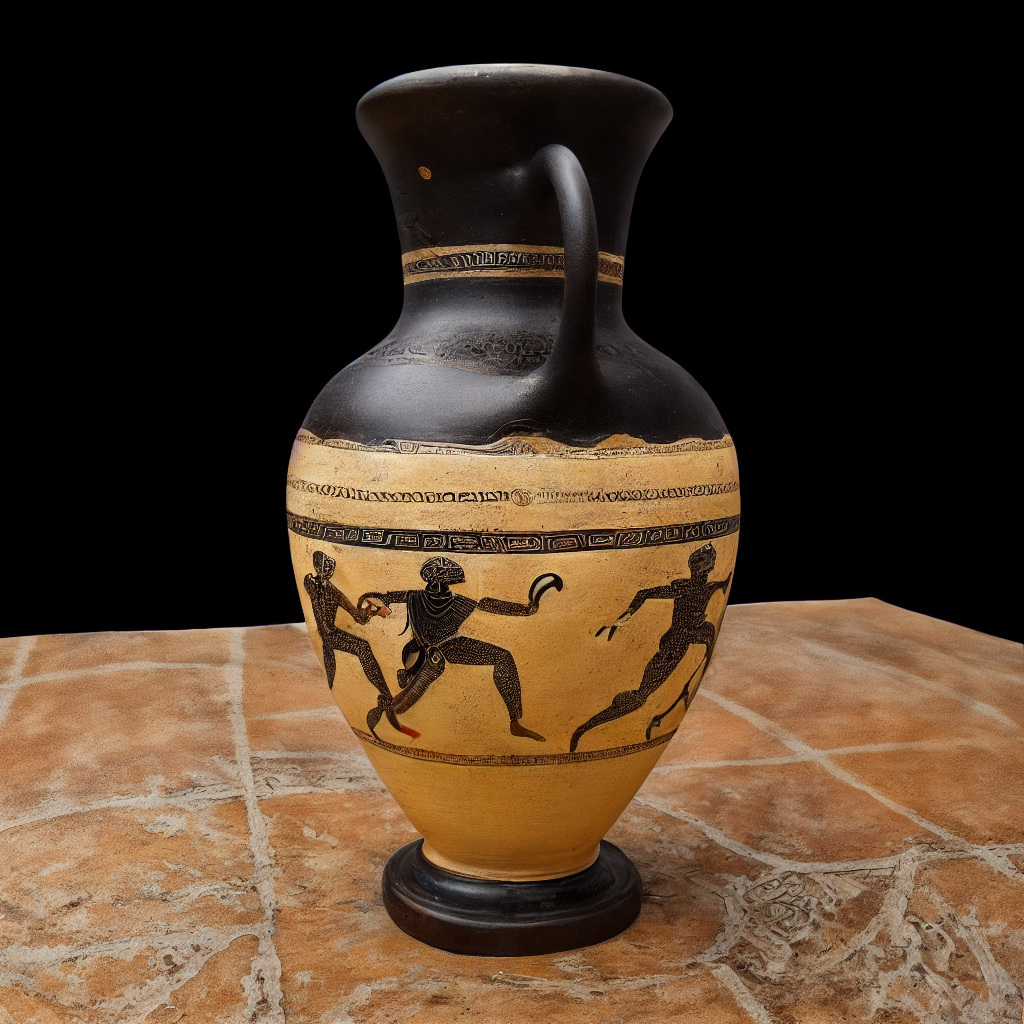}&
        \includegraphics[width=0.161\textwidth, trim=0 50 0 0, clip]{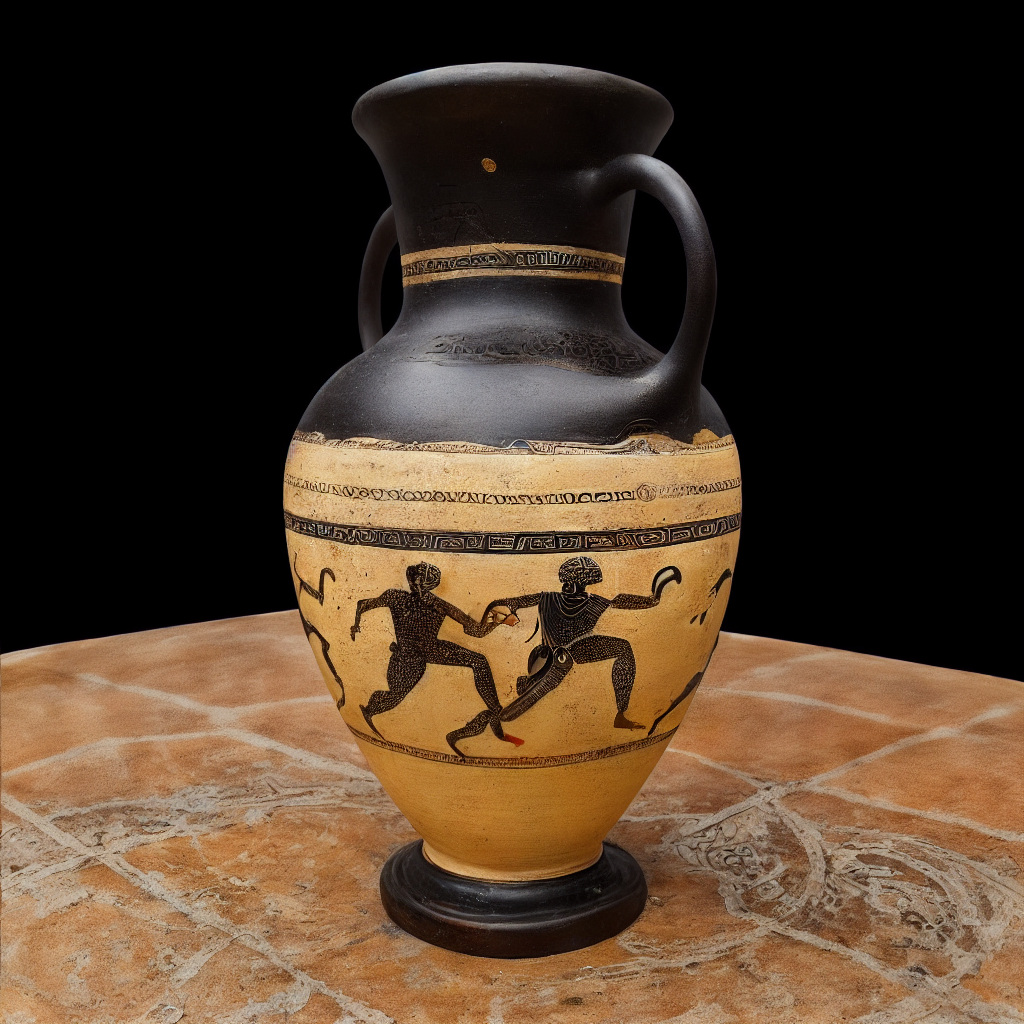}\\
        &
        \includegraphics[width=0.161\textwidth]{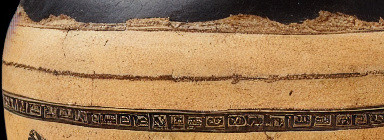}&
        \includegraphics[width=0.161\textwidth]{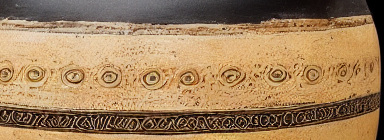}&
        \includegraphics[width=0.161\textwidth]{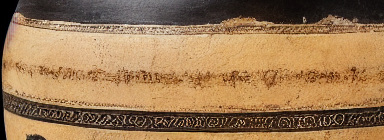}&
        \includegraphics[width=0.161\textwidth]{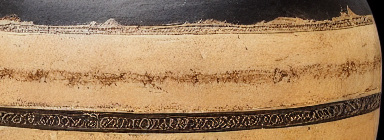}&
        \includegraphics[width=0.161\textwidth]{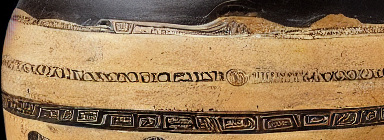}&
        \includegraphics[width=0.161\textwidth]{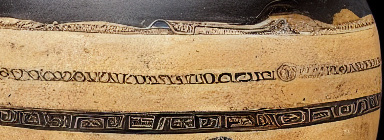}\\
        &
        \includegraphics[width=0.161\textwidth, trim=0 40 0 0, clip]{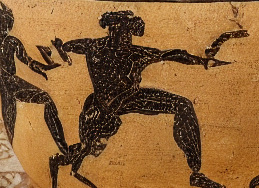}&
        \includegraphics[width=0.161\textwidth, trim=0 40 0 0, clip]{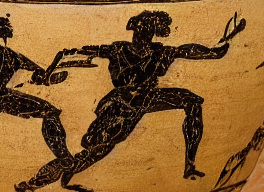}&
        \includegraphics[width=0.161\textwidth, trim=0 40 0 0, clip]{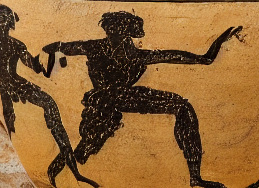}&
        \includegraphics[width=0.161\textwidth, trim=0 40 0 0, clip]{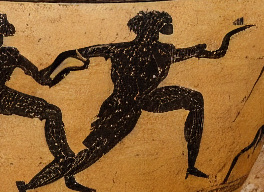}&
        \includegraphics[width=0.161\textwidth, trim=0 40 0 0, clip]{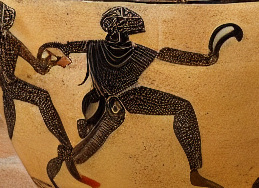}&
        \includegraphics[width=0.161\textwidth, trim=0 40 0 0, clip]{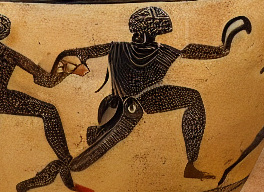}\\[2pt]
        \multirow{3}{*}{\rotatebox{90}{Renderings with reconstructed material\hspace*{-8em}}}&
        \includegraphics[width=0.161\textwidth, trim=0 50 0 0, clip]{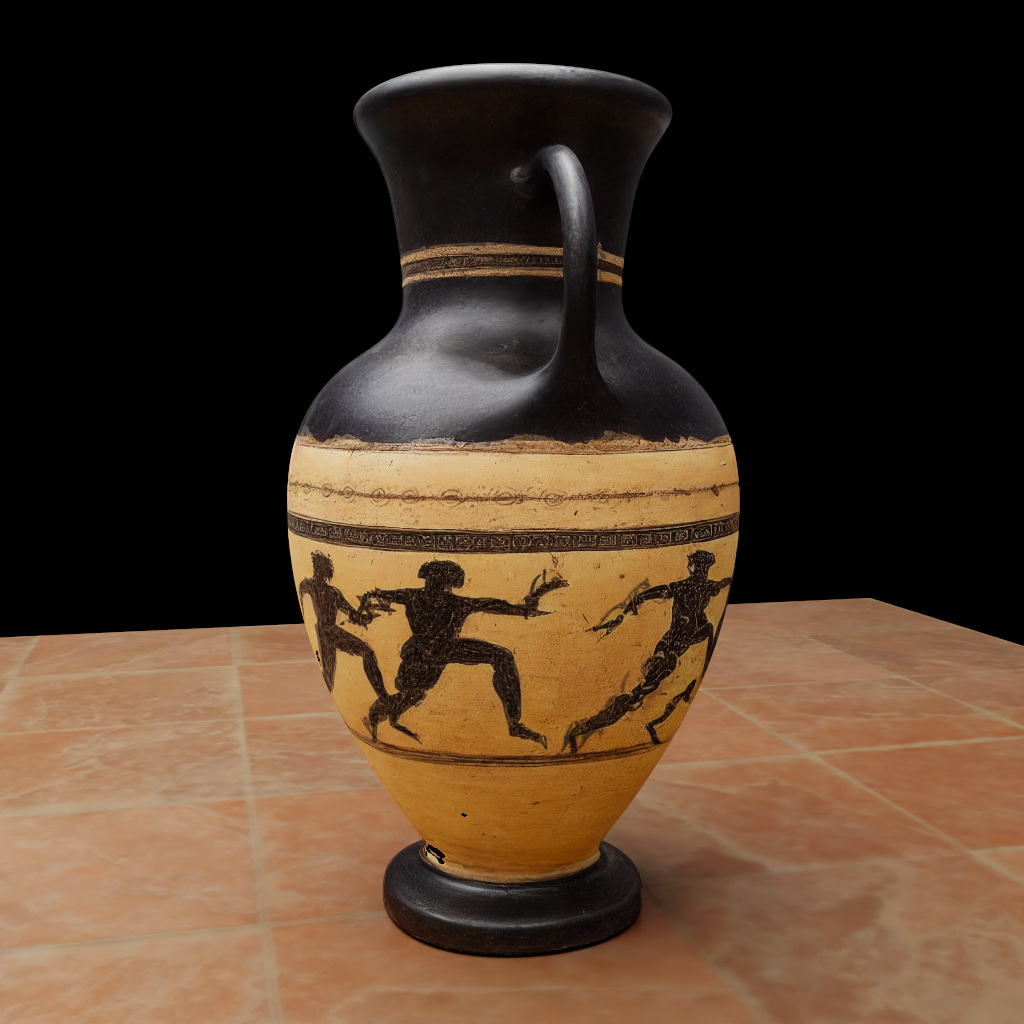}&
        \includegraphics[width=0.161\textwidth, trim=0 50 0 0, clip]{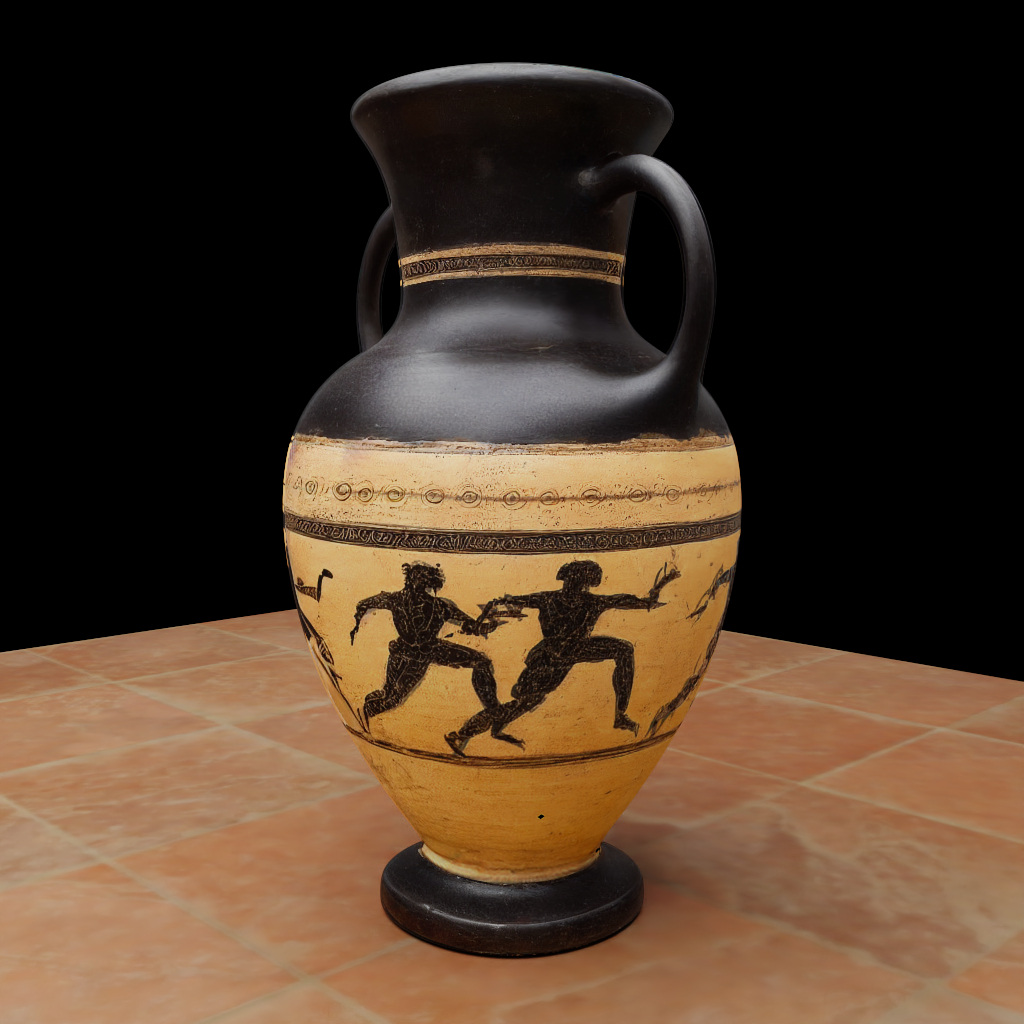}&
        \includegraphics[width=0.161\textwidth, trim=0 50 0 0, clip]{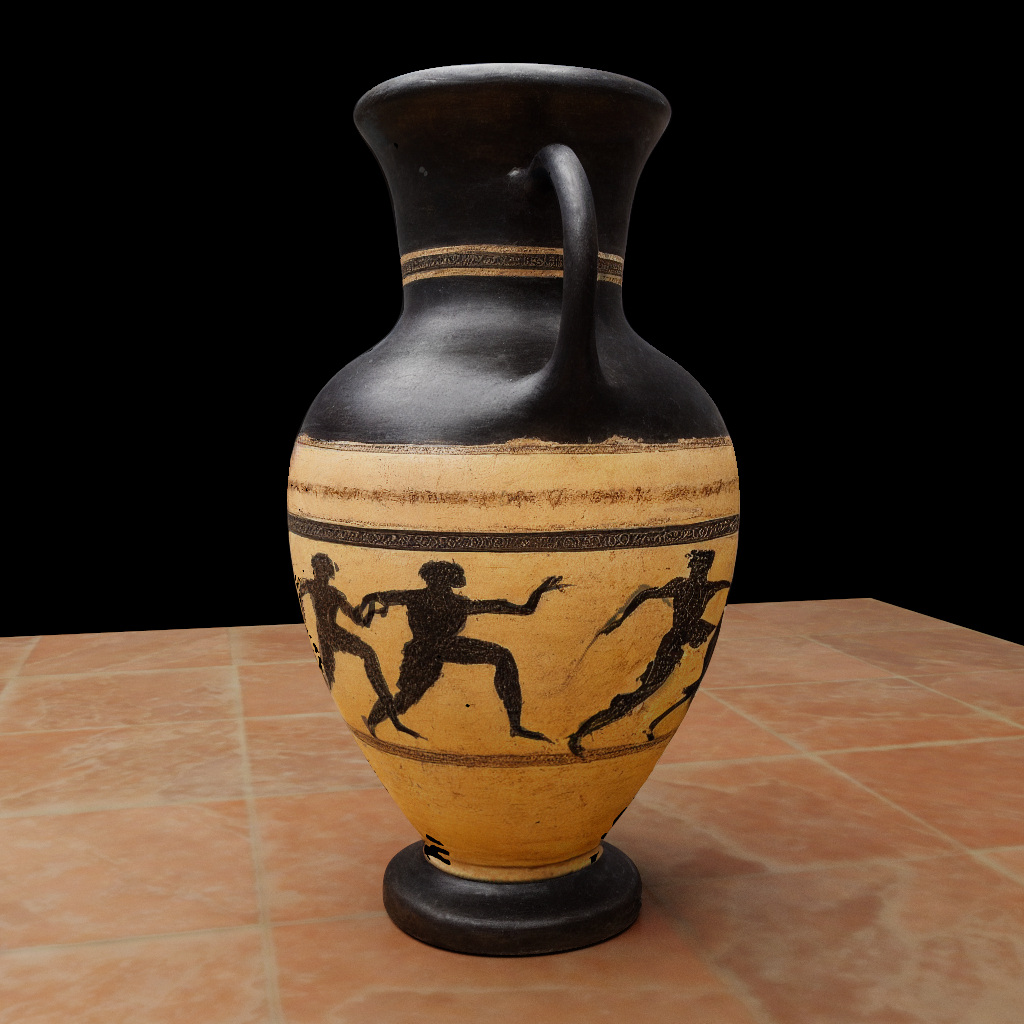}&
        \includegraphics[width=0.161\textwidth, trim=0 50 0 0, clip]{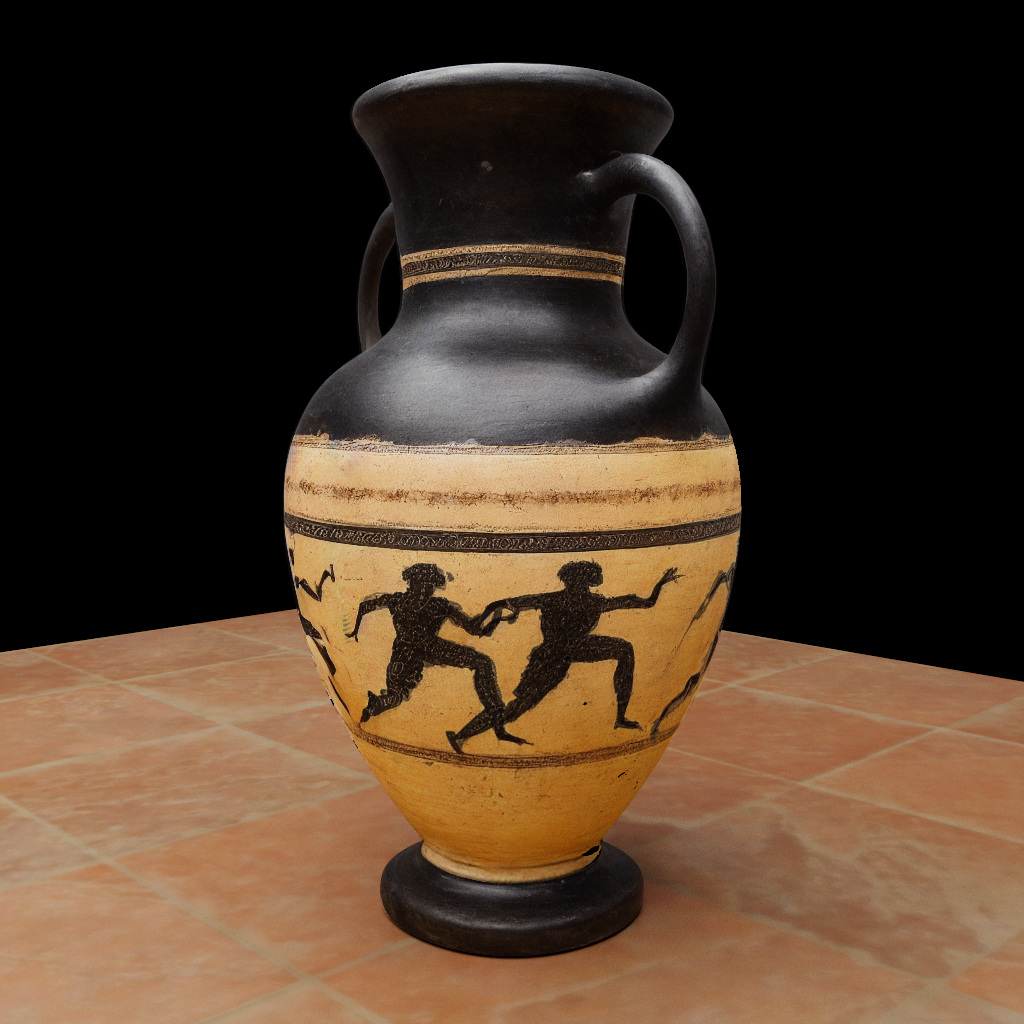}&
        \includegraphics[width=0.161\textwidth, trim=0 50 0 0, clip]{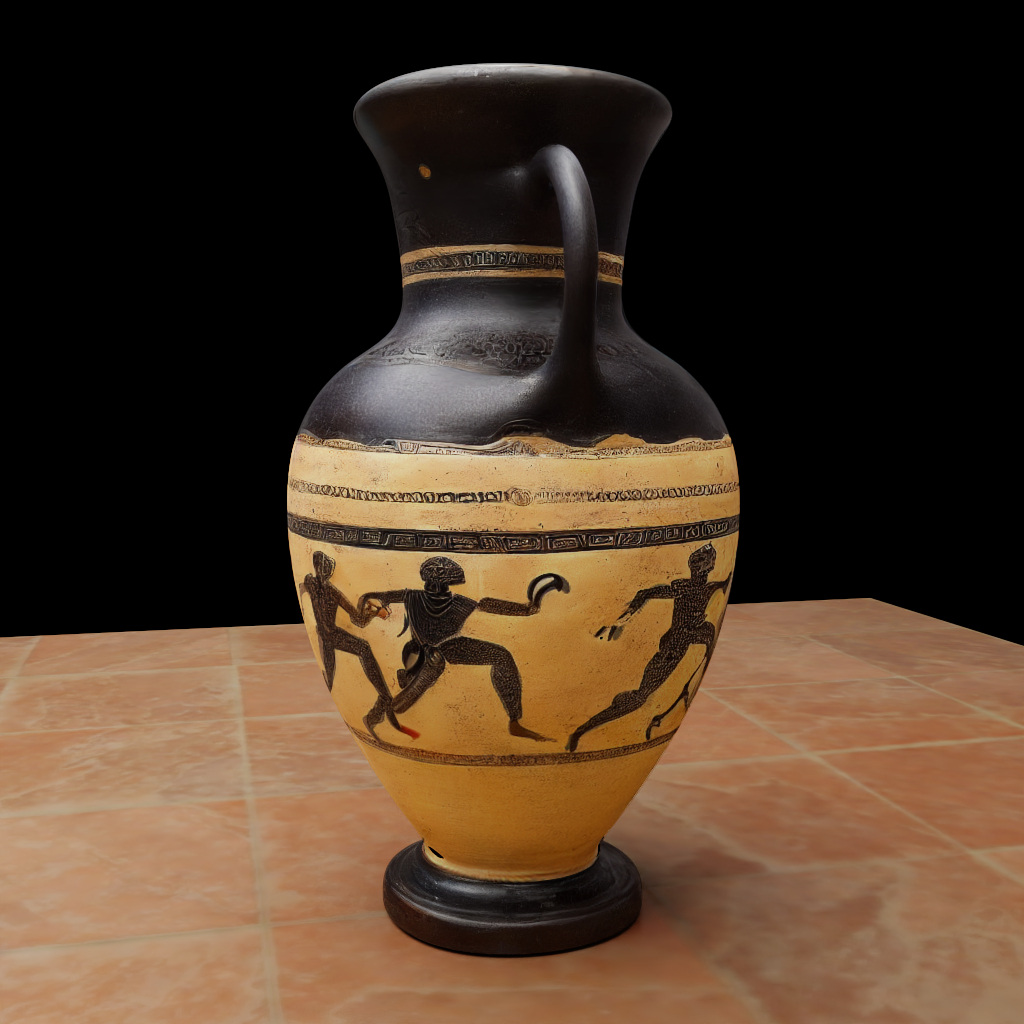}&
        \includegraphics[width=0.161\textwidth, trim=0 50 0 0, clip]{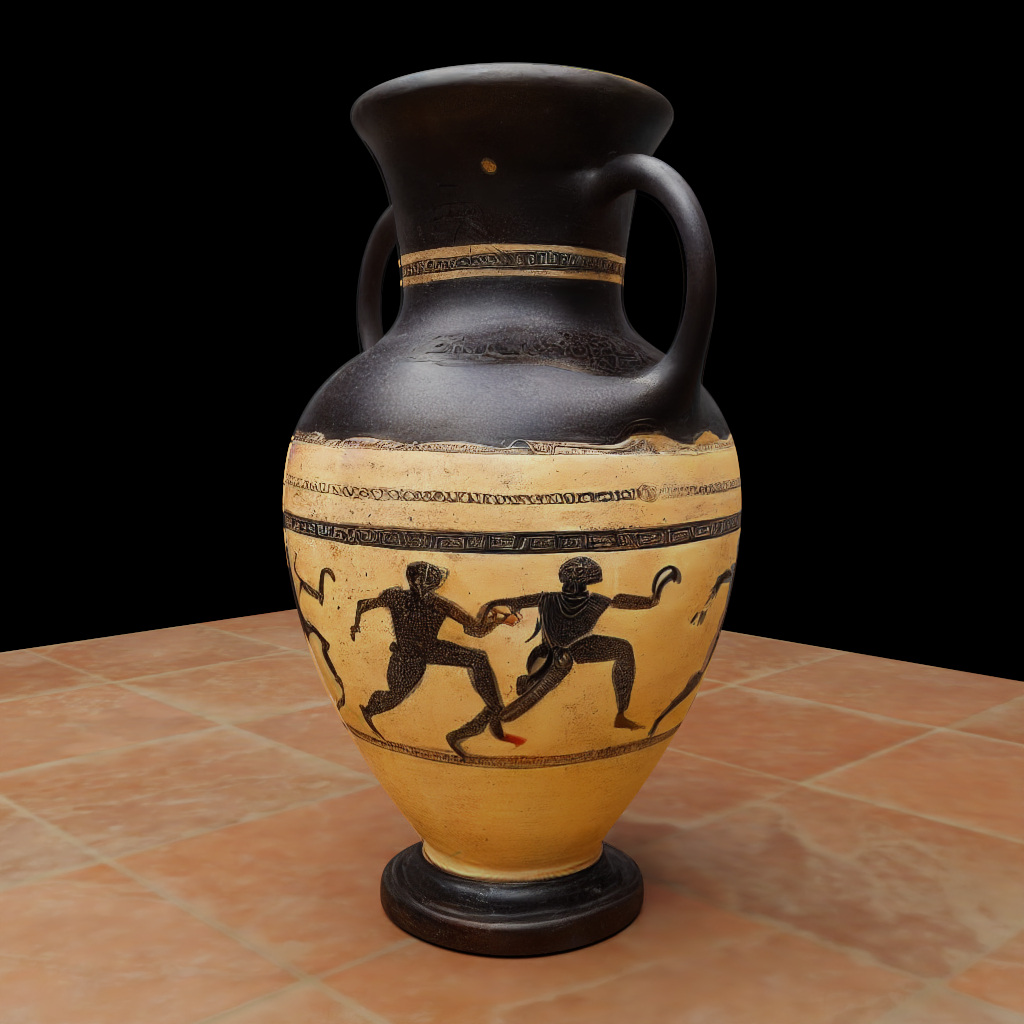}\\
        &
        \includegraphics[width=0.161\textwidth]{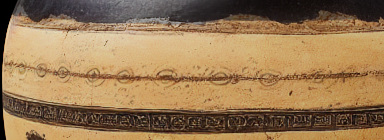}&
        \includegraphics[width=0.161\textwidth]{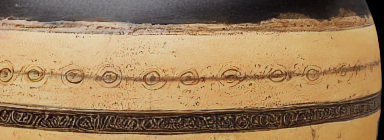}&
        \includegraphics[width=0.161\textwidth]{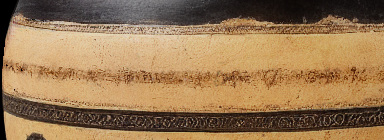}&
        \includegraphics[width=0.161\textwidth]{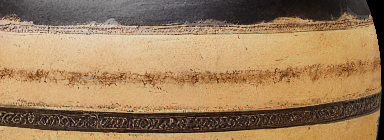}&
        \includegraphics[width=0.161\textwidth]{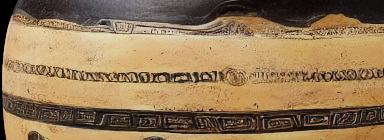}&
        \includegraphics[width=0.161\textwidth]{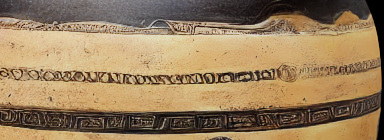}\\
        &
        \includegraphics[width=0.161\textwidth, trim=0 40 0 0, clip]{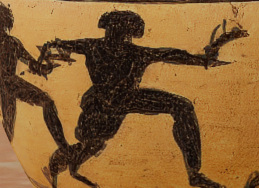}&
        \includegraphics[width=0.161\textwidth, trim=0 40 0 0, clip]{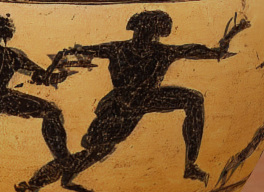}&
        \includegraphics[width=0.161\textwidth, trim=0 40 0 0, clip]{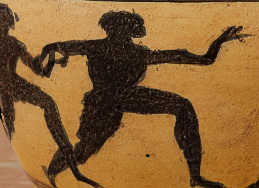}&
        \includegraphics[width=0.161\textwidth, trim=0 40 0 0, clip]{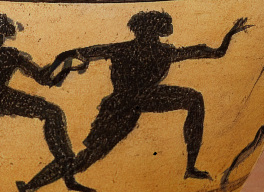}&
        \includegraphics[width=0.161\textwidth, trim=0 40 0 0, clip]{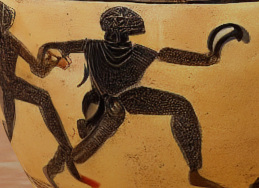}&
        \includegraphics[width=0.161\textwidth, trim=0 40 0 0, clip]{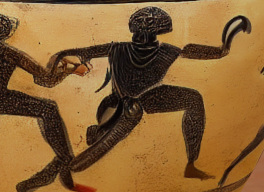}\\
        & View 1 & View 2 & View 1 & View 2 & View 1 & View 2
    \end{tabular}
    \vspace{-2mm}
    \caption{We compare results with our model, first without the multi-view visual prompting (i.e., assembling the conditioning images into a grid)~\cite{flashtex} (\textbf{a}), with it (\textbf{b}), and also with our two techniques for improving multi-view consistency.
    }
    \label{fig:invrender_inconsistent}
\end{figure*}

\newcommand{\padthree}[1]{%
  \ifnum#1<10 00#1\else%
  \ifnum#1<100 0#1\else%
  #1%
  \fi\fi%
}

\newcommand{\newsceneblock}[6]{%
\begin{tabular}{cccccc}
    \multicolumn{6}{c}{#2} \\
    \multicolumn{2}{c}{(a) Conditioning renders} &
    \multicolumn{2}{c}{(b) Diffusion outputs} &
    \multicolumn{2}{c}{(c) Reconstructed material} \\
    \begin{overpic}[width=0.16\textwidth]{figures/results/#1/initial_rendering_\padthree{#3}.jpg}
        \put(2,2){\tiny\textcolor{white}{View #3}}
    \end{overpic}
    &
    \begin{overpic}[width=0.16\textwidth]{figures/results/#1/initial_rendering_\padthree{#4}.jpg}
        \put(2,2){\tiny\textcolor{white}{View #4}}
    \end{overpic}
    &
    \begin{overpic}[width=0.16\textwidth]{figures/results/#1/diffused_\padthree{#3}.jpg}
        \put(2,2){\tiny\textcolor{white}{View #3}}
    \end{overpic}
    &
    \begin{overpic}[width=0.16\textwidth]{figures/results/#1/diffused_\padthree{#4}.jpg}
        \put(2,2){\tiny\textcolor{white}{View #4}}
    \end{overpic}
    &
    \begin{overpic}[width=0.16\textwidth]{figures/results/#1/masked/recovered_default_\padthree{#3}.jpg}
        \put(2,2){\tiny\textcolor{white}{View #3}}
    \end{overpic}
    &
    \begin{overpic}[width=0.16\textwidth]{figures/results/#1/masked/recovered_default_\padthree{#4}.jpg}
        \put(2,2){\tiny\textcolor{white}{View #4}}
    \end{overpic}
    \\
    \begin{overpic}[width=0.16\textwidth]{figures/results/#1/initial_rendering_\padthree{#5}.jpg}
        \put(2,2){\tiny\textcolor{white}{View 4}}
    \end{overpic}
    &
    \begin{overpic}[width=0.16\textwidth]{figures/results/#1/initial_rendering_\padthree{#6}.jpg}
        \put(2,2){\tiny\textcolor{white}{View #6}}
    \end{overpic}
    &
    \begin{overpic}[width=0.16\textwidth]{figures/results/#1/diffused_\padthree{#5}.jpg}
        \put(2,2){\tiny\textcolor{white}{View #5}}
    \end{overpic}
    &
    \begin{overpic}[width=0.16\textwidth]{figures/results/#1/diffused_\padthree{#6}.jpg}
        \put(2,2){\tiny\textcolor{white}{View #6}}
    \end{overpic}
    &
    \begin{overpic}[width=0.16\textwidth]{figures/results/#1/masked/recovered_default_\padthree{#5}.jpg}
        \put(2,2){\tiny\textcolor{white}{View #5}}
    \end{overpic}
    &
    \begin{overpic}[width=0.16\textwidth]{figures/results/#1/masked/recovered_default_\padthree{#6}.jpg}
        \put(2,2){\tiny\textcolor{white}{View #6}}
    \end{overpic}
    \\
    \midrule
    (d) Initial albedo & (e) Reconstructed albedo & (f) Initial normal & (g) Reconstructed normal & (h) Initial roughness & (i) Reconstructed roughness \\
    \includegraphics[width=0.16\textwidth]{figures/results/#1/texture_albedo_initial.jpg} &
    \includegraphics[width=0.16\textwidth]{figures/results/#1/texture_albedo_edited.jpg} &
    \includegraphics[width=0.16\textwidth]{figures/results/#1/texture_normals_initial.jpg} &
    \includegraphics[width=0.16\textwidth]{figures/results/#1/texture_normals_edited.jpg} &
    \includegraphics[width=0.16\textwidth]{figures/results/#1/texture_roughness_initial.jpg} &
    \includegraphics[width=0.16\textwidth]{figures/results/#1/texture_roughness_edited.jpg}
    \vspace*{4mm}
\end{tabular}
}

\begin{figure*}
    \centering%
    \setlength{\tabcolsep}{0.002\textwidth}%
    \renewcommand{\arraystretch}{1}%
    \footnotesize%
    \newsceneblock{AC}{\small Air conditioner, prompt: \prompt{Overused, rusty, old air-conditioning unit}}{1}{2}{5}{6}
    \newsceneblock{statue}{\small Statue, prompt \prompt{Mossy stone statue}}{1}{2}{6}{7}
    \vspace*{-6mm}
    \caption{We use renderings of the original asset (two pairs of two adjacent views are shown in (a)) to condition the diffusion model to produce images with enhanced appearance (b), which is then back-propagated into the original material definition. In (c), we show the resulting asset on four frames from a turntable animation (we picked frames that correspond to the conditioning views); see the accompanying video for the full animation.
    Columns (d) to (i) show albedo, normal, and roughness textures of the initial and reconstructed material.
    }
    \label{fig:results}
\end{figure*}

\begin{acks}

This material is based upon work supported by the National Science Foundation under Grant No. IIS2126407. 
We would like to thank Aaron Lefohn for supporting this research, and NVIDIA for funding the work with an NVIDIA academic partnership. 
We would also like to thank Arash Vahdat and Weili Nie for the highly insightful discussions on multi-view generation using diffusion models.

\end{acks}

\bibliographystyle{ACM-Reference-Format}
\bibliography{bibliography}

\appendix
\section{Additional results}

We present full results of materials from the main paper as well as  additional results in Figures~\ref{fig:results-a}--\ref{fig:results-c}. The presentation follows the main paper and shows results of our complete pipeline starting from basic assets all the way to the recovered material parameters.

Figure~\ref{fig:attention_viz_suppl} shows an uncropped version of Figure~\ref{fig:attention_viz} in the main paper, and visualizes exemplary attention scores before and after adding the bias for a single specific
surface point.

Finally, Figure~\ref{fig:hparams_usercontrol_suppl} is an extended version of Figure~\ref{fig:hparams_usercontrol} from the main paper and demonstrates the full set of user-controlled parameters and their impact on the generated visuals. The \emph{classifier-free guidance} parameter controls the amount of adherence to the prompt; the \emph{CNet Tile scale} and the added noise parameters control how much the generated material is allowed to deviate from the input views, and trades off the amount of detail vs preservation of the original design intent.

\newcommand{\supplpadthree}[1]{%
  \ifnum#1<10 00#1\else%
  \ifnum#1<100 0#1\else%
  #1%
  \fi\fi%
}

\newcommand{\supplnewsceneblock}[6]{%
\begin{tabular}{cccccc}
    \multicolumn{6}{>{\centering\arraybackslash}p{\textwidth}}{#2} \\
    \multicolumn{2}{c}{(a) Conditioning renders} &
    \multicolumn{2}{c}{(b) Diffusion outputs} &
    \multicolumn{2}{c}{(c) Reconstructed material} \\
    \begin{overpic}[width=0.16\textwidth]{figures/results/#1/initial_rendering_\supplpadthree{#3}.jpg}
        \put(2,2){\tiny\textcolor{white}{View #3}}
    \end{overpic}
    &
    \begin{overpic}[width=0.16\textwidth]{figures/results/#1/initial_rendering_\supplpadthree{#4}.jpg}
        \put(2,2){\tiny\textcolor{white}{View #4}}
    \end{overpic}
    &
    \begin{overpic}[width=0.16\textwidth]{figures/results/#1/diffused_\supplpadthree{#3}.jpg}
        \put(2,2){\tiny\textcolor{white}{View #3}}
    \end{overpic}
    &
    \begin{overpic}[width=0.16\textwidth]{figures/results/#1/diffused_\supplpadthree{#4}.jpg}
        \put(2,2){\tiny\textcolor{white}{View #4}}
    \end{overpic}
    &
    \begin{overpic}[width=0.16\textwidth]{figures/results/#1/masked/recovered_default_\supplpadthree{#3}.jpg}
        \put(2,2){\tiny\textcolor{white}{View #3}}
    \end{overpic}
    &
    \begin{overpic}[width=0.16\textwidth]{figures/results/#1/masked/recovered_default_\supplpadthree{#4}.jpg}
        \put(2,2){\tiny\textcolor{white}{View #4}}
    \end{overpic}
    \\
    \begin{overpic}[width=0.16\textwidth]{figures/results/#1/initial_rendering_\supplpadthree{#5}.jpg}
        \put(2,2){\tiny\textcolor{white}{View 4}}
    \end{overpic}
    &
    \begin{overpic}[width=0.16\textwidth]{figures/results/#1/initial_rendering_\supplpadthree{#6}.jpg}
        \put(2,2){\tiny\textcolor{white}{View #6}}
    \end{overpic}
    &
    \begin{overpic}[width=0.16\textwidth]{figures/results/#1/diffused_\supplpadthree{#5}.jpg}
        \put(2,2){\tiny\textcolor{white}{View #5}}
    \end{overpic}
    &
    \begin{overpic}[width=0.16\textwidth]{figures/results/#1/diffused_\supplpadthree{#6}.jpg}
        \put(2,2){\tiny\textcolor{white}{View #6}}
    \end{overpic}
    &
    \begin{overpic}[width=0.16\textwidth]{figures/results/#1/masked/recovered_default_\supplpadthree{#5}.jpg}
        \put(2,2){\tiny\textcolor{white}{View #5}}
    \end{overpic}
    &
    \begin{overpic}[width=0.16\textwidth]{figures/results/#1/masked/recovered_default_\supplpadthree{#6}.jpg}
        \put(2,2){\tiny\textcolor{white}{View #6}}
    \end{overpic}
    \\
    \midrule
    (d) Initial albedo & (e) Reconstructed albedo & (f) Initial normal & (g) Reconstructed normal & (h) Initial roughness & (i) Reconstructed roughness \\
    \includegraphics[width=0.16\textwidth]{figures/results/#1/texture_albedo_initial.jpg} &
    \includegraphics[width=0.16\textwidth]{figures/results/#1/texture_albedo_edited.jpg} &
    \includegraphics[width=0.16\textwidth]{figures/results/#1/texture_normals_initial.jpg} &
    \includegraphics[width=0.16\textwidth]{figures/results/#1/texture_normals_edited.jpg} &
    \includegraphics[width=0.16\textwidth]{figures/results/#1/texture_roughness_initial.jpg} &
    \includegraphics[width=0.16\textwidth]{figures/results/#1/texture_roughness_edited.jpg}
    \vspace*{4mm}
\end{tabular}
}

\begin{figure*}
    \centering%
    \setlength{\tabcolsep}{0.002\textwidth}%
    \renewcommand{\arraystretch}{1}%
    \footnotesize%
    \supplnewsceneblock{briefcase}{\small Briefcase, prompt: \prompt{Old, overused, weathered, scratched, leather briefcase}}{0}{1}{4}{5}
    \supplnewsceneblock{david}{\small David bust, prompt \prompt{Old, cracked statue of David}}{0}{1}{4}{5}
    \vspace*{-6mm}
    \caption{We use renderings of the original asset (two pairs of two adjacent views are shown in (a)) to condition the diffusion model to produce images with enhanced appearance (b), which is then back-propagated into the original material definition. In (c), we show the resulting asset on four frames from a turntable animation (we picked frames that correspond to the conditioning views); see the accompanying video for the full animation.
    Columns (d) to (i) show albedo, normal, and roughness textures of the initial and reconstructed material.
    }
    \label{fig:results-a}
\end{figure*}

\begin{figure*}
    \centering%
    \setlength{\tabcolsep}{0.002\textwidth}%
    \renewcommand{\arraystretch}{1}%
    \footnotesize%
    \supplnewsceneblock{pots}{\small Pottery Vases, prompt: \prompt{Weathered pottery vases}}{0}{2}{10}{12}
    \supplnewsceneblock{pot-simple}{\small Cooking Pot, prompt \prompt{rusty scratched old cooking pot}}{0}{1}{7}{8}
    \vspace*{-6mm}
    \caption{We use renderings of the original asset (two pairs of two adjacent views are shown in (a)) to condition the diffusion model to produce images with enhanced appearance (b), which is then back-propagated into the original material definition. In (c), we show the resulting asset on four frames from a turntable animation (we picked frames that correspond to the conditioning views); see the accompanying video for the full animation.
    Columns (d) to (i) show albedo, normal, and roughness textures of the initial and reconstructed material.
    }
    \label{fig:results-b}
\end{figure*}

\begin{figure*}
    \centering%
    \setlength{\tabcolsep}{0.002\textwidth}%
    \renewcommand{\arraystretch}{1}%
    \footnotesize%
    \supplnewsceneblock{woodenbox}{\small Wooden Box, prompt: \prompt{Aged, old, weathered wooden box}}{1}{2}{14}{15}
    \supplnewsceneblock{greekvase}{\small Greek Vase, prompt \prompt{A highly detailed ancient Greek vase, decorated with intricate mythological patterns and scenes of warriors, partially cracked and chipped from age, with areas of fading paint and earthy dust coating its surface. The vase is surrounded by a soft warm light. The style is realistic and cinematic, emphasizing texture and detail, with a focus on the cracks and aged patina on the vase.}}{4}{5}{10}{11}
    \vspace*{-6mm}
    \caption{We use renderings of the original asset (two pairs of two adjacent views are shown in (a)) to condition the diffusion model to produce images with enhanced appearance (b), which is then back-propagated into the original material definition. In (c), we show the resulting asset on four frames from a turntable animation (we picked frames that correspond to the conditioning views); see the accompanying video for the full animation.
    Columns (d) to (i) show albedo, normal, and roughness textures of the initial and reconstructed material.
    }
    \label{fig:results-c}
\end{figure*}

\begin{figure*}[h!]
    \centering%
    \setlength{\tabcolsep}{0.002\textwidth}%
    \renewcommand{\arraystretch}{1}%
    \footnotesize%
    \begin{tabular}{ccc}
        Unmodified attention scores
        &
        Latent pixel correspondences
        &
        Biased attention scores
        \\[0.8mm]
        \begin{overpic}[width=0.332\textwidth]{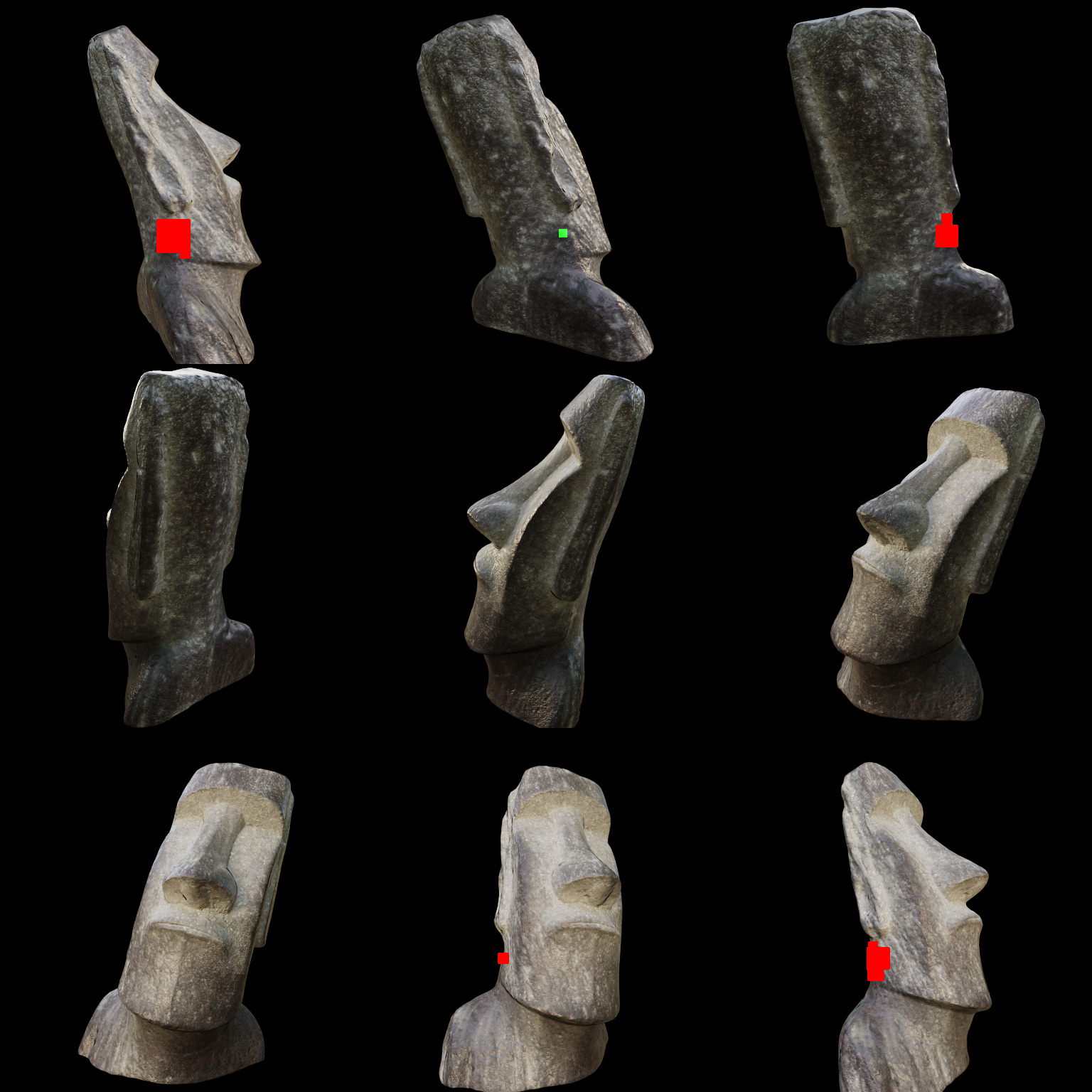}
            \begin{tikzpicture}[overlay, remember picture]
                \draw[-{Stealth[length=2mm]},green,thick,] (3.50,5.20) -- (3.12,4.73);
                \draw[-{Stealth[length=2mm]},red,thick,]   (1.45,5.25) -- (1.07,4.78);
                \draw[-{Stealth[length=2mm]},red,thick,]   (5.63,5.25) -- (5.25,4.78);
                \draw[-{Stealth[length=2mm]},red,thick,]   (4.27,1.31) -- (4.65,0.84);
                \draw[-{Stealth[length=2mm]},red,thick,]   (2.30,1.26) -- (2.68,0.79);
            \end{tikzpicture}
        \end{overpic}
        &
        \begin{overpic}[width=0.332\textwidth]{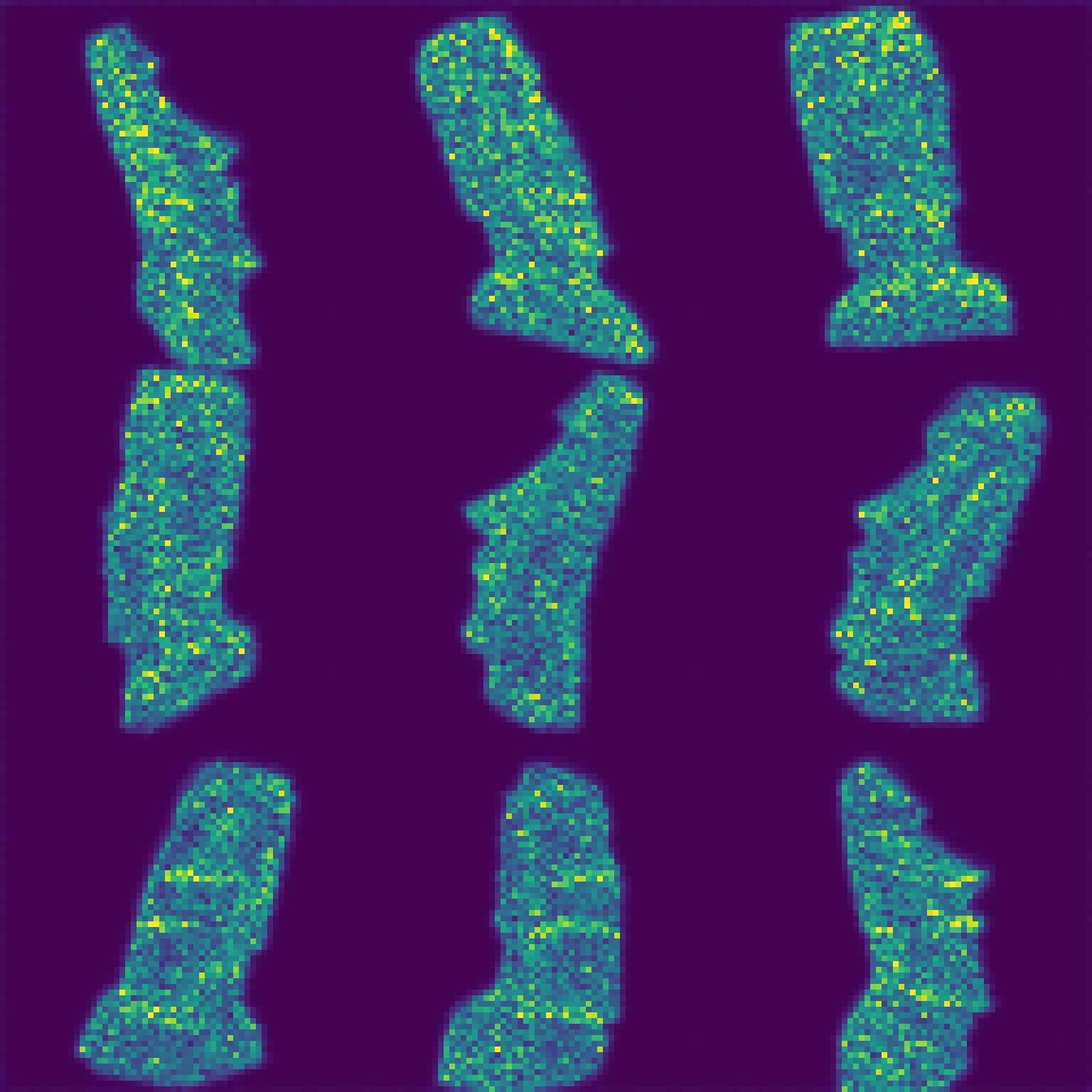}
        \end{overpic}
        &
        \begin{overpic}[width=0.332\textwidth]{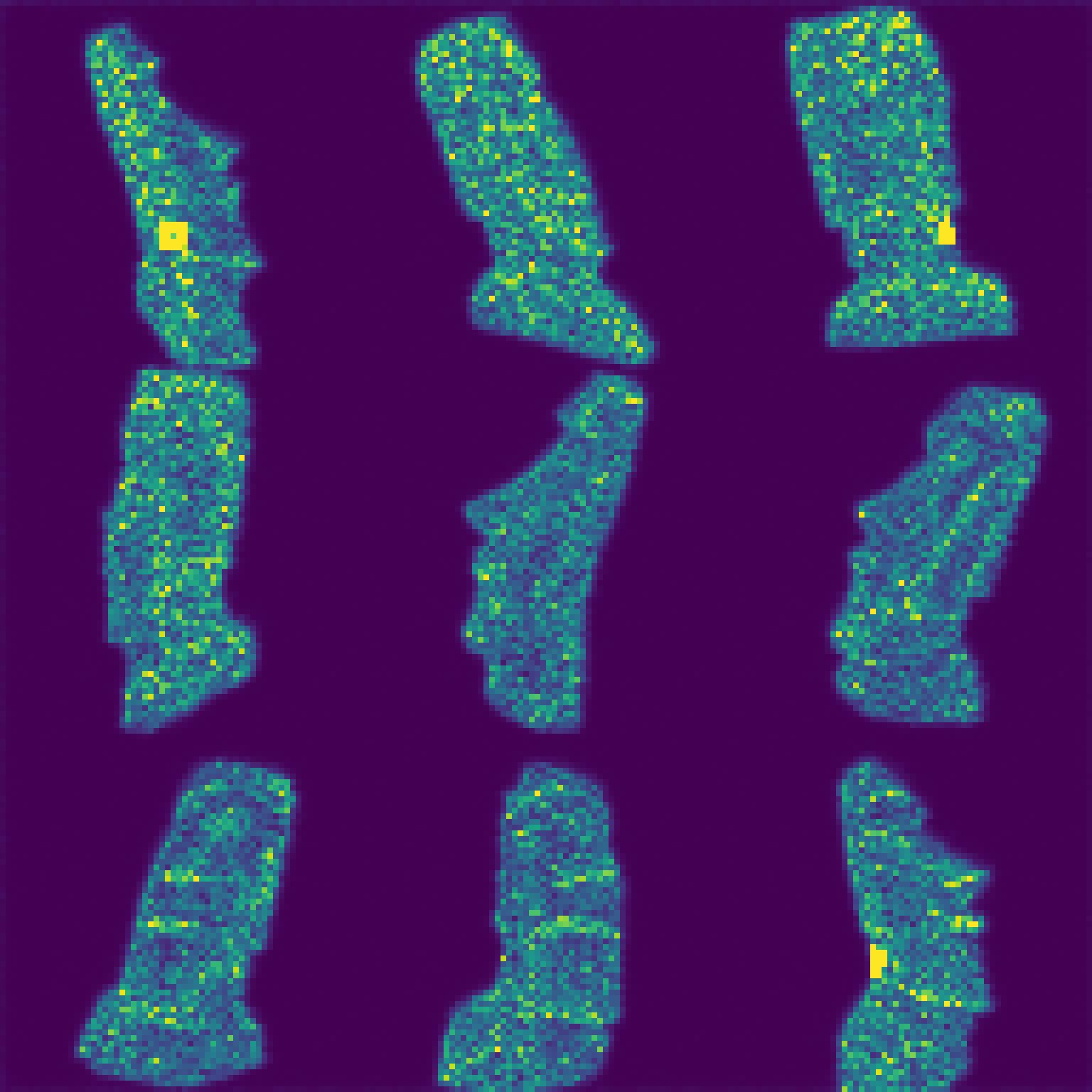}
            \begin{tikzpicture}[overlay, remember picture]
                \draw[-{Stealth[length=2mm]},red,thick,]   (1.45,5.25) -- (1.07,4.78);
                \draw[-{Stealth[length=2mm]},red,thick,]   (5.63,5.25) -- (5.25,4.78);
                \draw[-{Stealth[length=2mm]},red,thick,]   (4.27,1.31) -- (4.65,0.84);
                \draw[-{Stealth[length=2mm]},red,thick,]   (2.30,1.26) -- (2.68,0.79);
            \end{tikzpicture}
        \end{overpic}
    \end{tabular}
    \caption{
        An uncropped version of Figure~\ref{fig:attention_viz} from the main paper.
        Left:
        For one latent pixel (green), we highlight the corresponding neighborhoods in the other conditioning views in red; we use ray tracing to check for occlusions.
        Middle:
        One attention score matrix \emph{row} related to that green latent pixel can be rearranged into an image showing how much it attends to all other latent pixels in one stage of the diffusion model.
        Right:
        We bias the matrix entries in \emph{columns} that correspond to the identified red regions to promote attention---and hence consistency---between these latents.
    }
    \label{fig:attention_viz_suppl}
\end{figure*}
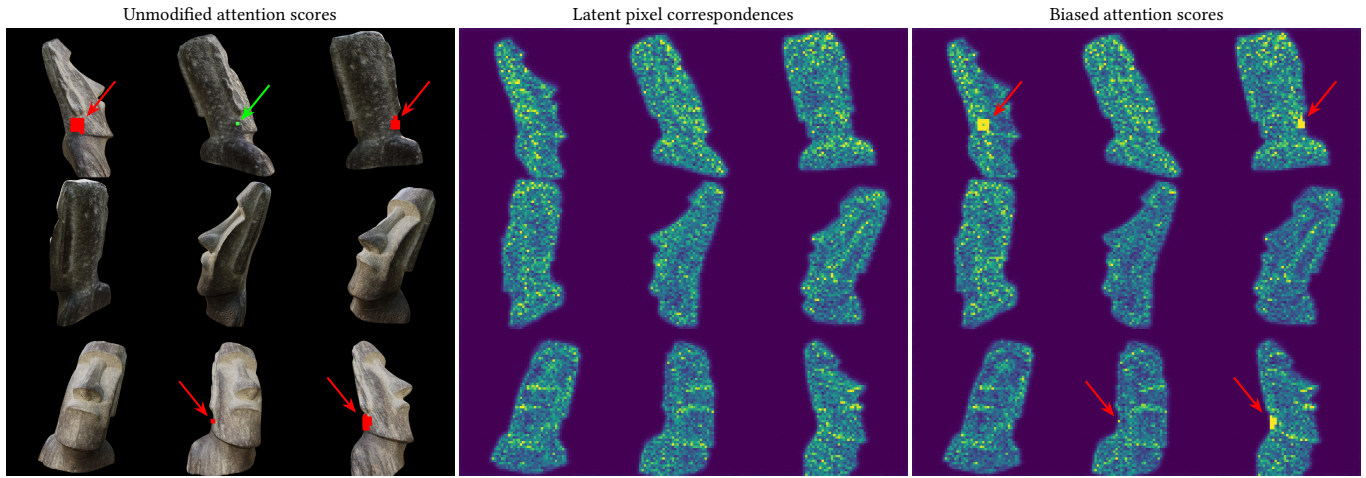

\begin{figure*}[htbp]
    \centering%
    \setlength{\tabcolsep}{0.002\textwidth}%
    \renewcommand{\arraystretch}{1}%
    \footnotesize%

    \begin{tabular}{cccccccc}
        &Initial asset& \multicolumn{6}{c}{
            \begin{tikzpicture}
                \draw[<-] (0,0) -- (0.45,0);
                \draw[-] (3.2,0) -- (6.0,0);
                \draw[-] (8.8,0) -- (11,0);
                \draw[->] (14.4,0) -- (15,0);
                \node[above] at (1.8,-0.2) {less emphasis on prompt};
                \node[above] at (7.5,-0.2) {Classifer free guidance};
                \node[above] at (13,-0.2) {more emphasis on prompt};
            \end{tikzpicture}
        }\\[-4pt]%
        && 4.0 & \textbf{7.5} & 11.0 & 14.5 & 18.0 & 21.5\\%
        \rotatebox{90}{\hspace*{4em}View 1}&%
        \includegraphics[height=0.14\linewidth]{figures/secondary_hparams/initial/0001.png}&%
        \includegraphics[height=0.14\linewidth]{figures/secondary_hparams/guidance_4.0_001.jpg}&%
        \includegraphics[height=0.14\linewidth]{figures/secondary_hparams/guidance_7.5_001.jpg}&%
        \includegraphics[height=0.14\linewidth]{figures/secondary_hparams/guidance_11.0_001.jpg}&%
        \includegraphics[height=0.14\linewidth]{figures/secondary_hparams/guidance_14.5_001.jpg}&%
        \includegraphics[height=0.14\linewidth]{figures/secondary_hparams/guidance_18.0_001.jpg}&%
        \includegraphics[height=0.14\linewidth]{figures/secondary_hparams/guidance_21.5_001.jpg}\\%
        \rotatebox{90}{\hspace*{4em}View 2}&
        \includegraphics[height=0.14\linewidth]{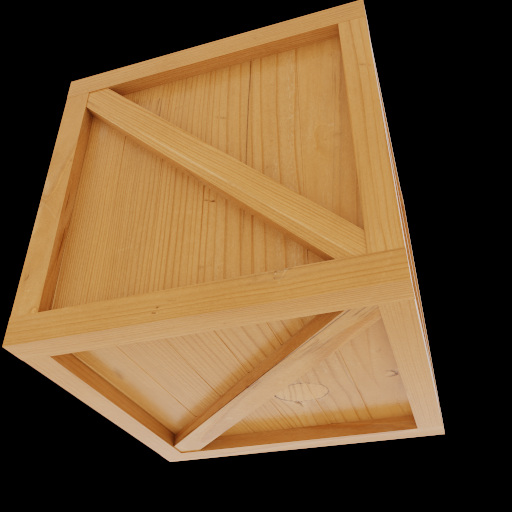}&%
        \includegraphics[height=0.14\linewidth]{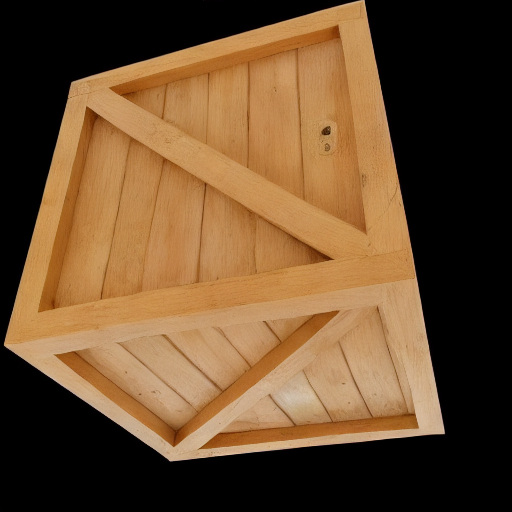}&%
        \includegraphics[height=0.14\linewidth]{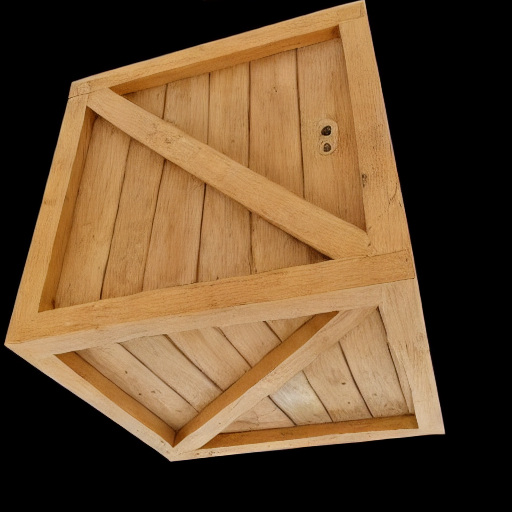}&%
        \includegraphics[height=0.14\linewidth]{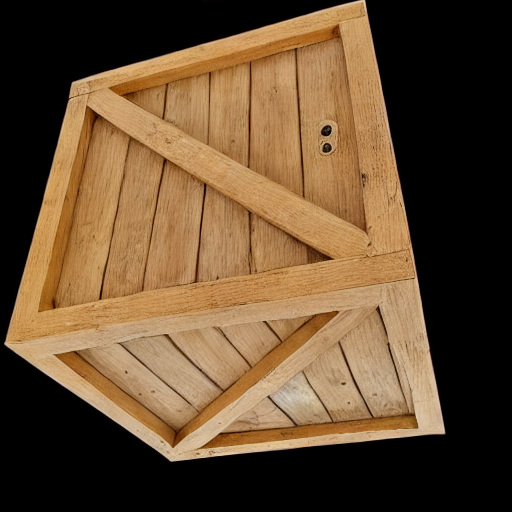}&%
        \includegraphics[height=0.14\linewidth]{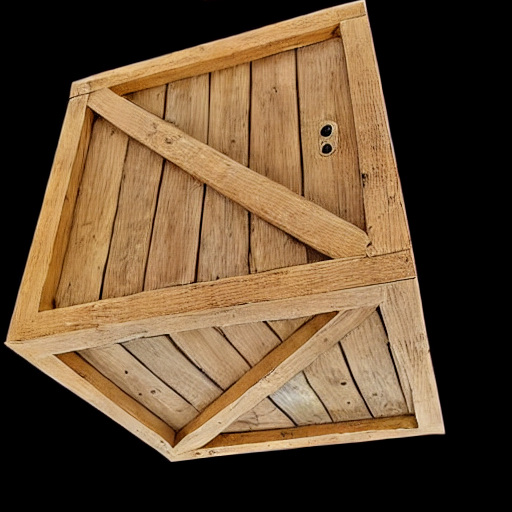}&%
        \includegraphics[height=0.14\linewidth]{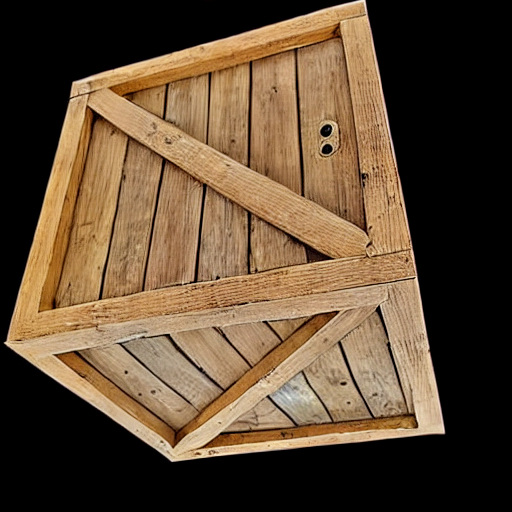}&%
        \includegraphics[height=0.14\linewidth]{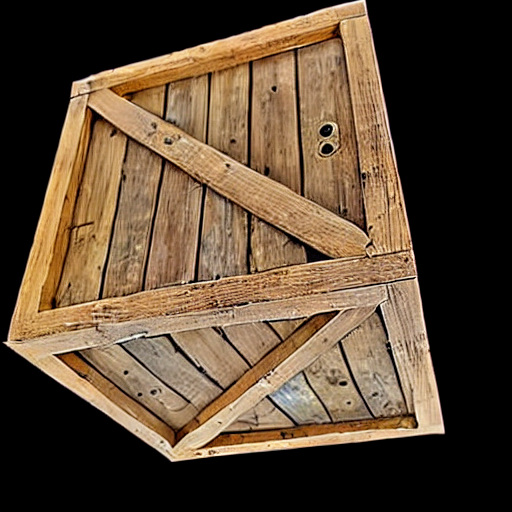}\\%
    \end{tabular}%

    \begin{tabular}{cccccccc}
        &Initial asset& \multicolumn{6}{c}{
            \begin{tikzpicture}
                \draw[<-] (0,0) -- (1,0);
                \draw[-] (2.6,0) -- (6.7,0);
                \draw[-] (8.3,0) -- (12,0);
                \draw[->] (14,0) -- (15,0);
                \node[above] at (1.8,-0.2) {initial lighting};
                \node[above] at (7.5,-0.2) {CNet Tile scale};
                \node[above] at (13,-0.2) {arbitrary lighting};
            \end{tikzpicture}
        }\\[-4pt]%
        && \textbf{1.0} & \textbf{0.8} & 0.6 & 0.4 & 0.2 & 0.0\\%
        \rotatebox{90}{\hspace*{4em}View 1}&%
        \includegraphics[height=0.14\linewidth]{figures/secondary_hparams/initial/0001.png}&%
        \includegraphics[height=0.14\linewidth]{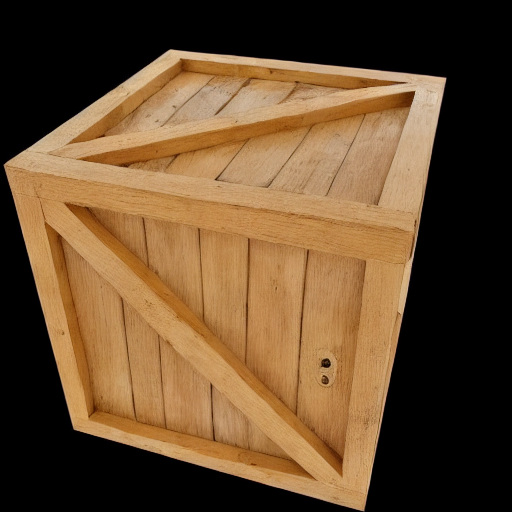}&%
        \includegraphics[height=0.14\linewidth]{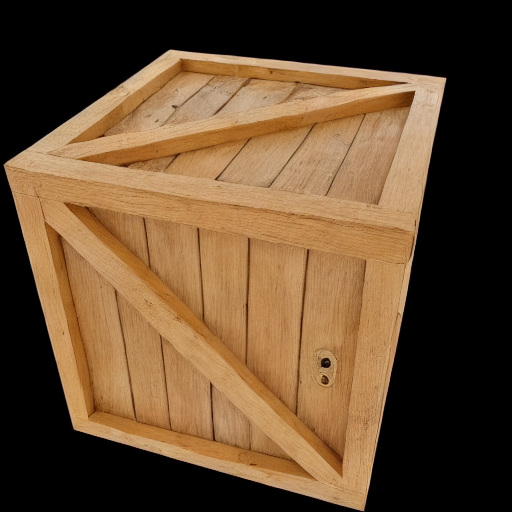}&%
        \includegraphics[height=0.14\linewidth]{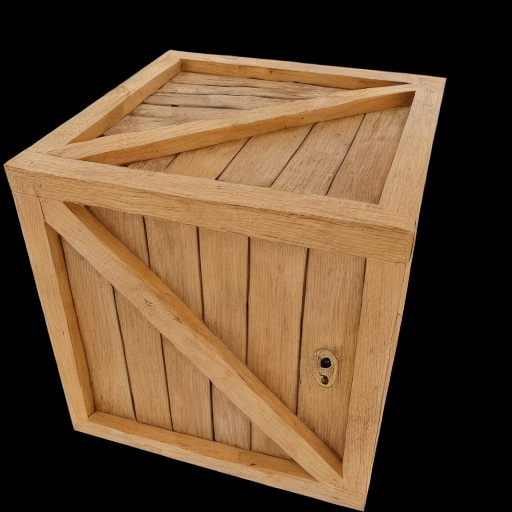}&%
        \includegraphics[height=0.14\linewidth]{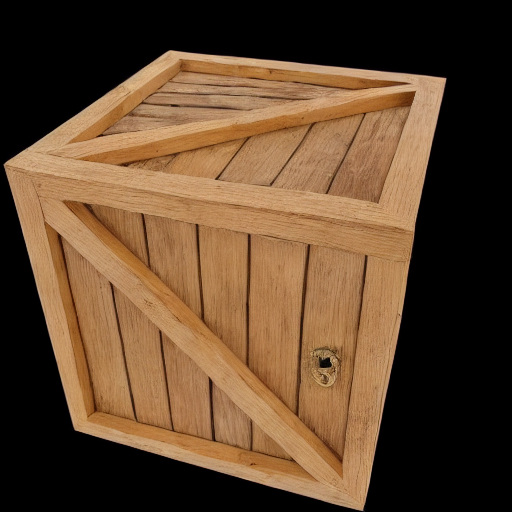}&%
        \includegraphics[height=0.14\linewidth]{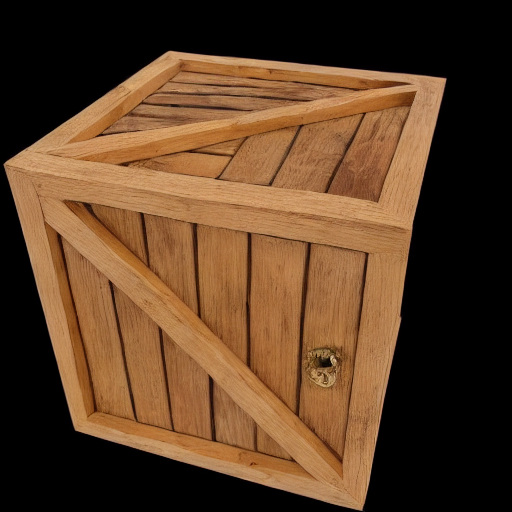}&%
        \includegraphics[height=0.14\linewidth]{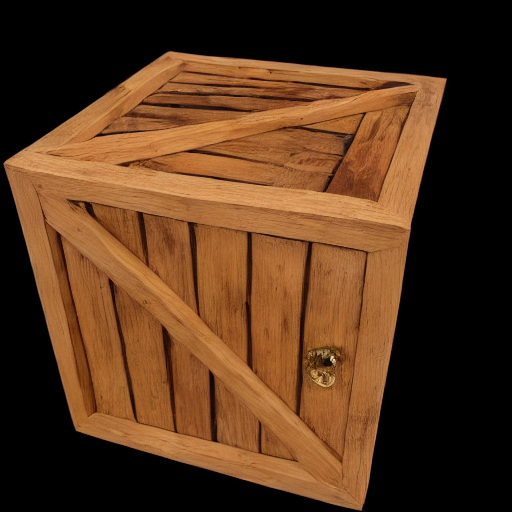}\\%
        \rotatebox{90}{\hspace*{4em}View 2}&
        \includegraphics[height=0.14\linewidth]{figures/secondary_hparams/initial/0006.png}&%
        \includegraphics[height=0.14\linewidth]{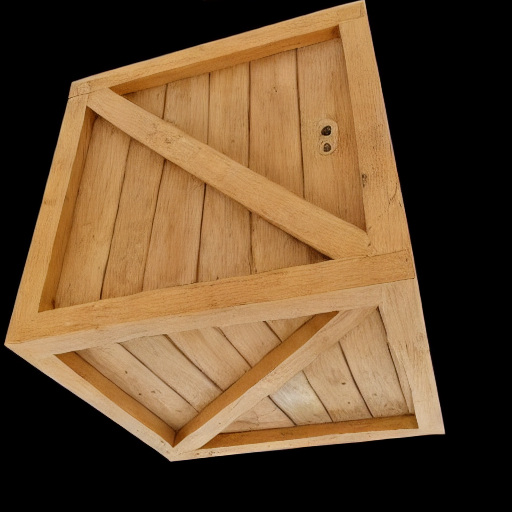}&%
        \includegraphics[height=0.14\linewidth]{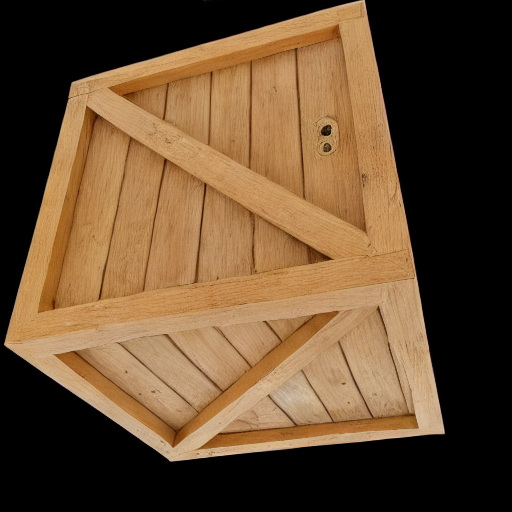}&%
        \includegraphics[height=0.14\linewidth]{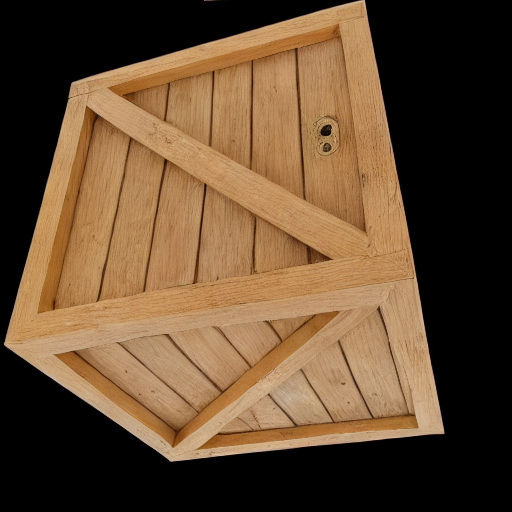}&%
        \includegraphics[height=0.14\linewidth]{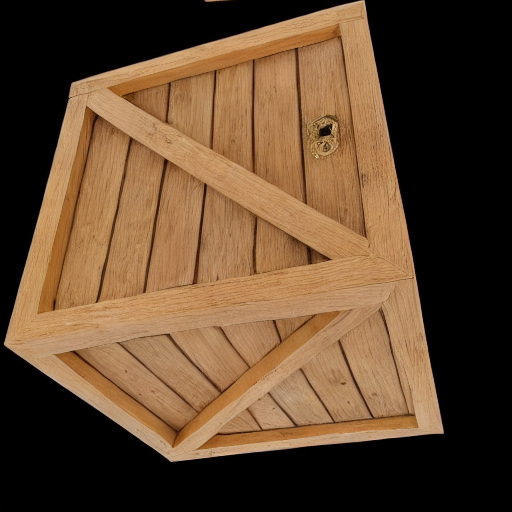}&%
        \includegraphics[height=0.14\linewidth]{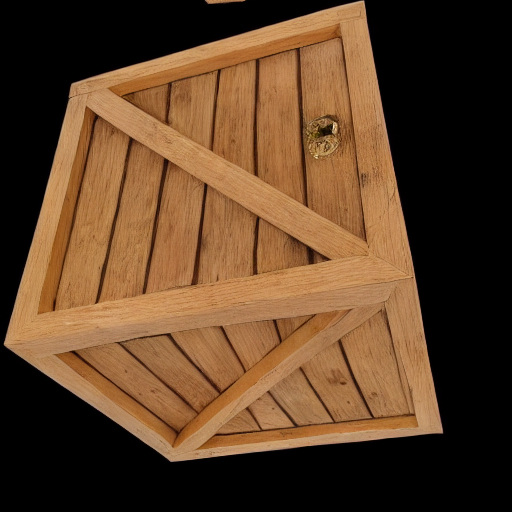}&%
        \includegraphics[height=0.14\linewidth]{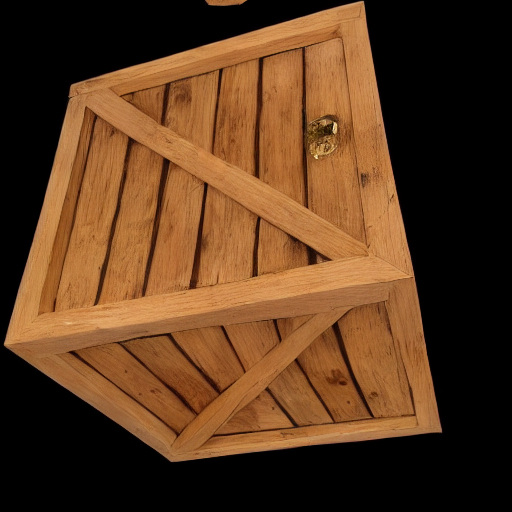}\\%
    \end{tabular}%

    \begin{tabular}{cccccccc}
        &Initial asset& \multicolumn{6}{c}{
            \begin{tikzpicture}
                \draw[<-] (0,0) -- (1,0);
                \draw[-] (2.6,0) -- (6.7,0);
                \draw[-] (8.3,0) -- (12,0);
                \draw[->] (14,0) -- (15,0);
                \node[above] at (1.8,-0.2) {weak edits};
                \node[above] at (7.5,-0.2) {Added noise};
                \node[above] at (13,-0.2) {strong edits};
            \end{tikzpicture}
        }\\[-4pt]%
        && 0.0 & 0.2 & 0.4 & 0.6 & 0.8 & \textbf{1.0}\\%
        \rotatebox{90}{\hspace*{4em}View 1}&%
        \includegraphics[height=0.14\linewidth]{figures/secondary_hparams/initial/0001.png}&%
        \includegraphics[height=0.14\linewidth]{figures/secondary_hparams/initial/0001.png}&%
        \includegraphics[height=0.14\linewidth]{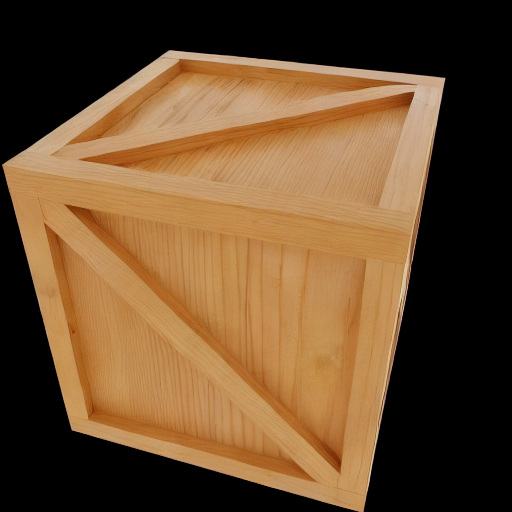}&%
        \includegraphics[height=0.14\linewidth]{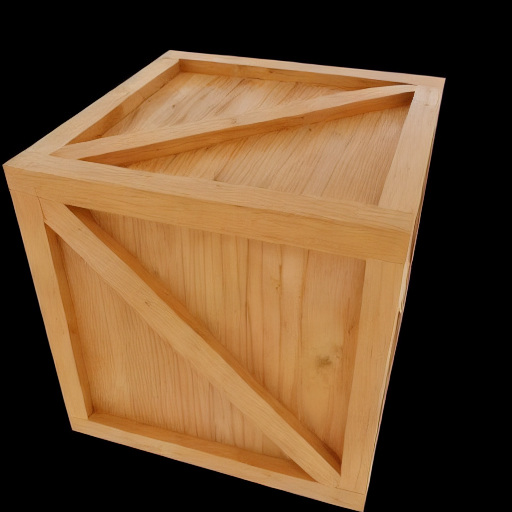}&%
        \includegraphics[height=0.14\linewidth]{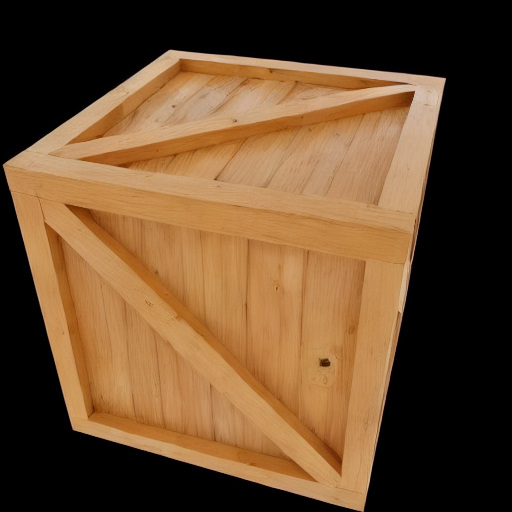}&%
        \includegraphics[height=0.14\linewidth]{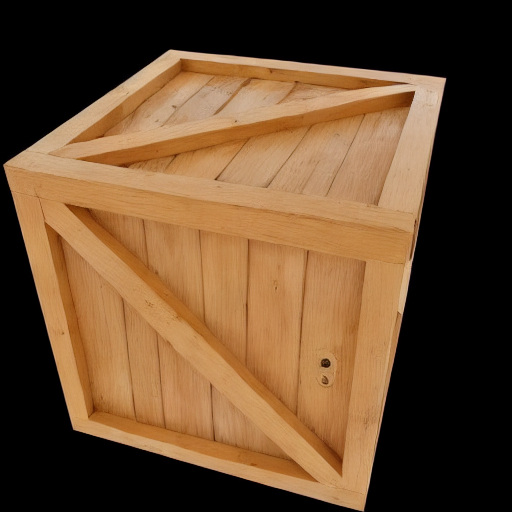}&%
        \includegraphics[height=0.14\linewidth]{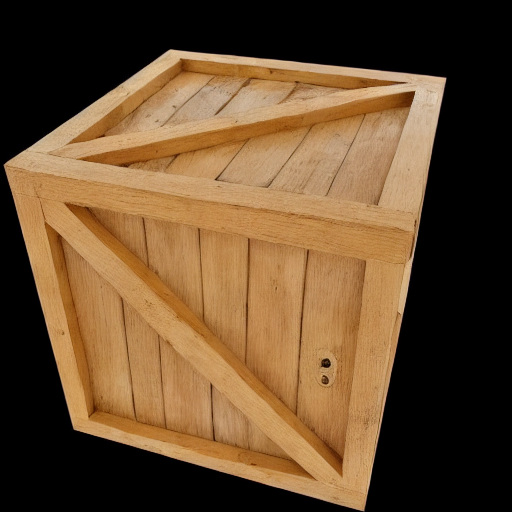}\\%
        \rotatebox{90}{\hspace*{4em}View 2}&
        \includegraphics[height=0.14\linewidth]{figures/secondary_hparams/initial/0006.png}&%
        \includegraphics[height=0.14\linewidth]{figures/secondary_hparams/initial/0006.png}&%
        \includegraphics[height=0.14\linewidth]{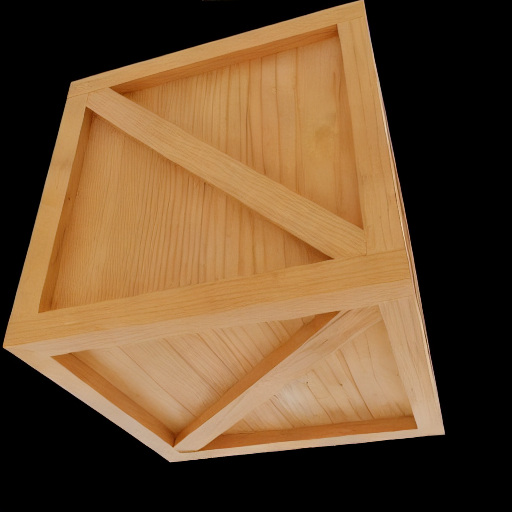}&%
        \includegraphics[height=0.14\linewidth]{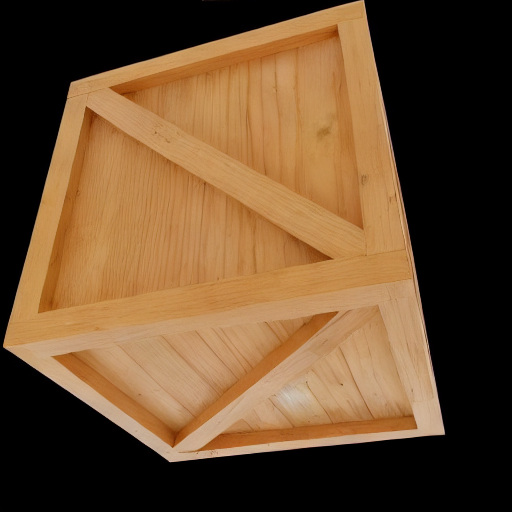}&%
        \includegraphics[height=0.14\linewidth]{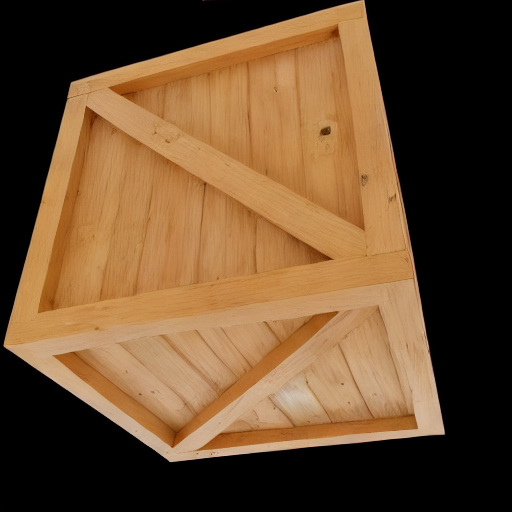}&%
        \includegraphics[height=0.14\linewidth]{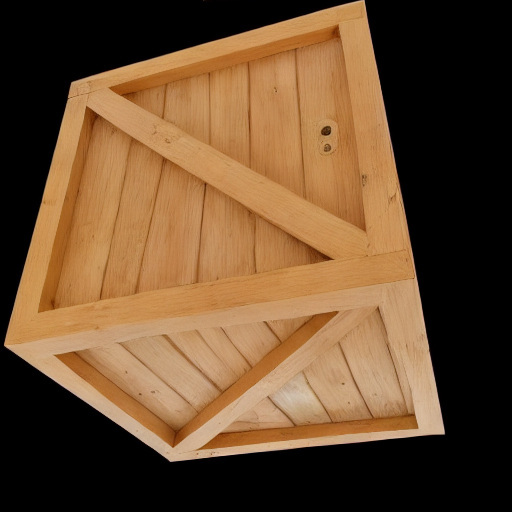}&%
        \includegraphics[height=0.14\linewidth]{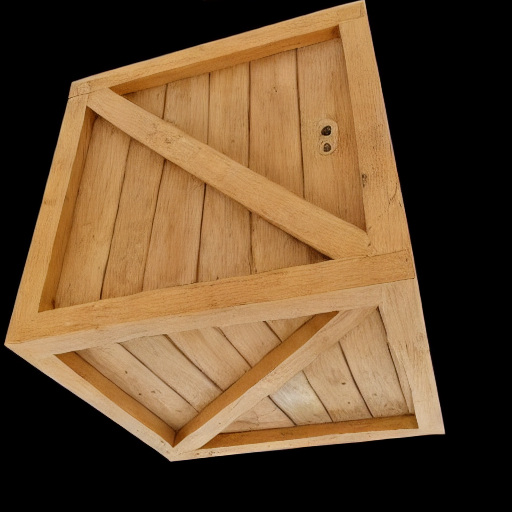}\\%
    \end{tabular}%

    \caption{Expanded version of Figure~\ref{fig:hparams_usercontrol} from the main paper. Apart from the \emph{classifier-free guidance} parameter (top), the underlying diffusion models of our system expose two further parameters that can be tweaked by users. Both the strength at which the \emph{ControlNet tile} (middle) and the \emph{input noise} (bottom) are applied allow tweaking how much the generated material details will deviate from the base material. Numbers highlighted in bold indicate parameter ranges we use throughout the results in the paper.}
    \label{fig:hparams_usercontrol_suppl}
\end{figure*}

\end{document}